\documentclass[acmlarge]{acmart}
\AtBeginDocument{%
  }

\setcopyright{acmcopyright}
\copyrightyear{2023}
\acmYear{2023}
\acmDOI{XXXXXXX.XXXXXXX}

\acmJournal{TOCHI}
\acmVolume{0}
\acmNumber{0}
\acmArticle{0}
\acmMonth{0}

\graphicspath{{./Figures/}}

\usepackage{fancyvrb,fvextra}

\newcommand{\revise}[1]{\textcolor{black}{#1}}

\begin{document}

\title{The Value of Interaction in Data Intelligence}

\author{Min Chen}
\email{min.chen@oerc.ox.ac.uk}
\orcid{0000-0001-5320-5729}
\affiliation{%
  \institution{University of Oxford}
  \country{UK}
}

\renewcommand{\shortauthors}{Min Chen}

\begin{abstract}
In human computer interaction (HCI), it is common to evaluate the value of HCI designs, techniques, devices, and systems in terms of their benefit to users.
It is less common to quantify the benefit of HCI to computers. Every HCI task allows a computer to receive some data from the user.
In many situations, the data received by the computer embodies human knowledge and intelligence in handling complex problems, and/or some critical information without which the computer cannot proceed.
In this paper, we present an information-theoretic framework for quantifying the knowledge received by the computer from its users via HCI.
We apply information-theoretic measures to some common HCI tasks as well as HCI tasks in complex data intelligence processes.
We formalize the methods for estimating such quantities analytically and measuring them empirically.
Using theoretical reasoning, we can confirm the significant but often undervalued role of HCI in data intelligence workflows.
\end{abstract}

\ccsdesc[500]{Human-centered computing~HCI theory, concepts and models}
\ccsdesc[500]{Mathematics of computing~Information theory}
\ccsdesc[300]{Information systems~Data analytics}
\ccsdesc[100]{Human-centered computing~Visualization theory, concepts and paradigms}

\keywords{Human-computer interaction, information theory, cost-benefit, interaction, knowledge, visualization, data intelligence.}


\maketitle

\begin{figure}[ht]
  \centering
    \hspace{10mm}\includegraphics[width=140mm]{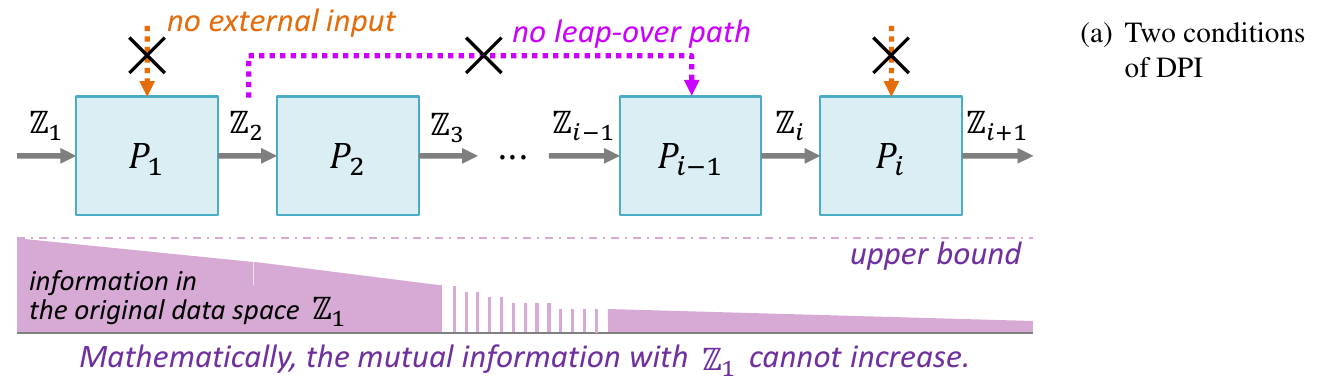}\\
    \hspace{10mm}\includegraphics[width=140mm]{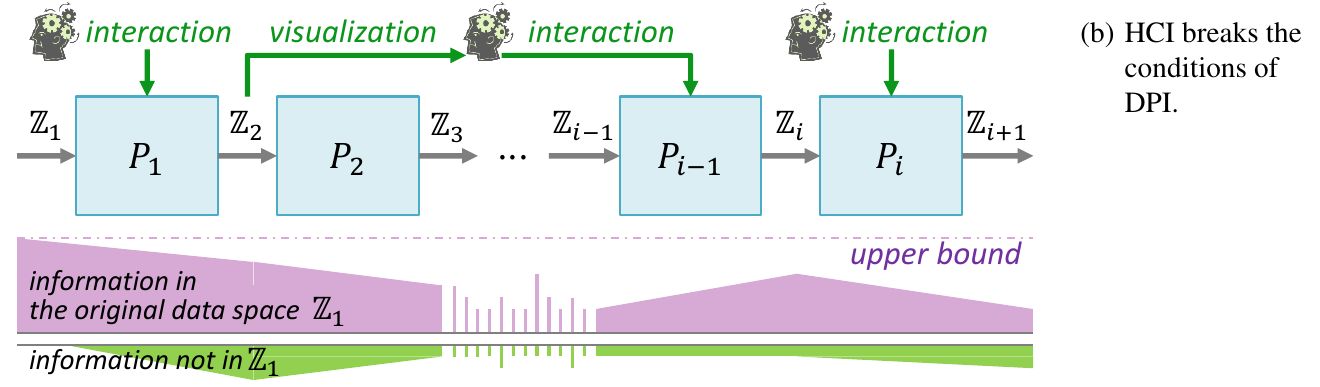}
  \caption{\emph{Data processing inequality} (DPI) is a mathematical theorem that confirms and underlines the challenge of information loss
in almost all fully automatic data intelligence workflows except some simple ones with very small data spaces.
Human computer interaction (HCI) can provide cost-beneficial means to alleviate the problems due to DPI.}
  \label{fig:DPI}
\end{figure}

\section{Introduction}
\label{sec:Intro}
\emph{Data intelligence} is an encompassing term for processes that transform data to decisions or knowledge, such as statistical inference,  algorithmic analysis, data visualization, machine learning, business intelligence, numerical modelling, computational simulation, prediction, and decision making.
While many of these processes are propelled by the desire for automation, human-computer interaction (HCI) has been and is still playing valuable roles in almost all nontrivial data intelligence workflows.
However, the benefits of HCI in a data intelligence workflow are often much more difficult to measure and quantify than its costs and disadvantages.
This inevitably leads to a more fervent drive for replacing humans with machines in data intelligence.

In \emph{information theory}, the \emph{Data Processing Inequality} (DPI) is a proven theorem. It states that fully automated processing of data can only lose but not increase information.
As Cover and Thomas explained \cite{Cover:2006:book}, ``\emph{No clever manipulation of data can improve the inferences that can be made from the data}.''
In most data intelligence workflows, since the original data space contains much more variations (in terms of entropy) than the decision space, the loss of information is not only inevitable but can also be very significant \cite{Chen:2016:TVCG}.

In the context of data visualization, Chen and J{\"a}nicke first pointed out that HCI alleviates the undesirable bottleneck of DPI because the mathematical conditions for proving DPI are no longer satisfied with the presence of HCI.
As illustrated in Fig. \ref{fig:DPI}(a), the proof of DPI assumes that (i) each process in a workflow must receive data only from its proceeding process, and (ii) the output of a process must depend only on the output of its proceeding process.
As illustrated in Fig. \ref{fig:DPI}(b), any human inputs based on human knowledge (e.g., the variation of context and task) violate the first condition.
Meanwhile any human inputs based on observing the previous processes in the workflow (e.g., the details being filtered out or aggregated) violate both conditions.
Therefore, if we can quantitatively estimate or measure the amount of information passing from humans to the otherwise automated processes, we can better appreciate the \textbf{value} of interaction in data intelligence.

In this paper, we present an information-theoretic framework for measuring the knowledge received by a computational process from human users via HCI.
It includes several fundamental measures that can be applied to a wide range of HCI modalities as well as the recently discovered cost-benefit measure for analyzing data intelligence workflows \cite{Chen:2016:TVCG,Chen:2022:E1,Chen:2022:E2}.
We describe the general method for estimating the amount of human knowledge delivered \revise{to computers} using HCI.
\revise{We demonstrate that the method works for knowledge featured in both low- and high-level HCI operations, including those in complex data intelligence workflows (e.g., machine learning).}  
We outline the general design for an empirical study to detect and measure human knowledge used in data intelligence.
\revise{By exploring the quantifiable value of HCI from the perspective of the benefits received by computers, we can appreciate the necessity and importance of HCI better.} 


\section{Related Work}
\label{sec:RelatedWork}
\revise{As Hornb{\ae}k and Oulasvirta discussed in their article \cite{Hornbaek:2017:CHI}, there are many types of interaction, and many different definitions of interaction. In this paper, we focused on one major category of HCI, namely the information transactions from humans to computers \cite{Saffer:2009:book}. The other major category of HCI is the information transactions from computers to humans, for which a related paper by Chen and Golan \cite{Chen:2016:TVCG} provides an information-theoretic discussion on data visualization.} 

In the field of HCI, the term of ``value'' has several commonly-used referents \revise{\cite{Friedman:1996:I,Light:2005:CHI,Gilmore:2008:CHI,Rotondo:2010:CHI}}, including
(a) worth in usefulness, utility, benefit, merit, or importance \revise{(e.g., \cite{Gilmore:1995:book,Cockton:2004:NordicCHI,Shneiderman:2010:book,Smith:2014:OzCHI})},
(b) monetary, material, developmental, or operational cost \revise{(e.g., \cite{Donoghue:2002:book,Bias:2005:book})},
(c) a quantity that can be measured, estimated, calculated, or computed \revise{(e.g., \cite{Shneiderman:2010:book})}, and
(d) a principle or standard in the context of moral or ethics (e.g., \revise{\cite{Light:2005:CHI})}.
In this paper, we examine the value of HCI processes primarily in terms of (a) and (c) with some supplemental discussions on (b).
Readers who are interested in (d) may consult other works in the literature (e.g., \cite{Shilton:2018:FT,Giaccardi:2011:I,Smith:2014:OzCHI,Rotondo:2010:CHI}).

Most research effort in HCI has been devoted to bring about the usefulness and benefits to humans.
The goals of HCI and the criteria for good HCI are typically expressed as
``\emph{support people so that they can carry out their activities productively and safely}'' \cite{Preece:1994:book};
``\emph{effective to use, efficient to use, safe to use, having good quality, easy to learn, easy to remember how to use}'' \cite{Preece:2015:book};
``\emph{time to learn, speed of performance, rate of errors by users, retention over time, subjective satisfaction}'' \cite{Shneiderman:2010:book}; and
``\emph{useful, usable, used}'' \cite{Dix:2003:book}.

\revise{Yi et al. examined different types of interaction in information visualization and considered seven categories of interaction (i.e., select, explore, reconfigure, encode, abstract, filter, and connect) \cite{Yi:2007:TVCG}. Most definitions started with the phrase ``show me'', implying the benefits to users. In the visualization literature, there are other categorizations or taxonomies of interaction, including those by Tweedie \cite{Tweedie:1997:CHI}, Dix and Ellis \cite{Dix:1998:AVI}, Sedig and Parsons \cite{Sedig:2013:THCI}, von Landesberger et al. \cite{Landesberger:2014:book}, and Wybrow et al. \cite{Wybrow:2014:book}. This work is concerned with information transactions from humans to computers, which include, but are not limited to, interaction in visualization processes.}

In general, it is less common to discuss the usefulness and benefits of HCI to computers.
While there is little doubt that the ultimate goal is for computers to assist humans, it will be helpful to understand and measure how much a computer needs to be assisted by its users before becoming useful, usable, and used.
It is hence desirable to complement the existing discourses on value-centered designs (e.g., \cite{Friedman:1996:I,Cockton:2004:NordicCHI,Bias:2005:book,Light:2005:CHI,Rotondo:2010:CHI}) by examining the value of HCI from the perspective of computers.

\revise{When we consider information received by computers, it becomes easier to estimate such ``value'' quantitatively.}
We can develop quantitative methods for measuring and estimating how much a computer needs to know since we can investigate the inner ``mind'' of a computer program more easily than that of human users.
The field of HCI has benefited from a wide-range of quantitative and qualitative research methods \cite{Helander:1997:book,Diaper:2003:book,Cairns:2008:book,Lazar:2010:book,Purchase:2012:book,Jacko:2012:book,MacKenzie:2013:book,Oulasvirta:2018:book,Norman:2018:book}, including quantitative methods such as formal methods, statistical analysis, cognitive modelling, and so on.
This work explores the application of information theory \cite{Shannon:1948:BSTJ,Cover:2006:book} in HCI.

Claude Shannon's landmark article in 1948 \cite{Shannon:1948:BSTJ} signifies the birth of information theory.
It has been underpinning the fields of data communication, compression, and encryption since.
Its applications include physics, biology, neurology, psychology, and computer science (e.g., visualization, computer graphics, computer vision, data mining, and machine learning).
\revise{Inspired by Fitts's model of human movement \cite{Fitts:1954:JEP,Fitts:1967:book}, HCI researchers have utilized a few information-theoretic measures to define quantitative models that can be estimated based on experimental data and used for making predictions (e.g., \cite{Card:1978:E,MacKenzie:1992:CHI}). Soukoreff et al. analyzed the analogical semantics of three measures, namely self-information $\log_2(x)$, Shannon's index of difficulty $\log_2(x+1)$, and Shannon entropy $-\sum (x \log_2 x)$ \cite{Soukoreff:2011:INTERACT}. Gori et al. \cite{Gori:2018:TOCHI} conducted further analysis on a number of variants of these three measures. 
Because of the analogy of a simple communication model, this family of models focused mainly on physiological phenomena in HCI, as it would be challenging to relate the analogy to many cognitive phenomena in HCI (e.g., selecting an option in response to a prompt, a colleague's advice and/or other contextual information).}

\revise{Since Bayes' theorem was formulated some 260 years ago, 
Bayesian inference has become a widely used statistical method in many disciplines. In HCI, researcher proposed to use Bayesian models in supporting HCI, e.g., onboard
a robotic wheelchair \cite{Atrash:2009:IUI}, navigation \cite{Liu:2017:CHI}, file retrieval \cite{Liu:2018:CHI}, and pointing movement \cite{Zhao:2022:UIST}. Researchers also studied whether and how HCI could play a role in semi-automated Bayesian inference processes \cite{Tsai:2011:PHFESAM,Micallef:2012:TVCG,Ottley:2016:TVCG,Mosca:2021:CHI}.
For some machine learning models, such as decision trees and Bayesian networks, one type of relative entropy (referred to as information gain in recent years) has been used to quantify the amount of entropy reduction in each processing step. This measure is indeed part of the formulation and proof of the DPI theorem as mentioned in Section \ref{sec:Intro}. Its use is subject to the conditions of a Markov chain \cite{Cover:2006:book}, which were broken in a typical chain of HCI actions \cite{Chen:2010:TVCG}.
}

\revise{This work focuses on a relatively new information-theoretic measure, cost-benefit ratio, which was} first proposed in 2016 \cite{Chen:2016:TVCG} in the context of visualization and visual analytics. An improvement was proposed recently to make the interpretation of the numerical quantification more consistent with practical observations \cite{Chen:2022:E1,Chen:2022:E2}. \revise{The measure does not assume the conditions of a Markov chain, facilitating informative measurement of a chain of human-centric processes in conjunction with machine-centric processes. The measure also includes sub-measures about errors and costs, offering a new tool for measuring, estimating, and evaluating the cost-benefit of HCI actions.}

The cost-benefit measure was used to prove mathematically the correctness of a major wisdom in HCI, ``\emph{Overview first, zoom, details-on-demand}'' \cite{Shneiderman:1996:VL,Chen:2010:TVCG,Chen:2016:book}, and to analyze the cost-benefit of different virtual reality applications \cite{Chen:2019:TVCG}.
Two pieces of previous work showed that the measure can be estimated in practical applications \cite{Tam:2017:TVCG} and be measured using empirical studies \cite{Kijmongkolchai:2017:CGF}. Other work demonstrated the uses of the measure in qualitative analysis of data intelligence workflows in general \cite{Chen:2019:CGF}, and in specific applications, e.g., in improving the visualization for ensemble machine learning \cite{Ye:2023:TVCG} and sensitive analysis of epidemiological models \cite{Rydow:2023:TVCG}.
While information theory has been applied successfully to visualization (i.e., interaction from computers to humans), this work focuses on the other direction of HCI (i.e., from humans to computers).
\revise{In particular, we demonstrate the applications this measure in different aspects of HCI, including low-level HCI actions (e.g., option selection) and high-level HCI actions (e.g., human decisions in machine learning workflows).
We present methods for obtaining such measures in empirical studies and estimating such measures analytically.}

\revise{Jankun-Kelly et al. proposed a model for interactive and exploratory visualization \cite{Jankun-Kelly:2007:TVCG}. Jansen and Dragicevic proposed a model for interaction with physical visualization objects and apparatus \cite{Jansen:2013:TVCG}. Lam provided a categorization of the cost during interaction and visualization \cite{Lam:2008:TVCG}. The cost-benefit measure discussed in the work can be applied to these interaction models and these types of costs. Nevertheless, conducting such model-specific analysis will require more focused research effort in the future.}

\section{Fundamental Measures}
\label{sec:Measures}
From every human action through a user interface or an HCI device, a computer receives some data, which typically encodes information that a running process on the computer wishes to know, and cannot proceed without. Through such interactions, humans transfer their knowledge to computers. In some cases, the computers learn and retain part of such knowledge (e.g., preference setting and annotation for machine learning). In many other cases, the computers asked blithely for the same or similar information again and again.

In HCI, we all appreciate that measuring the usefulness or benefits of HCI to humans is not a trivial undertaking.
In comparison, the amount of knowledge received by a computer from human users via HCI can be measured relatively easily.
Under an information-theoretic framework, we can first define several fundamental measures about the information that a computer receives from an input action.
We can then use these measures to compare different types of interaction mechanisms, e.g., in terms of the capacity and efficiency for a computer to receive knowledge from users.

\subsection{Alphabet and Letter}
When a running process on a computer pauses to expect an input from the user, or a thread of the process continuingly samples the states of an input device, all possible input values that can be expected or sampled are valid values of a univariate or multivariate variable.
In information theory, this mechanism can be considered in abstraction as a communication \emph{channel} from a user to a computational process.
This variable is referred to as an \emph{alphabet}, $\mathbb{Z}$, and these possible values are its \emph{letters}, $\{z_1, z_2, \ldots, z_n\}$.

In a given context (e.g., all uses of an HCI facility), each letter $z_i \in \mathbb{Z}$ is associated with a probability of occurrence, $p(z_i)$.
Before the process receives an input, the process is unsure about which letter will arrive from the channel.
The amount of uncertainty is typically measured with the Shannon entropy \cite{Shannon:1948:BSTJ} of the alphabet:
\begin{equation}%
\label{eq:Shannon}
	\mathcal{H}(\mathbb{Z}) = - \sum_{i=1}^n p(z_i) \log_2 p(z_i) \quad [\text{unit: bit}]
\end{equation}
%

We can consider alphabets broadly from two different perspectives, the \emph{input device} and the \emph{input action}, which are detailed in the following two subsections.

\subsection{Input Device Alphabet}
An \emph{input device alphabet} enumerates all possible states of a physical input device, which can be captured by a computational process or thread through sampling.
Such devices include keyboard, mouse, touch screen, joystick, game controller, VR glove, camera (e.g., for gestures), microphones (e.g., for voices), and many more.
Most of these devices feature multivariate states, each of which is a letter in the input device alphabet corresponding to the device.

For example, the instantaneous state of a simple 2D pointing device may record four values: its current location in $x$-$y$ relative to a reference point, the activation status of its left and right buttons.
The instantaneous state of a conventional keyboard may consist of 80-120 variables, one for each of its keys, assuming that simultaneous key activations are all recorded before being sequentialized and mapped to one or more key inputs in an input action alphabet (to be discussed in the next subsection).

The design of an input device may involve many human and hardware factors.
Among them, the frequency or probability, in which a variable (e.g., a key, a button, a sensor, etc.) changes its status, is a major design consideration.
Hence, this particular design consideration is mathematically underpinned by the measurement of Shannon entropy.
The common wisdom is to assign a lower operational cost (e.g., speed and convenience) to a variable that is more frequently changed (e.g., a more frequently-used button).
This is conceptually similar to entropic coding schemes such as Huffman encoding in data communication \cite{Huffman:1952:IRE}.

However, the sampling mechanism for an input device usually assumes an equal probability for all of its variables.
From the perspective of the device, any variables may change at any moment, and all letters (i.e., states) in the alphabet have the same probability.
This represents the maximal uncertainty about what is the next state of the input device, as well as the maximal amount of information that the device can deliver.
For an input device alphabet $\mathbb{Z}$ with $n$ letters and each letter has a probability of $1/n$, this maximal quantity is the maximum of the Shannon entropy in Eq.\,(\ref{eq:Shannon}), i.e., $\mathcal{H}_\text{max} = \log_2 n = \log_2 \|\mathbb{Z}\|$.
We call this quantity the \emph{instantaneous device capacity} and we denote it as $\mathcal{C}_\text{dev}$.

For example, for a simple 2D mouse with 2 on-off buttons, operating in conjunction with a display at a 1920$\times$1080 resolution, its instantaneous device capacity is:
\begin{align*}
	\mathcal{C}_\text{dev} = \mathcal{H}_\text{max} =&
	\log_2 1920 + \log_2 1080 + \log_2 2 + \log_2 2 \\
	\approx& 10.907 + 10.077 + 1 + 1 = 22.984 \; \text{bits}
\end{align*}

While the notion of instantaneous device capacity may be useful for characterizing input devices with which a sampling process has to be triggered by a user's action, it is not suited for input devices with a continuing sampling process (e.g., a video camera for gesture recognition).
Hence a more general and useful quantity for characterizing all input devices is to define the maximal device capacity over a unit of time.
We use ``unit: \emph{second}'' for the unit of time in the following discussions.
Let $\tau$ be the sampling rate, that is, maximal number of samples that a process can receive from an input device within a second.
Assuming that the instantaneous device capacity of the device is invariant for each sample, the \emph{bandwidth} (cf. bandwidth in data communication) of the device is defined as:
\[
	\mathcal{W}_\text{dev} = \tau \times \text{instantaneous device capacity} \quad [\text{unit: \emph{bit/s}}].
\]
\noindent Note: while the instantaneous device capacity is measured in \emph{bit}, the bandwidth, $\mathcal{W}_\text{dev}$, is measured in \emph{bit per second}.

For example, if the sampling rate of the aforementioned mouse is 100 Hz, then its bandwidth is $\mathcal{W}_\text{dev} \approx 2,298.4$ bits/s.
Similarly, consider a data glove with 7 sensors with a sampling rate of 200Hz. If its five sensors for finger flexure have 180 valid values each and the pitch and roll sensors have 360 valid values each, its bandwidth is:
\[
	\mathcal{W}_\text{dev} = 200 \big( 5 \log_2 180 + 2 \log_2 360 \big) \approx 10,888.6 \; \text{bits/s}. 
\]

\subsection{Input Action Alphabet}
An \emph{input action alphabet} enumerates all possible actions that a user can perform for a specific HCI task in order to yield an input meaningful to the computer.
Here the phrase ``a specific HCI task'' stipulates the condition in which the user is aware of what the computer wants to know, e.g., through a textual instruction or a visual cue on the screen or through context awareness based on previous experience or acquired knowledge.
The phrase ``meaningful to the computer'' stipulates the condition in which an action that the computer is not programmed to handle for the specific HCI task should not be included in the input action alphabet.

Consider a simple HCI task of making a selection out of $k$ \textbf{radio buttons}. (Multiple-choice buttons can also be included in this consideration.)
Assuming that selecting nothing is not meaningful to the computer, the corresponding input action alphabet is: $\mathbb{A}_{\text{radio}} = \{\text{option}_1, \text{option}_2, \ldots, \text{option}_k \}$.
When each option is associated with a binary bit in an implementation, the letters in the alphabet can be encoded as a set of $k$-bit binary codewords:
$\{00{\cdot\cdot}001, 00{\cdot\cdot}010, \ldots, 10{\cdot\cdot}000\}$.
If all $k$ options are equally probable, the entropy of the alphabet is $\log_2 k$.
A selection action by the user thus allows the computer to remove $\log_2 k$ bits of uncertainty, or in other words, to gain $\log_2 k$ bits of knowledge from the user for this specific HCI task.
We call this quantity the \emph{action capacity} of the HCI task, which is denoted as $\mathcal{C}_\text{act}$.

\begin{figure}[ht]
  \centering
  \begin{tabular}{@{}c@{\hspace{8mm}}c@{}}
    \includegraphics[height=36mm]{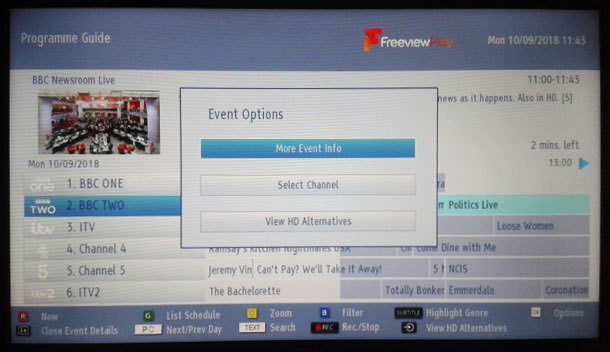} &
    \includegraphics[height=36mm]{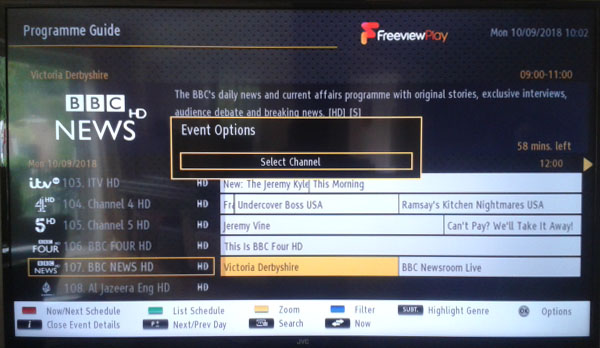}\\
    (a) a 3-letter alphabet & (b) a 1-letter alphabet
  \end{tabular}
  \caption{After selecting a channel from a TV listing, a TV set typically prompts a few options for a user to decide.}
  \label{fig:TVList}
\end{figure}

In practice, radio buttons featured in many HCI tasks do not have the same probability of being selected.
For example, as shown in Fig. \ref{fig:TVList}(a), after selecting a channel from a list of current shows, the TV displays an input action alphabet $\mathbb{A}$ with three options, $a_1$: ``More Event Info'', $a_2$: ``Select Channel'', and $z_3$: ``View HD Alternatives''.
The probability of $a_1$ depends on several statistical factors, e.g., how informative is the title in the list, how many users prefer to explore a less-known program via investigational viewing verse how many prefer reading detailed information, and so on.
The probability of $a_3$ depends on how often a user selects a non-HD channel from a TV listing with an intention to view the corresponding HD channel.
Different probability distributions for $\mathbb{A}$ will lead to different amount of knowledge $\mathcal{H}(\mathbb{A}) = \mathcal{C}_\text{act}$.
As exemplified by the instance below, the more skewed a distribution is, the less the knowledge is worth or the less action capacity that the HCI task has:
\begin{align}
\label{eq:CactTV}
p(a_1) = p(a_2) = p(a_3) = 1/3 \longrightarrow&\, \mathcal{C}_\text{act}(\mathbb{A}_a) \approx 1.58 \notag \\
p(a_1) = 0.2, p(a_2) = 0.7, p(a_3) = 0.1 \longrightarrow&\, \mathcal{C}_\text{act}(\mathbb{A}_b) \approx 1.16\\
p(a_1) = 0.09, p(a_2) = 0.9, p(a_3) = 0.01 \longrightarrow&\, \mathcal{C}_\text{act}(\mathbb{A}_c) \approx 0.52 \notag
\end{align}
\noindent When the probability of a letter in an alphabet becomes 1, the alphabet no longer has any uncertainty.
As shown in Fig. \ref{fig:TVList}(b), if a TV offers only one optional answer, the device capacity of the corresponding alphabet, $\mathcal{C}_\text{act}$, is of 0 bits.
We will revisit this example in Section \ref{sec:EstimateCBR}.

Similarly, we can apply entropic analysis to \textbf{check boxes}. Consider an input action alphabet $\mathbb{A}_{\text{checkbox}}$ that consists of $k$ check boxes.
There are $m = 2^k$ possible combinations: $\mathbb{A}_{\text{checkbox}} =$ $\{ \text{combination}_1, \text{combination}_2, \ldots, \text{combination}_m \}$.
The alphabet can be encoded using a $k$-bit code, ${b_1}{b_2}{b_3}\cdots{b_k}$, where each bit, $b_j, 1 \le j \le k$, indicates whether the corresponding checkbox is on or off.
If all combinations have the equal probability, the amount of knowledge that computer can gain from the user is $\mathcal{C}_\text{act}=k$ bits, which is also the maximum entropy of the alphabet. Note that $k > \log_2 k$ when $k > 1$, indicating that $k$ checkboxes have more action capacity than $k$ radio buttons except that they have the same action capacity when $k=1$ (assuming that both are allowed to have on-off states).

We now examine a more complicated type of input actions.
Consider an HCI task for drawing a \textbf{freehand path} using a 2D pointing device.
Assume the following implementation constraints:
(i) the computer can support a maximum $m$ sampling points for each path;
(ii) the drawing canvas is a rectangular area $[x_\text{min}, x_\text{max}] \times [y_\text{min}, y_\text{max}]$;
(iii) the points along the path are sampled at a regular time interval $\Delta t$, though the computer does not store the time of each sample; and
(iv) the sampling commences with the first button-down event and terminates with the subsequent button-up event.

Let $\mathbb{A}_\text{freehand}$ be all possible paths that a user may draw using the 2D pointing device, and $\mathbb{A}^{(k)}_\text{freehand}$ be a subset of $\mathbb{A}_\text{freehand}$, consisting of all paths with $k$ points ($k \in [1,m]$).
The sub-alphabet $\mathbb{A}^{(k)}_\text{freehand}$ thus enumerates all possible paths in the form of
$(x_1, y_1), (x_2, y_2), \ldots, (x_k, y_k)$
where each point $(x_i, y_i)$ is within the rectangular area $[x_\text{min}, x_\text{max}] \times [y_\text{min}, y_\text{max}]$.
If it is possible to select any pixel in the rectangular area for every point input, the total number of possible paths, is
$[(x_\text{max}-x_\text{min}+1) \times (y_\text{max}-y_\text{min}+1)]^k$, which is also the size of the sub-alphabet $\mathbb{A}^{(k)}_\text{freehand}$.

For example, given a $512 \times 512$ rectangular area, the grand total number of possible paths is:
\[
	\|\mathbb{A}_\text{freehand}\| =
	\sum_{k=1}^m \|\mathbb{A}^{(k)}_\text{freehand}\| =
	\sum_{k=1}^m 2^{18k} \ge 2^{18m}
\]
\noindent If all paths have an equal probability, the maximal amount of knowledge that the computer can gain from a user's freehand drawing action is thus $\mathcal{C}_\text{act}=18$ bits when $m=1$, or slightly more than $\mathcal{C}_\text{act}=18m$ bits when $m>1$.
For an alphabet of possible paths with up to $m=20$ points, the maximal amount of knowledge, $\mathcal{C}_\text{act}$,  is slightly more than 360 bits.
This is similar to the amount of knowledge that a computer would gain from an HCI action involving $2^{360}=2.3\times10^{108}$ radio buttons or 360 checkboxes.
%

Many \textbf{data glove} devices come with a built-in gesture recognition facility.
The gestures that can be recognized by such a device are letters of an input action alphabet $\mathbb{A}_\text{gesture}$.
For example, an alphabet may consist of 16 elementary gestures (1 fist, 1 flat hand, and 14 different combinations of figure pointing).
The maximum entropy of this alphabet, i.e., the maximal amount of knowledge $\mathcal{C}_\text{act}$ that can be gained, is $\mathcal{H}_\text{max}(\mathbb{A}_\text{gesture}) = 4$ bits.
If a system using the data glove can recognize a more advanced set of gestures, each of which is comprised of one or two elementary gestures, the advanced alphabet consists of $16\times16$ letters.\footnote{Note: Repeating the same elementary gesture, e.g., ``fist'' + ``fist'', is considered as one elementary gesture due to the ambiguity in recognition.}
The maximum entropy is increased to 8 bits.
When we begin to study the probability distributions of the elementary gestures and the composite gestures, this is very similar to Shannon's study of English letters and their compositions in data communication \cite{Shannon:1951:BSTJ}.

\subsection{Input Device Utilization}
\begin{figure}
    \centering
    \includegraphics[width=\linewidth]{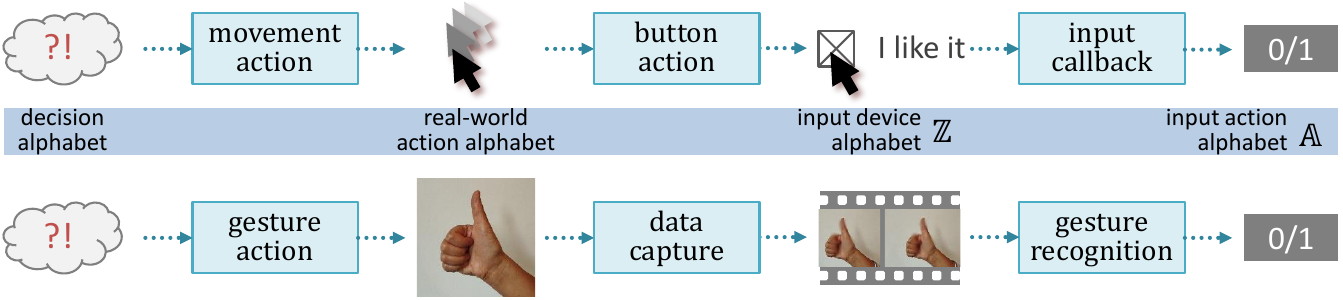}
    \caption{\revise{Two example sequences of transformations from human decisions and actions to a piece of binary data that represents the intended action for the computer know. The information space of this intended action is referred to as \emph{input action alphabet}. The human actions captured by the computer can sometimes be much more complex than the intended action (e.g., a video of a gesture). The information space of such data is referred to as \emph{input device alphabet}.}}
    \label{fig:Transform}
\end{figure}

\revise{Fig. \ref{fig:Transform} illustrates the difference between input device alphabet and input action alphabet. Sometimes, the data captured by an input device may be exactly the same as the data that represents the intended action precisely. For example, for a checkbox mechanism, the input device alphabet may be considered as the same as the input action alphabet. The input callback function is analogically similar to a communication \emph{channel}. Sometimes, the intended ``meaning'' of an action is much more complex. For example, for a gesture-based mechanism, a video may be captured. The gesture recognition function no longer behaves like a channel. In information theory, a more general term, \emph{transformation}, is used to encapsulate all processes for transforming one alphabet to another, i.e., including all those cyan rectangular boxes. The combination of a sequence of transformations is also referred to as a transformation (e.g., gesture action $\rightarrow$ data capture $\rightarrow$ gesture recognition). The individual processing steps in a transformation are also transformations (e.g., the small movement actions within a movement action).}

\begin{figure}[t]
  \centering
  \includegraphics[width=120mm]{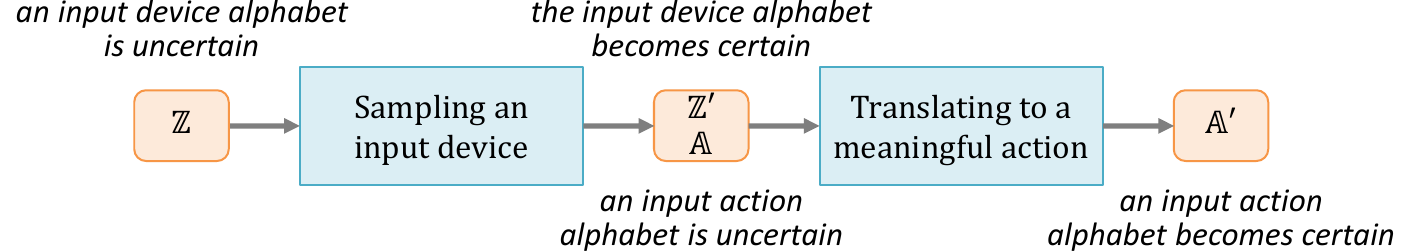}\\
  (a) The two alphabets during a computer receives an input.\\[2mm]
  \includegraphics[width=120mm]{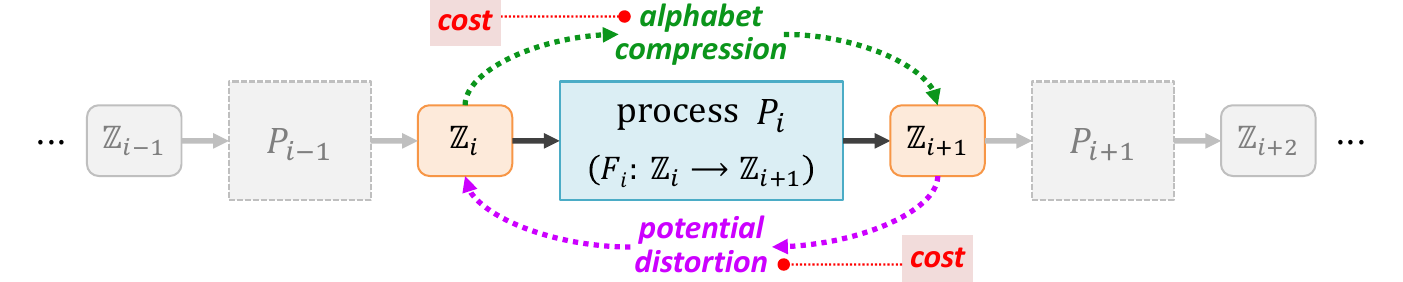}\\
  (b) The three abstract measures about a process.\\[2mm]
  \includegraphics[width=120mm]{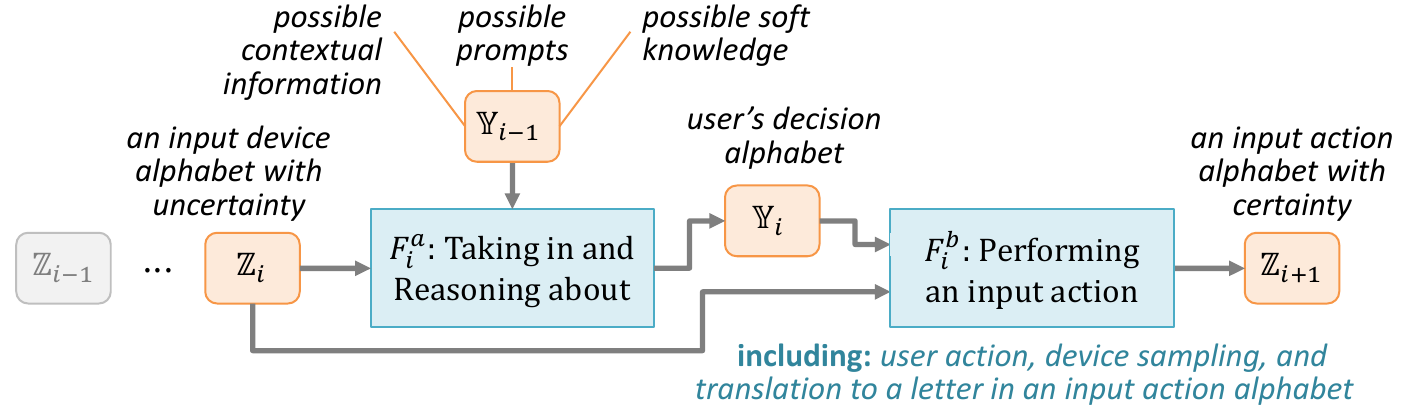}\\
  (c) The two transformations from knowledge to action data.
  \caption{Schematic illustrations of the fundamental components and measures in performing an HCI task.}
  \label{fig:CBR}
\end{figure}

As illustrated in Fig. \ref{fig:CBR}(a), performing an HCI task involves two interrelated transformations:
one is associated with an input device alphabet and another with an input action alphabet; and
one characterizes the resources used by the HCI task, and another for the amount of knowledge delivered for the HCI task.
The level of utilization of an input device can thus be measured by:
\[
	\text{DU} = \frac{\text{Action Capacity}}{t \times \text{Device Bandwidth}} = \frac{\mathcal{C}_\text{act}}{t \times \mathcal{W}_\text{dev}} 
\]
\noindent where $t$ is the time (in unit: \emph{second}) taken to perform the HCI task.
In general, instead of using an accurate time $t$ for each particular HCI task, one can use an average time $t_\text{avg}$  estimated for a specific category of HCI tasks.

Using the examples in the above two subsections, we can estimate the DU for HCI tasks using radio buttons, check boxes, freehand paths, and gestures.
Consider a set of four radio buttons with a uniform probability distribution, a portion of a display screen of $128\times128$ pixels, and a simple 2D mouse with 2 on-off buttons and 100Hz sampling rate.
Assume that the average input time is 2 seconds, we have:
\[
	\text{DU}_\text{radio} = \frac{2\,\text{bits}}{2\,\text{s} \times 100 \times (7+7+1+1)\,\text {bits/s}} = \frac{2}{3200} \approx 0.06\%
\]
Following on the previous discussion on different probability distributions of an input action alphabet, we can easily observe that the more skewed distribution, the lower the action capacity $\mathcal{C}_\text{act}$ and thereby the lower the DU. 

If the same mouse and the same portion of the screen device are used for a set of four check boxes with a uniform probability distribution and we assume that the average input time is 4 seconds, we have:
\[
	\text{DU}_\text{checkbox} = \frac{2^4\,\text{bits}}{4\,\text{s} \times 100 \times (7+7+1+1)\,\text{bits/s}} = \frac{16}{6400} = 0.25\%
\]

Consider that the same 100 Hz mouse and the same portion of the screen device are used for drawing a freehand path with a uniform probability distribution for all possible paths.
We assume that on average, a freehand path is drawn in 1 second, yielding 100 points along the path.
The DU is thus:
\[
	\text{DU}_\text{freehand} \approx \frac{(7+7)\,\text{bits} \times 100}{1\,\text{s} \times 100 \times (7+7+1+1)\,\text{bits/s}} = \frac{1400}{1600} = 87.5\%
\]

For gesture input using the aforementioned 200 Hz data glove, if an elementary gesture on average takes 2 seconds to be recognized with a reasonable certainty, the DU is:
\[
	\text{DU}_\text{gesture} \approx \frac{4\,\text{bits}}{2\,\text{s} \times 10888.6\,\text{bits/s}} = \frac{4}{21777.2} = 0.02\%
\]

Some HCI tasks require additional display space or other resources for providing users with additional information (e.g., multiple-choice questions) and some do not (e.g., keyboard shortcuts).
In the former cases, the extra resources should normally be included in the consideration of device bandwidth $\mathcal{W}_\text{dev}$.
At the same time, the varying nature of the information should be included in the consideration of the input action alphabet.
For instance, for a set of 10 different yes-no questions, there are two variables, one represents the options for the questions and one for the options of ``yes'' or ``no''. 
Hence the corresponding alphabet actually consists of 20 letters (e.g., $\{$``Q1-yes'', ``Q1-no'', $\ldots$, ``Q10-yes'', ``Q10-no''$\}$) rather than just ``yes'' and ``no''.
For more complicated additional information, such as different visualization images, we can extend the definitions in this work, e.g., by combining the (input) device utilization herein with the display space utilization defined in \cite{Chen:2010:TVCG}. 

From the discussions in this section, we can observe that the quantitative measurement allows us to compare the capacity and efficiency of different input devices and HCI tasks in a sense that more or less correlates with our intuition in practice.
However, there may also be an uncomfortable sense that the device utilization is typically poor for many input devices and HCI tasks.
One cannot help wonder if this would support an argument about having less HCI.
\revise{This doubt will be discussed in the next section.}

\section{Cost-benefit of HCI}
\label{sec:CBR}
Chen and J\"{a}nicke raised a similar question about display space utilization \cite{Chen:2010:TVCG} when they discovered that quantitatively the better utilization of the display space did not correlate to the better visual design.
While they did identify the implication of visualization and interaction upon the mathematical proof of DPI as discussed in Section \ref{sec:Intro}, the effectiveness and efficiency of visualization was not addressed until 2016 when Chen and Golan proposed their information-theoretic measure for analyzing the cost-benefit of data intelligence processes \cite{Chen:2016:TVCG}.
As HCI plays a valuable role in almost all nontrivial data intelligence workflows, we hereby use this measure to address the question about the cost-benefit of HCI.

The \emph{cost-benefit measure} by Chen and Golan considers three abstract measures that summarize a variety of factors that may influence the effectiveness and efficiency of a data intelligence workflow or individual machine- or human-centric processes in the workflow. Recently Chen et al. proposed an improvement to the original formula by replacing the unbounded divergence term with an bounded one \cite{Chen:2022:E1,Chen:2022:E2}. In this work, we adopt this new formula, which is:
\begin{equation}
\begin{split}
\label{eq:CBR}
\frac{\text{Benefit}}{\text{Cost}} =&
\frac{\text{Alphabet Compression} - \text{Potential Distortion}}{\text{Cost}} 
= \frac{\mathcal{H}(\mathbb{Z}_i) - \mathcal{H}(\mathbb{Z}_{i+1}) - \mathcal{H}_\text{max}(\mathbb{Z}_i) \mathcal{D}_\text{cs}(\mathbb{Z}'_i\|\mathbb{Z}_i)}{\text{Cost}}\\
=& \dfrac{-\displaystyle\sum_{j=1}^{n_i} \phi_j \log_2 \phi_j +
\sum_{k=1}^{n_{i+1}} \theta_k \log_2 \theta_k -
\dfrac{\log_2 n_i}{2} \sum_{j=1}^{n_i} \bigl( \phi_j + \psi_j \bigr)
\log_2 \big( |\phi_j - \psi_j|^2 + 1 \bigr)
}{\text{cost}}
\end{split}
\end{equation}
\noindent where $\mathbb{Z}_i = \{z_{i,1}, z_{i,2}, \ldots, z_{i,n_i} \}$ is the input alphabet to a process $P_i$, and $\mathbb{Z}_i$ is associated with a probability distribution $\Phi(\mathbb{Z}_i) = \{ \phi_1, \phi_2, \ldots, \phi_{n_i}\}$. $\mathbb{Z}_{i+1} = \{z_{i+1,1}, z_{i+1,2}, \ldots, z_{i+1,n_{i+1}} \}$ is the output alphabet of the process $F_i$, and $\mathbb{Z}_{i+1}$ is associated with a distribution $\Theta(\mathbb{Z}_{i+1}) = \{ \theta_1, \theta_2, \ldots, \theta_{n_{i+1}}\}$.
$\mathbb{Z}'_i$ is an alphabet reconstructed by an inverse process $G_i$, which may be considered as an approximation of $F_i^{-1}$. $\mathbb{Z}'_i$ has the same letters as $\mathbb{Z}_i$ but is likely associated with a different probability distribution $\Psi(\mathbb{Z}'_i) = \{ \psi_1, \psi_2, \ldots, \psi_{n_i}\}$.   

Given an HCI process, which may represent the completion of an HCI task from start to finish, a micro-step during the execution of an HCI task, or macro-session comprising several HCI tasks, the measure first considers the transformation from an alphabet before the processing to the alphabet after the processing.
As given in Eq.\,(\ref{eq:CBR}), this abstract measure is referred to as \emph{Alphabet Compression}.

Consider the process $P_i$ as a function, $F_i: \mathbb{Z}_i \rightarrow \mathbb{Z}_{i+1}$, which consists of all actions from the point when a user starts executing a HCI task to the point when a computer stores the information about the input (in terms of the input action alphabet $\mathbb{A}_\text{act}$) and is ready to forward this information to the subsequent computational processes, such as $P_{i+1}, P_{i+2},$ and so on.
In information theory, such a function is often referred to as a \emph{transformation} from one alphabet to another.
Alphabet compression measures the entropic difference between the two alphabets, $\mathcal{H}(\mathbb{Z}_i) - \mathcal{H}(\mathbb{Z}_{i+1})$.

As discussed in Section \ref{sec:Measures}, every HCI task is defined by an input action alphabet that captures the essence what a computer would like to know.
The computer is uncertain before the transformation, and becomes certain after the transformation.
The amount of uncertainty to be removed by a user's interaction equals to the action capacity $\mathcal{C}_\text{act}$ of the input action alphabet $\mathbb{A}_\text{act}$.
In terms of Eq.\,(\ref{eq:CBR}), we have $\mathcal{C}_\text{act} = \mathcal{H}(\mathbb{Z}_i)$ since $\mathbb{A}_\text{act} = \mathbb{Z}_i$.

At the end of the HCI task, the computer receives an answer from the user, the subsequent alphabet $\mathbb{Z}_{i+1}$ usually consists of only one letter (e.g., selecting a radio button).
Therefore the entropy $\mathcal{H}(\mathbb{Z}_{i+1})$ is 0, and the alphabet compression $\mathcal{H}(\mathbb{Z}_i) - \mathcal{H}(\mathbb{Z}_{i+1}) = \mathcal{H}(\mathbb{Z}_i) = \mathcal{C}_\text{act}$.

As illustrated in Fig. \ref{fig:CBR}(b), alphabet compression measures an quantity about the forward mapping from $\mathbb{Z}_i$ to $\mathbb{Z}_{i+1}$.
The more entropy is removed, the higher amount of alphabet compression, and hence the higher amount of benefit according to Eq.\,(\ref{eq:CBR}).
If we did not have another measure to counter-balance alphabet compression, a computer randomly chooses a ratio button or fails to recognizes a gesture correctly would not have direct impact on the benefit of HCI.
Therefore it is necessary to introduce the second abstract measure \emph{Potential Distortion}, which is mathematically defined by the term $\mathcal{H}_\text{max}(\mathbb{Z}_i)\mathcal{D}_{cs}(\mathbb{Z}'_i\|\mathbb{Z}_i)$.

If a computer could intelligently think about the information provided by a user (e.g., a radio button or a gesture recognized from a video stream), the computer would not trust the information fully. The computer could doubt whether the user might have selected radio button C instead of B if the textual description were better, or a gesture of No. 3 might actually be No. 4.   
The potential distortion measures a quantity for the reverse mapping from $\mathbb{Z}_{i+1}$ to $\mathbb{Z}_i$.
We use $\mathbb{Z}'_i$ to denote the alphabet resulting from this reverse mapping.
$\mathbb{Z}'_i$ has the same set of letters as $\mathbb{Z}_i$, but usually a different probability distribution.
If a computer can always detect and stores a user's intended input correctly, the potential distortion $\mathcal{D}_\text{cs}(\mathbb{Z}'_i\|\mathbb{Z}_i)$ is 0.
A high value of the potential distortion indicates a high level of inconsistency between the probability distributions of $\mathbb{Z}'_i$ and $\mathbb{Z}_i$.
In information theory, there are many divergence measures for quantifying such inconsistency. Based on the theoretical and empirical evaluation by Chen et al. \cite{Chen:2022:E1,Chen:2022:E2}, we use $\mathcal{D}_{cs}(\mathbb{Z}'_i\|\mathbb{Z}_i)$ in this work.
As $0 \leq \mathcal{D}_{cs}(\mathbb{Z}'_i\|\mathbb{Z}_i) \leq 1$, the amount of potential distortion caused by a transformation is bounded by $[0, \mathcal{H}_\text{max}]$. 
Readers who are interested in the mathematical definitions of alphabet compression and potential distortion may also consult \cite{Cover:2006:book,Chen:2016:TVCG} for further details.

The third abstract measure is the \emph{Cost} of the process, which should ideally be a measurement of the energy consumed by a machine- or human-centric process.
In practice, this is normally approximated by using time, a monetary quantity, or any other more obtainable measurement.
For example, in HCI, we may use the average time, cognitive load, skill levels for a user to perform an HCI task, computational time, or monetary cost of computational resources for recognizing a human action.
\revise{The selection of an approximate cost measure depends several factors, including (i) the feasibility of obtaining such measures, (ii) the desired accuracy of such measures, and (iii) the consistency in using such measures to compare a set of scenarios (e.g., the performance of different users).   
In the study conducted by Kijmongkolchai et al. \cite{Kijmongkolchai:2017:CGF}, response time was used as a cost measure.
In the analysis of cost-benefit of using virtual reality technology in visualization, Chen et al. focused on cognitive load \cite{Chen:2019:TVCG}.}
If we use device bandwidth as the cost while assuming that the computer always detects and stores the user's input correctly, the cost-benefit measure is the same as the measure of input device utilization DU.

In fact, we have only examined the second transformation, $F^b_i$, for performing a HCI task as depicted in Fig. \ref{fig:CBR}(c) where there is less tangible and often unnoticeable first step.
Before a user considers an input action alphabet $\mathbb{A}_\text{act} = \mathbb{Z}_i$, the user has to take in and reason about various information that may affect an action of HCI.
Collectively all possible variations of any information that may be considered for a HCI task are letters in an alphabet, denoted as $\mathbb{Y}_{i-1}$ in Fig. \ref{fig:CBR}(c).
Hence the first step of ``taking in and reasoning about'' is, in abstract, a transformation, $F^a_i: \mathbb{Y}_{i-1} \times \mathbb{Z}_{i} \rightarrow \mathbb{Y}_i$.
As $F^a_i$ takes place in a user' mind, it is often unnoticeable.
Broadly speaking, $F^a_i$ may take in the following types of information.

\textbf{Explicit Prompt.} This includes any information that is purposely provided by a computer or a third party for the HCI task concerned, e.g., textual and visual prompts for radio buttons or checkboxes, audio questions asked prior to voice-activated commands, instructions from a trainer or a user manual to a trainee, and so forth.

\textbf{Situational Information.} This includes any information provided by a computer or an environment where interaction occurs.
The information is not specifically, but can be used, for the HCI task.
This may include the texts or drawings that a user is currently working on when the user issues a ``save as'' command, and the current sound or lighting quality in a video conference when the user issues a command to switch on or off the video stream. 

\textbf{Soft Knowledge.} This includes any information that resides in the user's mind, which can be called upon to support the HCI task. Tam et al. \cite{Tam:2017:TVCG} considered two main types of soft knowledge: \emph{soft alphabets} and \emph{soft models}. The former encompasses factual knowledge that is not available as explicit prompts or situational information, e.g., the knowledge about keyboard shortcuts, the knowledge about the reliability for the computer to recognize a gesture or voice. The latter encompasses analytical knowledge that can be used to derive information for the HCI task dynamically.
For example, a user may assess the levels of risk associated to each radio button (or in general, each letter in $\mathcal{A}_\text{act}$). While the levels of risk are letters of a soft alphabet, the alphabet exists only after a soft model has been executed.

\begin{figure}[t]
  \centering
  \includegraphics[width=100mm]{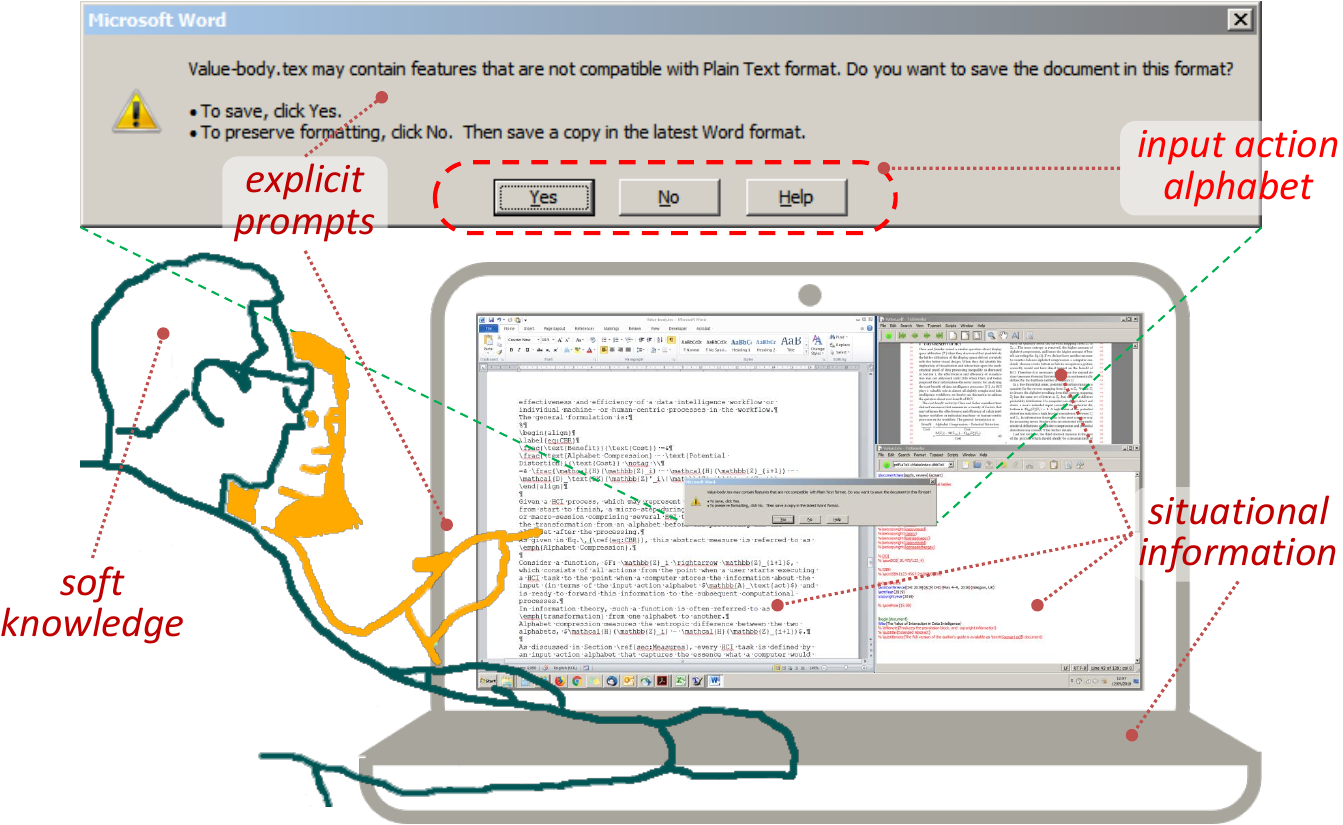}
  \caption{A simple HCI task may be affected by three types of variables, which are collectively a very complex alphabet.}
  \label{fig:Latex}
\end{figure}

Fig. \ref{fig:Latex} shows an example of an HCI task.
A user is editing a .tex file using a word processor (Microsoft Word) because it is easy to zoom in and out.
After the user issues a ``Ctrl-S'' command, the computer displayed a pop-up window of 734$\times$140 pixels, with a textual prompt.
The input action alphabet $\mathbb{A}_\text{act}$ has three multiple-choice buttons.
Hence the maximal benefit that can be brought by the transformation $F^b_i$ in Fig. \ref{fig:CBR}(c) for this case is about 1.58 bits.

Meanwhile, the word processor may have different explicit prompts following a ``Ctrl-S'' command according to, e.g., the file modification status, the existence of a file with the same name, access permission, etc.
A colleague may offer advice as to which button to choose.
The display screen may show different situational information, e.g., different documents being edited, and different concurrent windows that may or may not be related the file being processed.
The user may have the soft knowledge that a .tex file is a plain text file, the so-called ``features'' in the prompt cannot be processed by a \LaTeX\ compiler, the ``help'' button does not provide useful guidance to this particular way of using the word processor, and so on.
As we can see that $\mathbb{Y}_{i-1}$ is not a simple alphabet and has a non-trivial amount of entropy, we can conclude that the two transformations, $F^a_{i}$ and $F^b_{i}$, together bring about benefit much more than 1.58 bits.

\section{Estimating Cost-Benefit Analytically}
\label{sec:EstimateCBR}
The cost-benefit metric described in Section \ref{sec:CBR} provides HCI with a mean for probabilistic characterization of complex phenomena.
While it can be calculated from gathered data about every letter in an alphabet in some simple or highly controlled settings (e.g., see Section \ref{sec:MeasureCBR}), it is more practical to estimate the three measures in real-world applications, e.g., for comparing different user interface designs or evaluating HCI facilities in a data intelligence workflow.%
\footnote{In thermodynamics, the notion of \emph{entropy} provides a microscopic measure, reflecting the fundamental understanding about thermodynamic phenomena. It is typically estimated based on macroscopic quantities such as temperature, volume, and pressure that are more easily measureable.}

Let us first exemplify the estimation method by revisiting the channel selection scenario in Fig. \ref{fig:TVList} in Section \ref{sec:Measures}.
A coarse estimation can be made with an assumption that a user's selection is always correct.
In this case, the potential distortion in Eq.\,(\ref{eq:CBR}) is of 0 bits.
The amount of alphabet compression thus equals to the action capacity $\mathcal{C}_\text{act}$.
Meanwhile, from a usability perspective, the cost can be estimated based on the time, effort, or cognitive load required for selecting each option.
For example, in Fig. \ref{fig:TVList}(a), the top option $a_1 \in \mathbb{A}$ is the default selection, and requires only one [OK] action on the remote controller.
The second option $a_2$ requires a [$\blacktriangledown$] action followed by [OK], while the third option $a_3$ requires three actions: [$\blacktriangledown$], [$\blacktriangledown$], and [OK].
If the \textbf{time} for each button action is estimated to take 2 seconds, the average cost for this HCI task is:
\[
	\text{cost} = 2 \bigl( p(a_1) + 2 p(a_2) + 3 p(a_3) \bigr) \quad [\text{unit: second}]
\]
Using the three example probability distributions for the input action alphabet $\mathbb{A}$ in Section \ref{sec:Measures}, we can obtain, 
\begin{align*}
p(a_1) = p(a_2) = p(a_3) = 1/3 \longrightarrow&\, \text{cost}(\mathbb{A}_a) =4 \\
p(a_1) = 0.2, p(a_2) = 0.7, p(a_3) = 0.1 \longrightarrow&\, \text{cost}(\mathbb{A}_b) = 3.8\\
p(a_1) = 0.09, p(a_2) = 0.9, p(a_3) = 0.01 \longrightarrow&\, \text{cost}(\mathbb{A}_c) = 3.84
\end{align*}
Combining with the calculation of $\mathcal{C}_\text{act}$ in Eq.\,\ref{eq:CactTV} in Section \ref{sec:Measures}, we have the cost-benefits ratios for the three probability distributions are approximately
$\mathbb{A}_a: 0.40 (=1.58/4)$,
$\mathbb{A}_b: 0.30 (=1.16/3.8)$, and
$\mathbb{A}_c: 0.13 (=0.52/3.84)$ bits/s respectively.
Hence for the skewed distribution associated with $\mathbb{A}_c$, the cost-benefit is very low.

During the design or evaluation of the TV system, a UX expert may discover that users select $a_2$ ``Select Channel'' more frequently than the other two options.
The UX expert can consider an alternative design by swapping the position of $a_1$ and $a_2$.
With the changes of the corresponding probability distributions, the UX expert can value the improvement of the cost-benefit quantitatively, such as, $\mathbb{A}_b: 0.30 \nearrow 0.41$ and $\mathbb{A}_c: 0.13 \nearrow 0.23$.

A more detailed estimation may consider the factor that users may mistakenly press [OK] for the default option.
For example, if in 20\% cases, users are intended to select $a_2$ but select the default $a_1$ by mistake, there are both potential distortion and extra cost.
In the case of $\mathbb{A}_b$, the reconstructed probability distribution is $p'(a_1) = 0.4, p'(a_2) = 0.5, p'(a_3) = 0.1$.
The potential distortion can be calculated as $\mathcal{H}_\text{max}\mathcal{D}_\text{cs} \approx 1.58 \times 0.05 \approx 0.08$ bits.
In the case of $\mathbb{A}_c$, the reconstructed probability distribution is $p'(a_1) = 0.29, p'(a_2) = 0.7, p'(a_3) = 0.01$.
The potential distortion is $\mathcal{H}_\text{max}\mathcal{D}_\text{cs} \approx 1.58 \times 0.06 \approx 0.09$ bits.
Let the extra time for showing detailed information about a TV show and going back to the original three options is 4 seconds.
We can estimate that the extra time in the two cases are:
$20\% \times 4 = 0.8$ seconds on average.
The cost benefit ratio will be reduced as:
$\mathbb{A}_b: 0.30 \searrow 0.23$ and $\mathbb{A}_c: 0.13 \searrow 0.09$.
If the mistakes were to reach 51\% or more, the metric would return a negative value for $\mathbb{A}_c$.

Similarly, one may estimate the user's effort as the cost by counting the steps needed to perform an action, such as reading the screen, looking at the remote control, and pressing a button.
One may also weigh these steps differently based on pre-measured cognitive load for different types of elementary steps, which may be obtained, for instance, using electroencephalography (EEG) (e.g., \cite{Tan:2010:book}).

For the example in Fig. \ref{fig:TVList}(b), it is easy to observe that the cost-benefit is always 0 since $\mathcal{C}_\text{act} \equiv 0$ bits, though the cost for pressing [OK] on the remote control may not be considered high.
This quantitative measure is consistent with what most UX experts would conclude qualitatively.

\begin{figure*}[t]
  \centering
  \includegraphics[width=150mm]{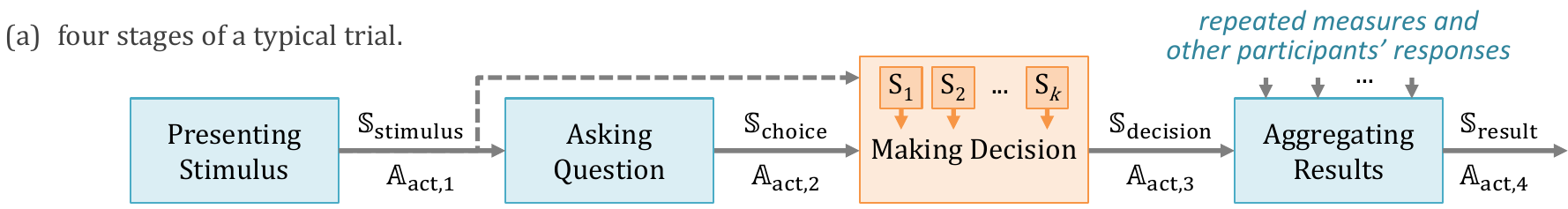}\\[2mm]
  \includegraphics[width=150mm]{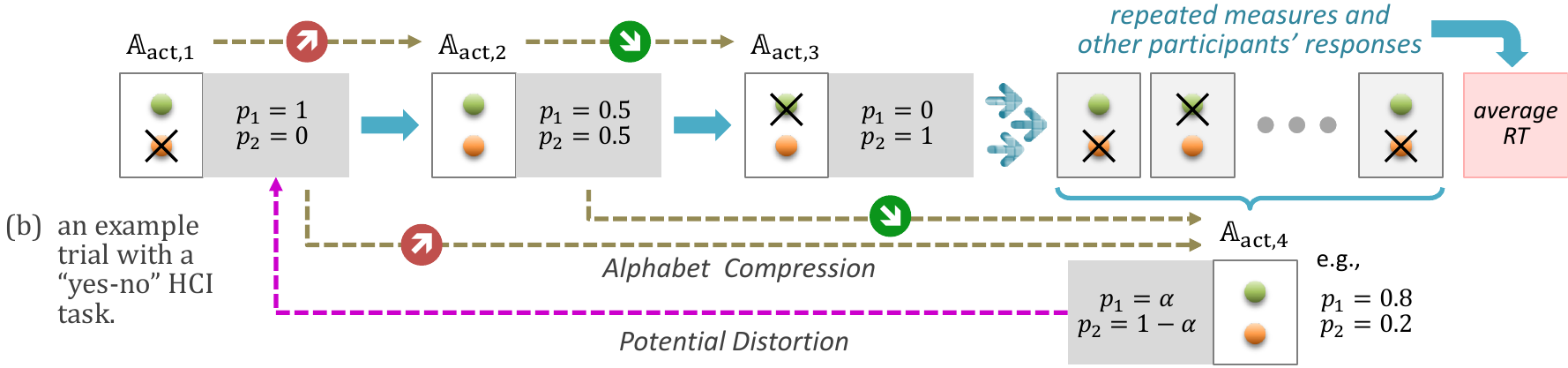} 
  \caption{(a) The alphabets in an abstract workflow representing a trial in a typical controlled empirical study. (b) The example of a simple ``yes-no'' trial for detecting and measuring human knowledge in HCI.}
  \label{fig:Study}
\end{figure*}

The estimation for the channel selection task does not consider any situational information or soft knowledge.
When such variables are considered as part of an HCI task, as illustrated in Fig. \ref{fig:CBR}(c), the amount of cost-benefit usually increase noticeably.
For example, consider the \LaTeX\ example in Fig. \ref{fig:Latex}.
If the word processor on the left has 5 different pop-up windows in responses to a ``save'', ``save as'', or ``Ctrl-S'' command, each with 3 options, the input action alphabet has 15 letters.
The maximum alphabet compression for the second process $F_2$ in Fig. \ref{fig:CBR}(c) is about 3.9 bits.

On the other hand, when given any one of 10 file types (e.g., .doc, .tex, .txt, .htm, etc.), the user has the knowledge about whether formatting styles matter.
There are 10 binary variables or 10 bits of knowledge available.
Consider conservatively that on average a user deletes or modifies 10 English letters independently before saving.
The user knows whether it is critical to overwrite the existing file when a pop-up window asks for a confirmation.
There are 10 nominal variables, each with some 26 valid values for English letters.
As the entropy of English alphabet is about 4.7 bits \cite{Shannon:1951:BSTJ}, the total amount of knowledge available is about 47 bits.
Without considering other factors (e.g., digits, symbols, etc.), we can conservatively estimate the amount of alphabet compression for the process $F_1$ in Fig. \ref{fig:CBR}(c) is about $(10+47) - 3.9$ bits.
Let us assume that selecting one of the three options takes 1 second.
The cost-benefit for such a simple HCI task ($F_1 + F_2$) is at the scale of 57 bits/s.

Tam et al. \cite{Tam:2017:TVCG} estimated the amount of human knowledge available to two interactive machine learning workflows.
Both workflows were designed to build decision tree models, one for classifying facial expression in videos and other for classifying types of visualization images.
They were curious by the facts that the interactive workflows resulted in more accurate classifiers than fully automated workflows.
They estimated the amount of human knowledge available to the two workflows.
Using the approach exemplified by the \LaTeX\ example, they identified 9 types of soft knowledge in the facial expression workflow and 8 types in the other.
In both cases, there were several thousands of bits of knowledge available to the computational processes in the workflows.

\section{Measuring Cost-Benefit Empirically}
\label{sec:MeasureCBR}
As the cost-benefit ratio described in Section \ref{sec:CBR} is relatively new, there has been only one reported empirical study attempting to measure the three quantities in the formula.
Kijmongkolchai et al. \cite{Kijmongkolchai:2017:CGF} conducted a study to measure the cost-benefit of three types of soft knowledge used during the visualization of time series data.
This includes the knowledge about (i) the context (e.g., about an electrocardiogram but not weather temperature or stock market data), (ii) the pattern to be identified (e.g., slowly trending down), and (iii) the statistical measure to be matched with the time series plot.

The knowledge concerned can be considered as the transformation $F^a_i$ in Fig. \ref{fig:CBR}(c), while the participants' answers to the trial questions can be considered the transformation $F^b_i$.
Kijmongkolchai et al. converted the conventional measures of accuracy and response time to that of benefit and cost in Eq.\,(\ref{eq:CBR}).
In \cite{Kijmongkolchai:2017:CGF}, Kijmongkolchai et al. described briefly the translation from (accuracy, response time) to (benefit, cost) with the support of a supplementary spreadsheet.
Here we generalize and formalize their study, and present a conceptual design that can be used as a template for other empirical studies for detecting and measuring humans' soft knowledge in HCI.

Consider a common design for a controlled experiment, in which an apparatus presents a stimulus to participants in each trial, poses a question or gives the input requirement, and asks them to make a decision or perform an HCI action.
The participants' action in response to the stimulus and input requirement is a human-centric form of data intelligence.
Fig. \ref{fig:Study}(a) illustrates the workflow of such a trial.

A stimulus may comprise of textual, visual, audio, and other forms of data as the input to the process.
Normally, one would consider that the alphabet, $\mathbb{S}_\text{stimulus}$, contains only the stimuli designed for a trial or a set of trials for which participants' responses can be aggregated.
However, if the pre-designed stimuli are unknown to participants, one must consider that $\mathbb{S}_\text{stimulus}$ consists of all possible stimuli that could be presented to the participants.
For instance, the design of a study may involve only 64 pairs of colors and ask users to determine which is brighter.
Since any pairing of two colors are possible, $\mathbb{S}_\text{stimulus}$ actually consists of $N \times N$ letters where $N$ is the number of colors that can be shown on the study apparatus.
On a 24-bit color monitor, $N = 2^{24}$ and $N \times N = 2^{48} \ggg 64$.
Hence the entropy of $\mathbb{S}_\text{stimulus}$ is usually very high.

On the other hand, the participants' inputs are commonly facilitated by multiple-choice buttons, radio buttons, or slide bars, which has a smaller alphabet $\mathbb{S}_\text{choice}$.
In some studies, more complicated inputs, e.g., spatial locations and text fields, are used, corresponding to large alphabets.
Nevertheless, for any quantitative analysis, such complicated inputs will be aggregated to, or grouped into, a set of post-processed letters in a smaller alphabet $\mathbb{S}'_\text{choices}$.
It is not difficult to notice that $\mathbb{S}_\text{choice}$ or $\mathbb{S}'_\text{choice}$ is essentially an input action alphabet $\mathbb{A}_\text{act}$. In the following discussion, we do not distinguish between $\mathbb{S}_\text{choice}$ and $\mathbb{S}'_\text{choice}$.

Once a participant has made a decision, only one letter in the alphabet $\mathbb{A}_\text{decision}$ has the probability value 1, while the other letters are of probability 0. The $\mathbb{A}_\text{decision}$ is thus of entropy 0 bits.
However, after one merges all the repeated measures and responses from different participants into an alphabet $\mathbb{S}_\text{result}$ (with the same letters of $\mathbb{A}_\text{decision}$), the letters in $\mathbb{S}_\text{result}$ are expected to be associated with difference numbers or frequencies of occurrence.

From the perspective of interaction, the alphabet $\mathbb{A}_\text{act}$ has different probability distributions at different stages as illustrated in Fig. \ref{fig:Study}(a).
Before and after the stimulus presentation stage, $\mathbb{A}_\text{act,1}$ has a ground truth for each trial, and thus one letter has the full probability 1.
After the question stage, the letters in $\mathbb{A}_\text{act,2}$ are pretended to have an equal probability $1/\|\mathbb{A}_\text{act}\|$.
After the decision stage, only one letter in $\mathbb{A}_\text{act,3}$ is chosen, which thus has the full probability 1.
After the aggregation stage, $\mathbb{A}_\text{act,4}$ has a probability distribution reflecting all repeated measures and all participants' responses.

The humans' soft knowledge used in the transformation from $\mathbb{S}_\text{stimulus}$ to $\mathbb{S}_\text{choice}$ and to $\mathbb{S}_\text{decision}$ can be very complicated.
The amount of alphabet compression can be huge.
Nevertheless, the essence of any controlled experiment is to investigate one or a few aspects of this soft knowledge while restricting the variations of many other aspects.
Here we refer one particular aspect under investigation as a \emph{sub-model}, $\mathbf{S}$, which may be a heuristic function for extracting a feature or factor from the stimulus or for retrieving a piece of information that is not in the stimulus.

Let us first examine a very simple ``yes-no'' trial designed to investigate if a sub-model $\mathbf{S}$ has role to play in the transformation from $\mathbb{S}_\text{stimulus}$ to $\mathbb{S}_\text{decision}$.
As illustrated in Fig. \ref{fig:Study}(b),
 at the beginning $\mathbb{A}_\text{act,1}$ has two letters $\{Y, N\}$.
Assuming that $Y$ is the ground truth, the probabilities are $p(Y) = 1, p(N) = 0$.
Note that in \cite{Kijmongkolchai:2017:CGF}, Kijmongkolchai et al. introduced a tiny small value $0 < \epsilon < 1$ to moderate $p(Y)$ and $p(N)$ in order to prevent the Kullback-Leibler divergence $\mathcal{D}_\text{KL}$ from handling the conditions of $\log 0$ and $\div 0$ because their calculation was based on the original formula for the cost-benefit measure by Chen and Golan \cite{Chen:2016:TVCG}. Here we adopt the new formula by Chen and Sbert \cite{Chen:2022:E1,Chen:2022:E2}, where the unbounded $\mathcal{D}_\text{KL}$ was replaced with the bounded $\mathcal{D}_\text{cs}$. Therefore, the awkward moderation by $\epsilon$ is no longer necessary.

Any alphabet with two letters can potentially has a maximal entropy value of $\;\mathcal{H}_\text{max}=1$ bit. 
When the question, ``yes'' or ``no'', is posed to each participant, $\mathbb{A}_\text{act,2}$ is associated with the probabilities $p(Y) =  p(N) = 0.5$.
$\mathbb{A}_\text{act,2}$ is thus of the maximal entropy 1 bit, indicating the maximal level of uncertainty.

When a decision is made by a participant, the probability distribution of $\mathbb{A}_\text{act,3}$ is either $p(Y) = 1, p(N) = 0$ or $p(Y) = 0, p(N) = 1$.
From an individual participant, the apparatus (e.g., the computer) receives an answer with certainty.  $\mathbb{A}_\text{act,3}$ is of minimal entropy 0 bits. 

After all related responses are collected, $\mathbb{A}_\text{act,4}$ has probabilities $p(Y) = \alpha, p(N) = 1-\alpha$.
$\mathbb{A}_\text{act,2}$ has the maximum amount of entropy of 1 bit, while that of $\mathbb{A}_\text{act,4}$ is between 0 and 1 bits depending on $\alpha$.
If $\alpha = 0.5$ (e.g., random choices), the sub-model $\mathbf{S}$ offers no alphabet compression.
If $\alpha = 1$ (i.e., all ``yes'' answers) or $\alpha = 0$ (i.e., all ``no'' answers), $\mathbf{S}$ enables $1$ bit alphabet compression from $\mathbb{A}_\text{act,2}$ to $\mathbb{A}_\text{act,4}$.
Without repeated measures, all participants individually achieve the same alphabet compression.
We will discuss the case of repeated measures towards the end of this section.

\begin{figure}
  \centering
  \begin{tabular}{@{}c@{\hspace{4mm}}c@{\hspace{4mm}}c@{}}
       \includegraphics[width=45mm]{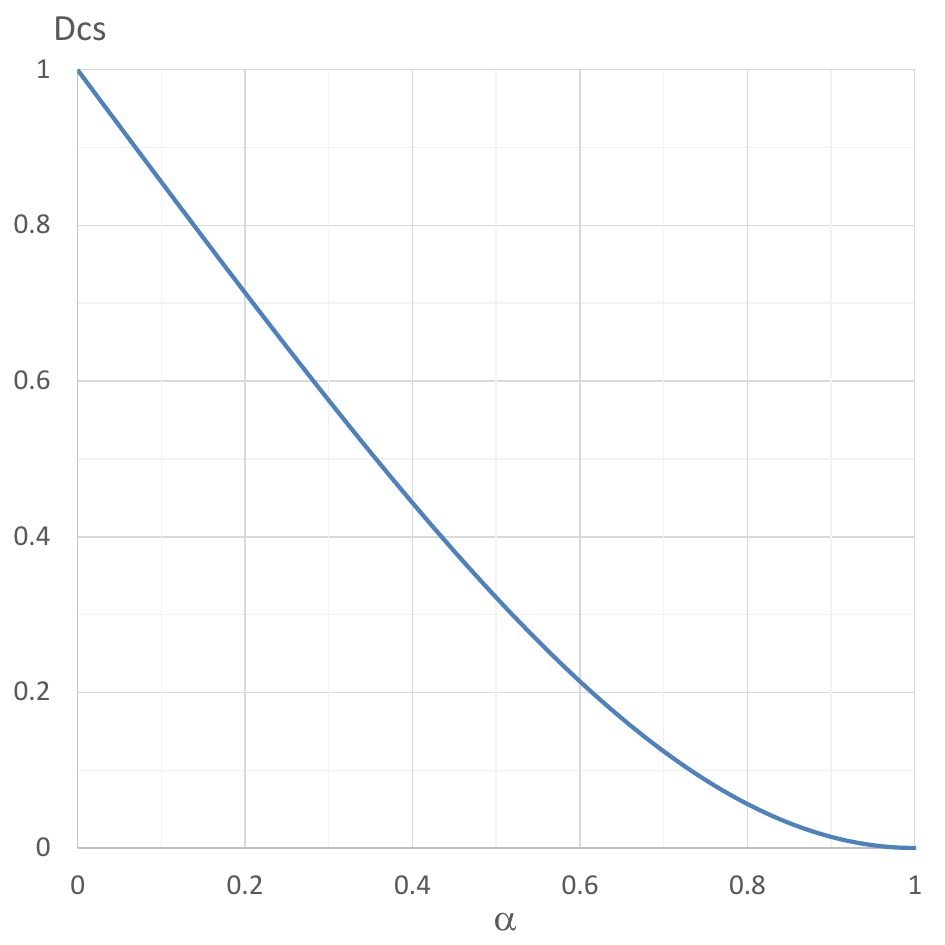} &
       \includegraphics[width=45mm]{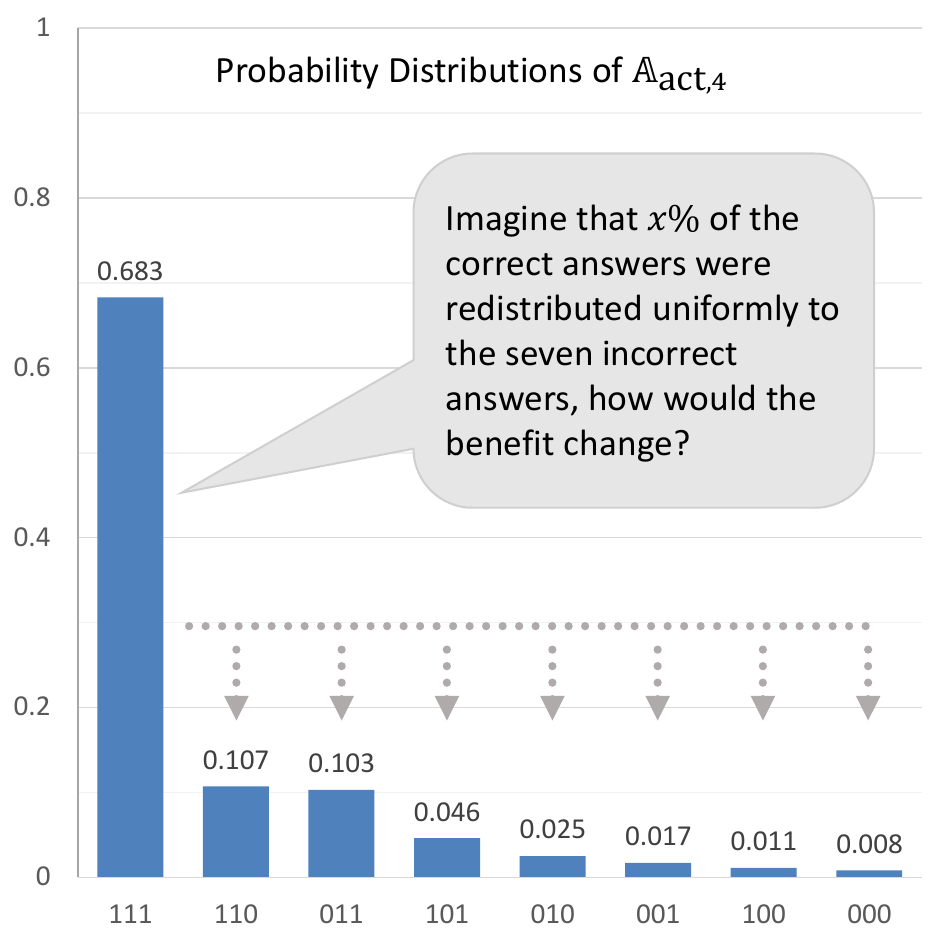} &
       \includegraphics[width=45mm]{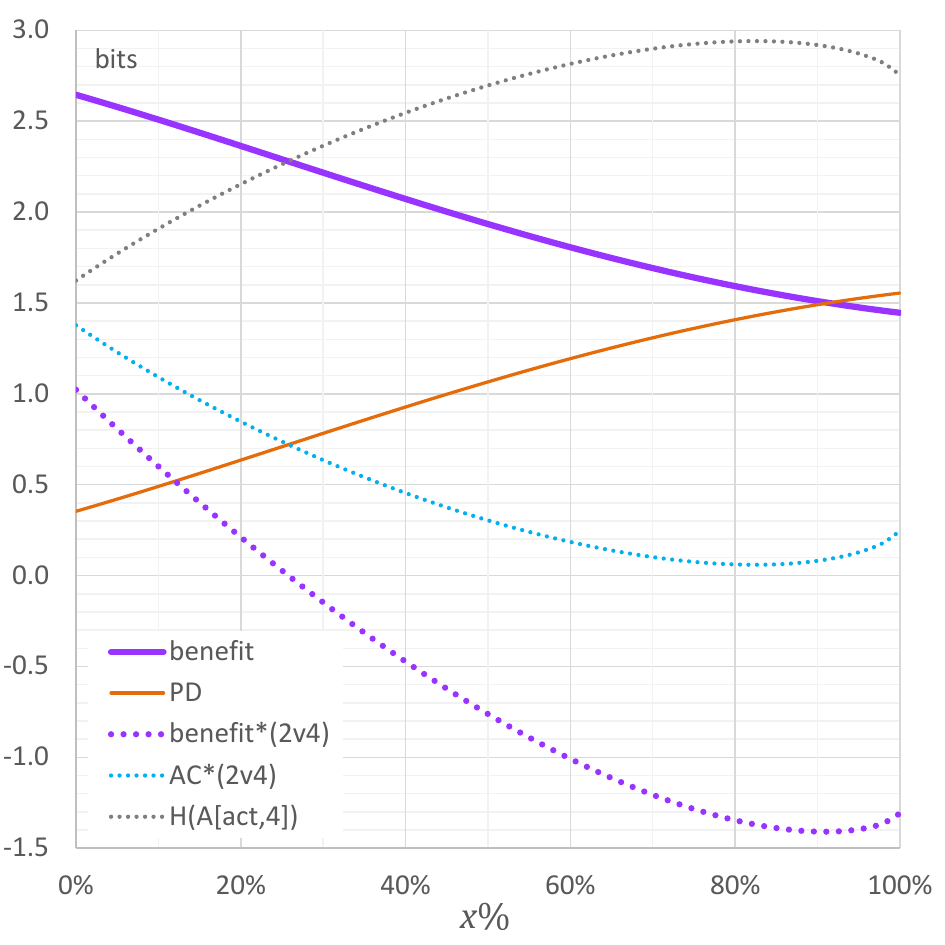} \\
       \small (a) $\mathcal{D}_\text{cs}$ decreases when $\alpha: 0 \rightarrow 1$ &
       \small (b) an example set of study results &
       \small (c) the impact of hypothesized results 
  \end{tabular}
  \caption{(a) For two simple probability distributions $\{1, 0\}$ and $\{ \alpha, 1-\alpha\}$, Divergence $\mathcal{D}_\text{cs}$ decreases when $\alpha$ changes from 0 to 1 (i.e., $\{ \alpha, 1-\alpha\}$ changes from completely opposite of $\{1, 0\}$ to the same as $\{1, 0\}$.
  (b) Given a distribution of participants' answers, we may wish to hypothesize different distributions. (c) We can estimate the benefit values for such hypothesized distributions.}
  \label{fig:SoftKnow}
\end{figure}

Meanwhile, without repeated measures, the potential distortion has to be estimated using the collective results from all participants.
As shown in Fig. \ref{fig:Study}(b), it is measured based on the reverse mapping from $\mathbb{A}_\text{act,4}$ to $\mathbb{A}_\text{act,1}$.
As simple ``yes-no'' alphabet can be coded using binary codewords as $\mathbb{A}_\text{act}=\{N, Y\}=\{0,1\}_2$, $\mathbb{A}_\text{act,1}$ and $\mathbb{A}_\text{act,4}$ have the same set of letters but may have different probability distributions. As mentioned earlier, $P(\mathbb{A}_\text{act,1}) = \{1.0, 0.0\}$. 
If all participants have answered ``yes'', we have $\alpha = 1$ and $P(\mathbb{A}_\text{act,4}) = \{1.0, 0.0\}$. Using the sub-formula for $\mathcal{D}_\text{cs}$ in Eq.\,\ref{eq:CBR}, we have $\mathcal{D}_\text{cs} = 0$.
If all participants have answered ``no'', $\alpha = 0$, $P(\mathbb{A}_\text{act,4}) = \{0.0, 1.0\}$, and $\mathcal{D}_\text{cs} = 1$.
Fig. \ref{fig:SoftKnow}(a) shows the trend of decreasing divergence $\mathcal{D}_\text{cs}$ when $\alpha$ changes from 0 (i.e., all incorrect) to 1 (i.e., all correct).



For an empirical study designed to examine a sub-model at a slightly higher resolution, we can assign $k>1$ bits to the input action alphabet $\mathbb{A}_\text{act}$.
For example, a 3-bit alphabet $\mathbb{A}_\text{act}=\{a_1, a_2, \ldots, a_8\}$ can be labelled as $\{000, 001, \ldots, 111\}_2$.
It is necessary to use all $2^k$ letters as choices in order to maximize the entropy of $\mathbb{A}_\text{act,2}$ at the question stage.
For examining the combined effects of several sub-models, we assign a bit string to each sub-model and then concatenate their bit strings together.
For example, to study one 2-bit sub-model $\mathbf{U}$ and two 1-bit sub-models $\mathbf{V}$ and $\mathbf{W}$, we can have
$\mathbb{A}_\text{act} =$ $\{u_1v_1w_1,\, u_1v_1w_2,\, u_1v_2w_1,\, \ldots,\, u_4v_2w_2\}$ and can be labelled as $\{0000, 0001, 0010, \ldots, 1111\}_2$.

Given an input action alphabet with $n=2^k$ letters (i.e., all possible answers in a trial), one assigns a ground truth in $\mathbb{A}_\text{act,1}$, e.g., $P(a_1)=1$ and $P(a_i) = 0,\; i=2,3,\ldots,n$.
In conjunction with a stimulus, one poses a question with $n$ choices in $\mathbb{A}_\text{act,2}$, which are pretended to have an equal probability of $1/n$.
After an individual participant has answered the question, only one letter $a_j$ in $\mathbb{A}_\text{act,3}$ is selected, i.e., $P(a_j)=1$ and $P(a_i) = 0, i \in [1..n] \wedge i \neq j$.
After collecting all related responses, the probability of each letter in $\mathbb{A}_\text{act,4}$ is computed based on its frequency in participants' responses.
One can then convert the accuracy and response time to cost-benefit as:
\begin{equation}
\label{eq:Study}
	\frac{\text{benefit}}{\text{cost}} \approx
	\frac{\mathcal{H}(\mathbb{A}_\text{act,2}) -
		\mathcal{H}(\mathbb{A}_\text{act,3}) - 
		\mathcal{H}_\text{max}(\mathbb{A}_\text{act,2})\mathcal{D}_\text{cs}%
		(\mathbb{A}_\text{act,4}\|\mathbb{A}_\text{act,1})}%
	{\text{average response time}}
\end{equation}

Using the experimental data in \cite{Kijmongkolchai:2017:CGF} as an example, we have a 3-bit input action alphabet for three sub-models (each with 1-bit resolution).
Each of the eight possible answers in $\mathbb{A}_\text{act}$ is encoded by three bits $b_1b_2b_3$, where $b_i = 1$ if the sub-model $\mathbf{S}_i$ functions correctly, and $b_i = 0$ otherwise.
In their experiment, the participants were presented with eight optional answers that are randomly ordered. Hence, 
$\mathcal{H}(\mathbb{A}_\text{act,2}) = 3$ bits and
$\mathcal{H}(\mathbb{A}_\text{act,3}) = 0$ bits.
The alphabet compression for an individual is thus 3 bits.
Their study obtained a set of accuracy data in terms of the percentages of eight possible answers in $\mathbb{A}_\text{act,4}$.
The values, which are depicted in Fig. \ref{fig:SoftKnow}(b), are as follows:
\begin{itemize}
\item 68.3\% for letter 111 --- $\mathbf{S}_1$, $\mathbf{S}_2$, $\mathbf{S}_3$ are all correct.
\item 10.7\% for letter 110 --- $\mathbf{S}_1$, $\mathbf{S}_2$ are correct.
\item 10.3\% for letter 011 --- $\mathbf{S}_2$, $\mathbf{S}_3$ are correct.
\item 4.6\% for letter 101 --- $\mathbf{S}_1$, $\mathbf{S}_3$ are correct.
\item 2.5\% for letter 010 --- only $\mathbf{S}_2$ is correct.
\item 1.7\% for letter 001 --- only $\mathbf{S}_3$ is correct.
\item 1.1\% for letter 100 --- only $\mathbf{S}_1$ is correct.
\item 0.8\% for letter 000 --- $\mathbf{S}_1$, $\mathbf{S}_2$, $\mathbf{S}_3$ are all incorrect.
\end{itemize}

\noindent The alphabet $\mathbb{A}_\text{act,4}$ in fact represents the reconstruction of $\mathbb{A}_\text{act,1}$ by the participants in response to $\mathbb{A}_\text{act,2}$.  
As $\mathbb{A}_\text{act,1}$ represents the ground truth, the potential distortion of $\mathbb{A}_\text{act,4}$ is therefore measured against $\mathbb{A}_\text{act,1}$, i.e.,
$\mathcal{H}_\text{max}\mathcal{D}_\text{cs}(\mathbb{A}_\text{act,4}\|\mathbb{A}_\text{act,1})$.
Using the above accuracy values for the probability distribution of $\mathbb{A}_\text{act,4}$, we can obtain $\mathcal{H}_\text{max}\mathcal{D}_\text{cs}(\mathbb{A}_\text{act,4}\|\mathbb{A}_\text{act,1})$ as $0.354$ bits.
Using the nominator of Eq.\,\ref{eq:Study}, the benefit is about 2.646 bits.

As suggested in Fig. \ref{fig:SoftKnow}(b), one may hypothesize how the benefit value would have changed if the study had returned different results. For example, one may imagine such changes by redistributing $x\%$ of the correct answers (labelled as 000) uniformly to the seven incorrect answers. Fig. \ref{fig:SoftKnow}(c) illustrates the impact of such hypothesized results.
When $x\%$ increases from $0\%$ to $100\%$, the benefit (solid purple line) decreases while the potential distortion (solid orange line) increases.
Note that the benefit calculation in Eq.\,\ref{eq:Study} is based on an assumption that the study is to simulate a scenario where a definite decision is made for each posed question. The probability distribution of $\mathbb{A}_\text{act,4}$ will not be available to any succeeding processes.

Occasionally, we may use a study to simulate a group scenario where multiple group members offer their own independent answers, which are available to the succeeding processes as votes distribution in percentages. In the group scenario, the information of votes distribution conveys more uncertainty than a definite answer in the first scenario. There is thus less alphabet compression in the group scenario. For such a scenario, we replace the term $\mathcal{H}(\mathbb{A}_\text{act,2}) - \mathcal{H}(\mathbb{A}_\text{act,3})$ in Eq.\,\ref{eq:Study} with $\mathcal{H}(\mathbb{A}_\text{act,2}) - \mathcal{H}(\mathbb{A}_\text{act,4})$.
As shown in Fig. \ref{fig:SoftKnow}(c), the benefit curve (dotted purple line) is lower than the solid purple curve for the scenario of a definite decision.
From Fig. \ref{fig:SoftKnow}(c), one may notice that when some 26\% (or more) correct answers are redistributed to incorrect answers, the benefit become negative. When about 90\% correct answers are redistributed, the benefit reaches the minimum. This is because the entropy $\mathcal{H}(\mathbb{A}_{act,4})$ (dotted gray line) reaches its maximum at $x\% \approx 82\%$, causing the alphabet compression (dotted cyan line) to reach its minimum. 


The method of repeated measures is typically for dealing with difficulties in capturing definite inputs from each participant. The repeated measures for each participant are either aggregated first to yield a quasi-consistent measure or fused into the overall statistical measure involving all participants. As long as repeated measures are used for addressing such difficulties, one should still use Eq.\,\ref{eq:Study}. Only in a rare scenario where repeated measures are used to simulates an inconsistent decision process by an individual and the probability distribution of $\mathbb{A}_\text{act,4}$ is calculated based on the repeated measures obtained from just one participant, one can use $\mathcal{H}(\mathbb{A}_\text{act,4})$ instead of $\mathcal{H}(\mathbb{A}_\text{act,3})$ in Eq.\,\ref{eq:Study}.

\section{Data Intelligence}
\label{sec:DataInt}
In general, a data intelligence process is a transformation from some input data to some output data. The process can be a complex workflow or an infinitesimally small step in a workflow. The output data can be a final decision at the end of a workflow, or some intermediate data in a workflow. Hence, simple HCI processes, e.g., selecting a radio button, are in principle a data intelligence process. Of course, most people would relate the term ``data intelligence processes'' to more complex processes, and some might interpret the term as artificial intelligence (AI). As Chen and Golan showed in their work \cite{Chen:2016:TVCG}, the cost-benefit measure is applicable to both machine-centric and human-centric processes. In this section, we show that the information-theoretic measures, which were discussed in relatively ``small'' contexts in the previous sections, are also applicable to HCI in complex data intelligence workflows. We will use the development of classification models using machine learning (ML) as an example.

A typical classification model, $M$, takes an input data object (e.g., an image, a sentence, a time series, etc.) and delivers a class label. In terms of information theory, $M$ is a transformation from a data alphabet $\mathbb{D}$ to a class alphabet $\mathbb{C}$, such that $\mathbb{D}$ contains all possible input data objects in the context that $M$ is used, and $\mathbb{C}$ contains all labels that $M$ may generate.
For example, $\mathbb{D}_a$ may contain all possible color images of $W \times H$ pixels with an animal. $\mathbb{C}_a$ contains $k$ labels for $k$ types of animals. Alternatively, $\mathbb{D}_b$ may contain all possible color images of $W \times H$ pixels, while $\mathbb{C}_b$ contains $k+1$ labels for $k$ animal classes plus one ``unknown'' class label. In terms of the number of letters, $\|\mathbb{D}_b\| \ggg \|\mathbb{D}_a\|$.

While $M$ may likely be an automated process, the development of $M$ is rarely automated, but involves a huge amount of HCI activities. We can estimate the cost-benefit of such HCI activities. In general, $M$ is an instance among all possible functions that can be executed on a computer, which is referred to as the space of Turing Machine. In other words, an ML workflow is a transformation about an alphabet $\mathbb{M}$ that contains all possible functions in the space of Turing Machine. The transformation starts with $\mathbb{M}_0$ where all functions are equally probable, gradually changes the associated probability distribution, and finally reaches an alphabet $\mathbb{M}_N$ where the trained model $M \in \mathbb{M}$ has the probability 1 and all other possible functions have the probability 0. The entropy of $\mathbb{M}_0$ is $\infty$ for a Turing Machine with infinite tape length, while the entropy of $\mathbb{M}_N$ is 0. The among of alphabet compression is $\mathcal{H}(\mathbb{M}_0) - \mathcal{H}(\mathbb{M}_N) = \infty$.

\begin{figure}[th]
    \centering
    \includegraphics[width=\linewidth]{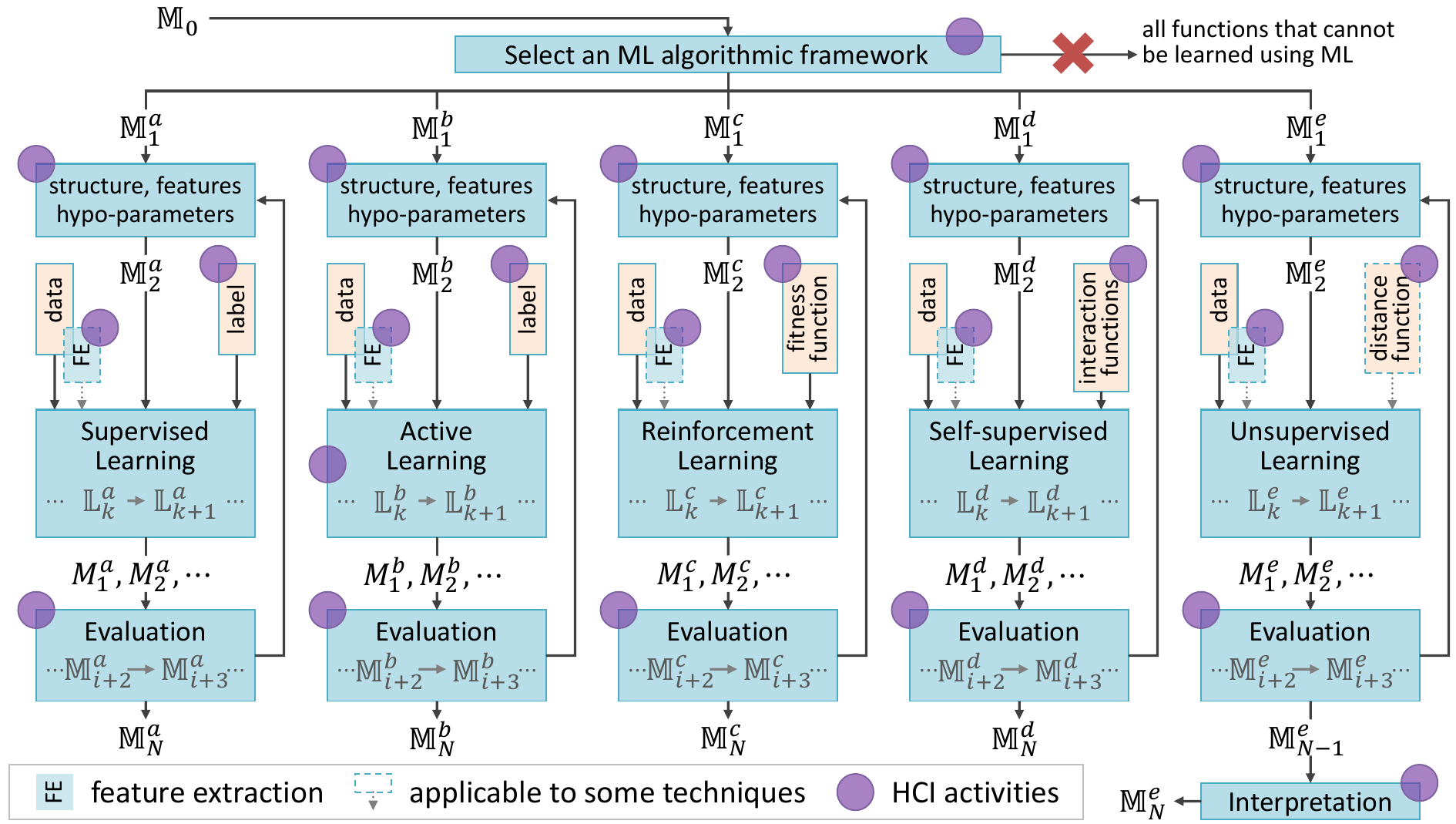}
    \caption{All ML workflows based on five major algorithmic frameworks feature a noticeable amount of HCI activities, and such activities enable a huge amount of alphabet compression and play a significant role in transforming $\mathbb{M}_0$ to $\mathbb{M}_N$.}
    \label{fig:HCI4ML}
\end{figure}

As soon as an ML developer decides to use a specific algorithmic framework of ML (e.g., convolutional neural network (CNN) or decision tree (DT), the initial alphabet $\mathbb{M}_0$ is transformed to $\mathbb{M}_1$, where the probability of many functions became 0. For example, the spaces of CNN and DT are known to be much smaller than the space of Turing Machine, so some functions, which can be written as a program in a conventional programming language, cannot be realized using a CNN or a DT. As illustrated in Fig. \ref{fig:HCI4ML}, if an ML developer selects ``Supervised Learning'', it changes the probabilities of all functions in $\mathbb{M}_0$, resulting in $\mathbb{M}_1^a$ such that those functions that cannot be realized using supervised learning will have a probability 0, while those functions that can be found using supervised learning will become more probable than they were in $\mathbb{M}_0$. An apparently simple decision by the ML developer results in a huge amount alphabet compression. 

When an ML developer further defines the template structure of an ML model, hypo-parameters, the feature set, many functions in $\mathbb{M}_1$ become impossible or less probable. Let $\mathbb{M}_2$ encompasses all remaining functions with their probabilities. The transformation from $\mathbb{M}_1$ to $\mathbb{M}_2$ involves many HCI activities, and the amount of alphabet compression is usually huge, hence the benefit of HCI can be huge.

For supervised learning, the labels assigned to data objects are typically entered by using HCI. For a classification scheme with $l$ classes, entering a label by an annotator enables $\log_2 l$ bits of alphabet compression. In other words, the computer receives $\log_2 l$ bits of information. For example, if there are 8 classes, $\log_2 l = 3$ bits. During the training process, each data label makes a small contribution, usually much less than $\log_2 l = 3$ bits due to the redundant, dependent, and conflicting information among the labels, to the alphabet compression from $\mathbb{M}_2^a$ that contains numerous candidate functions (i.e., models) to a specific model $M^a$. Therefore it is necessary to have a lot of data labels.

It is rare for an ML workflow to deliver a model with only one training process. The evaluation of a trained model by ML developers usually leads to a decision to invoke another iteration with modified hypo-parameters and sometimes the structure and feature selection.
Since each iteration delivers a model $M_i^a$ in alphabet $\mathbb{M}_{i+2}^a$, models from iterations $1, 2, \ldots, i$ (i.e., $M_1^a, M_2^a, \ldots, M_i^a$) are trained and evaluated, one of which is more likely to be selected than others. Hence the probability distribution of $\mathbb{M}_{i+2}^a$ changes with increasing $i$ during the iterative training processes.
In an ML workflow, there are usually many iterations, lasting for months. There are thus a lot of HCI activities. 

For active learning, there are also HCI activities during the training processes. Through HCI, ML developers are able to influence the evolution of the model space, $\mathbb{L}$, dynamically within a training process. Typically, the goal of HCI was to speed up the convergence (e.g., alphabet compression) by selecting more helpful training data.
 
The terms such as ``Self-supervised Learning'' or ``Unsupervised Learning'' may give an impression that these techniques do not rely on human input. It is mostly true that these techniques demand less HCI activities as it is not necessary to use HCI to label thousands or millions of data objects. However, through a smaller number of HCI activities, ML developers provide the ML processes with much critical knowledge. As shown in Fig. \ref{fig:HCI4ML}, in the case of ``Reinforcement Learning'', a fitness function needs to be defined, and in the case of ``Self-supervised Learning'', two or more interaction functions are needed. Some unsupervised learning techniques require a distance function, which demands a similar level of alphabet compression by using HCI.
Consider an alphabet $\mathbb{F}$ that includes all candidate functions, it would have numerous letters. Finding a letter $f \in \mathbb{F}$ to be a fitness function or interaction function is a transformation that enables a noticeable amount of alphabet compression. For example, if there were 1 million candidate functions in $\mathbb{F}$, the HCI activities for entering a particular function $f$ into the computer would enable nearly 20 bits alphabet compression. During the training process, the function $f$ would be used again and again in all iterations. If the training process converges, $f$ contributes to the alphabet compression from $\mathbb{M}_0$ to $\mathbb{M}_N$. 

As mentioned at the beginning of this section, we consider the development of classification models as an example of data intelligence workflows.  While the ML techniques under the four algorithmic frameworks on the left of Fig. \ref{fig:HCI4ML} produce models that can perform classification tasks, unsupervised learning techniques (under the fifth framework on the right) typically produce models that group the data objects in the training data into $k$ clusters. Such clustering models are not classification models yet, but can be transformed to classification models using HCI.

For example, it is common to project multivariate data objects onto a 2D scatter plot, where data objects are depicted as 2D points and color-coded according to their corresponding cluster labels. With such a scatter plot, an ML developer can divide the 2D space into $k$ partitions, e,g., by drawing polylines. The partitioning effort usually demands some complex human decisions as the clusters often overlap with each other in the projected space. Hence the HCI activities inject human knowledge that the computer do not have. Since there are numerous ways of divide the 2D space, HCI enables a huge amount of alphabet compression in the last step, ``Interpretation'', of the unsupervised learning workflow on the right of Fig. \ref{fig:HCI4ML}. For example, consider that the 2D space is at the $1024 \times 1024 = 2^{20}$ pixel resolution, and there are $k=8=2^3$ clusters. Theoretically, any one of the $2^{20}$ pixel may be associated with any one of the clusters. The alphabet $\mathbb{M}^e_{N-1}$ could have up to $3 \times 2^{20}$ bits of uncertainty (the maximal entropy).  

If an ML observes the visual patterns of the 8 clusters using the scatter plot and draws polylines to partition the 2D space, there are $2^{20}$ options for each point on these polylines. In other words, the input action alphabet for each input (see Section \ref{sec:Measures}) has $2^{20}$ letters.
If the ML developer managed to divide the 2D space into 8 partitions with $32=2^5$ inputs of 2D points, the computer would have received $2^{25}$ bits of information. With the information, the computer would be able to produce a classification model such that when any new data object is projected onto one of 8 partitions, it is classified with the same cluster label for that partition. In producing $\mathbb{M}^e_N$, the computer would have used the $2^{25}$ bits of human inputs to achieve alphabet compression of $3 \times 2^{20}$ bits.

\section{Conclusions}
\label{sec:Conclusions}
%

In this paper, we have shown that information-theoretic measures can provide a mathematical approach to  evaluate the cost-benefit of input actions. In conjunction with the previous discourse on the cost-benefit of visualization by Chen and Golan \cite{Chen:2016:TVCG}, we can now appreciate the value of the two-way HCI quantitatively. We have shown the applicability of this approach to basic input actions (e.g., selecting a radio button) as well as to complex input actions (e.g., defining a function in ML). The measures can be estimated analytically and measured empirically.

The information-theoretic approach presented in this paper is not a replacement for but an addition to the existing toolbox for supporting the design and evaluation of HCI devices, interfaces, and systems.
Because this approach allows us to examine the benefit of HCI to computers, i.e., from a perspective different from the commonly adopted focuses on the benefits to human users, it offers a new tool complementary to the existing qualitative and quantitative methods.

With estimated or measured quantitative values of HCI, we can appreciate more the necessity of HCI, especially in data intelligence workflows. We showed an analysis of a few ML workflows corresponding to widely-used algorithmic frameworks. The quantitative analysis confirms numerous qualitative arguments about humans' role in ML and AI, e.g., in the conversations in the HCAI Google Group \cite{HCAI:2022:web}.
To study the value of HCI is not in any way an attempt to forestall the advancement of technologies such as data mining, ML, and AI.
On the contrary, such research can help us understand better the transformation from human knowledge to computational models, and help us develop better automated processes to be used in data intelligence workflows.
As shown in an ontological map by Sacha et al. \cite{Sacha:2019:TVCG}, many steps in ML workflows have benefited, or can benefit, from visualization and interaction.
It is indeed not the time to reduce HCI in data intelligence, but to design and provide more cost-beneficial HCI.

In this paper, we have touched a few topics in a relatively broad spectrum of HCI in order to demonstrate the generality and applicability of the several information-theoretic measures discussed. Like all mathematical approaches, the proposed information-theoretic approach no doubt needs further experimentation, refinement, enrichment, and improvement before it can become a deployable tool in practical applications. Hopefully it will become part of the long-term endeavor of the HCI community.


\bibliographystyle{ACM-Reference-Format}
\bibliography{VoI}


\begin{thebibliography}{69}


\ifx \showCODEN    \undefined \def \showCODEN     #1{\unskip}     \fi
\ifx \showDOI      \undefined \def \showDOI       #1{#1}\fi
\ifx \showISBNx    \undefined \def \showISBNx     #1{\unskip}     \fi
\ifx \showISBNxiii \undefined \def \showISBNxiii  #1{\unskip}     \fi
\ifx \showISSN     \undefined \def \showISSN      #1{\unskip}     \fi
\ifx \showLCCN     \undefined \def \showLCCN      #1{\unskip}     \fi
\ifx \shownote     \undefined \def \shownote      #1{#1}          \fi
\ifx \showarticletitle \undefined \def \showarticletitle #1{#1}   \fi
\ifx \showURL      \undefined \def \showURL       {\relax}        \fi
\providecommand\bibfield[2]{#2}
\providecommand\bibinfo[2]{#2}
\providecommand\natexlab[1]{#1}
\providecommand\showeprint[2][]{arXiv:#2}

\bibitem[Atrash and Pineau(2009)]%
        {Atrash:2009:IUI}
\bibfield{author}{\bibinfo{person}{A. Atrash} {and} \bibinfo{person}{J.
  Pineau}.} \bibinfo{year}{2009}\natexlab{}.
\newblock \showarticletitle{A Bayesian Reinforcement Learning Approach for
  Customizing Human-Robot Interfaces}. In \bibinfo{booktitle}{\emph{Proc. 14th
  International Conference on Intelligent User Interfaces}}.
  \bibinfo{pages}{355–360}.
\newblock


\bibitem[Bias and Mayhew(2005)]%
        {Bias:2005:book}
\bibfield{author}{\bibinfo{person}{R.~G. Bias} {and} \bibinfo{person}{D.~J.
  Mayhew}.} \bibinfo{year}{2005}\natexlab{}.
\newblock \bibinfo{booktitle}{\emph{Cost Justifying Usability}
  (\bibinfo{edition}{2nd} ed.)}.
\newblock \bibinfo{publisher}{Elsevier, Oxford, UK.}
\newblock


\bibitem[Cairns and Cox(2008)]%
        {Cairns:2008:book}
\bibfield{editor}{\bibinfo{person}{P. Cairns} {and} \bibinfo{person}{A.~L.
  Cox}} (Eds.). \bibinfo{year}{2008}\natexlab{}.
\newblock \bibinfo{booktitle}{\emph{Research Methods for Human-Computer
  Interaction}}.
\newblock \bibinfo{publisher}{Cambridge University Press}.
\newblock


\bibitem[Card et~al\mbox{.}(1978)]%
        {Card:1978:E}
\bibfield{author}{\bibinfo{person}{S.~K. Card}, \bibinfo{person}{W.~K.
  English}, {and} \bibinfo{person}{B.~J. Burr}.}
  \bibinfo{year}{1978}\natexlab{}.
\newblock \showarticletitle{Evaluation of mouse, rate-controlled isometric
  joystick, step keys, and text keys for text selection on a CRT}.
\newblock \bibinfo{journal}{\emph{Ergonomics}} \bibinfo{volume}{21},
  \bibinfo{number}{8} (\bibinfo{year}{1978}), \bibinfo{pages}{601--613}.
\newblock


\bibitem[Chen(2020)]%
        {Chen:2020:book}
\bibfield{author}{\bibinfo{person}{M. Chen}.} \bibinfo{year}{2020}\natexlab{}.
\newblock \showarticletitle{Cost-Benefit Analysis of Data Intelligence -- Its
  Broader Interpretations}.
\newblock In \bibinfo{booktitle}{\emph{Advances in Info-Metrics: Information
  and Information Processing across Disciplines}}. \bibinfo{publisher}{Oxford
  University Press}.
\newblock


\bibitem[Chen et~al\mbox{.}(2022)]%
        {Chen:2022:E2}
\bibfield{author}{\bibinfo{person}{M. Chen}, \bibinfo{person}{A. Abdul-Rahman},
  \bibinfo{person}{D. Silver}, {and} \bibinfo{person}{M. Sbert}.}
  \bibinfo{year}{2022}\natexlab{}.
\newblock \showarticletitle{A bounded measure for estimating the benefit of
  visualization (Part II): case studies and empirical evaluation}.
\newblock \bibinfo{journal}{\emph{Entropy}} \bibinfo{volume}{24},
  \bibinfo{number}{2} (\bibinfo{year}{2022}), \bibinfo{pages}{282}.
\newblock


\bibitem[Chen and Ebert(2019)]%
        {Chen:2019:CGF}
\bibfield{author}{\bibinfo{person}{M. Chen} {and} \bibinfo{person}{D.~S.
  Ebert}.} \bibinfo{year}{2019}\natexlab{}.
\newblock \showarticletitle{An ontological framework for supporting the design
  and evaluation of visual analytics systems}.
\newblock \bibinfo{journal}{\emph{Computer Graphics Forum}}
  \bibinfo{volume}{38}, \bibinfo{number}{3} (\bibinfo{year}{2019}),
  \bibinfo{pages}{131--144}.
\newblock


\bibitem[Chen et~al\mbox{.}(2016)]%
        {Chen:2016:book}
\bibfield{author}{\bibinfo{person}{M. Chen}, \bibinfo{person}{M. Feixas},
  \bibinfo{person}{I. Viola}, \bibinfo{person}{A. Bardera},
  \bibinfo{person}{H.-W. Shen}, {and} \bibinfo{person}{M. Sbert}.}
  \bibinfo{year}{2016}\natexlab{}.
\newblock \bibinfo{booktitle}{\emph{Information Theory Tools for
  Visualization}}.
\newblock \bibinfo{publisher}{A K Peters/CRC Press}.
\newblock


\bibitem[Chen et~al\mbox{.}(2019)]%
        {Chen:2019:TVCG}
\bibfield{author}{\bibinfo{person}{M. Chen}, \bibinfo{person}{K. Gaither},
  \bibinfo{person}{N.~W. John}, {and} \bibinfo{person}{B. McCann}.}
  \bibinfo{year}{2019}\natexlab{}.
\newblock \showarticletitle{Cost-benefit analysis of visualization in virtual
  environments}.
\newblock \bibinfo{journal}{\emph{IEEE Transactions on Visualization and
  Computer Graphics}} \bibinfo{volume}{25}, \bibinfo{number}{1}
  (\bibinfo{year}{2019}), \bibinfo{pages}{32--42}.
\newblock


\bibitem[Chen and Golan(2016)]%
        {Chen:2016:TVCG}
\bibfield{author}{\bibinfo{person}{M. Chen} {and} \bibinfo{person}{A. Golan}.}
  \bibinfo{year}{2016}\natexlab{}.
\newblock \showarticletitle{What May Visualization Processes Optimize?}
\newblock \bibinfo{journal}{\emph{IEEE Transactions on Visualization and
  Computer Graphics}} \bibinfo{volume}{22}, \bibinfo{number}{12}
  (\bibinfo{year}{2016}), \bibinfo{pages}{2619--2632}.
\newblock


\bibitem[Chen and J\"anicke(2010)]%
        {Chen:2010:TVCG}
\bibfield{author}{\bibinfo{person}{M. Chen} {and} \bibinfo{person}{H.
  J\"anicke}.} \bibinfo{year}{2010}\natexlab{}.
\newblock \showarticletitle{An Information-theoretic Framework for
  Visualization}.
\newblock \bibinfo{journal}{\emph{IEEE Transactions on Visualization and
  Computer Graphics}} \bibinfo{volume}{16}, \bibinfo{number}{6}
  (\bibinfo{year}{2010}), \bibinfo{pages}{1206--1215}.
\newblock


\bibitem[Chen and Sbert(2022)]%
        {Chen:2022:E1}
\bibfield{author}{\bibinfo{person}{M. Chen} {and} \bibinfo{person}{M. Sbert}.}
  \bibinfo{year}{2022}\natexlab{}.
\newblock \showarticletitle{A bounded measure for estimating the benefit of
  visualization (Part I): theoretical discourse and conceptual evaluation}.
\newblock \bibinfo{journal}{\emph{Entropy}} \bibinfo{volume}{24},
  \bibinfo{number}{2} (\bibinfo{year}{2022}), \bibinfo{pages}{228}.
\newblock


\bibitem[Cockton(2004)]%
        {Cockton:2004:NordicCHI}
\bibfield{author}{\bibinfo{person}{G. Cockton}.}
  \bibinfo{year}{2004}\natexlab{}.
\newblock \showarticletitle{Value-Centred HCI}. In
  \bibinfo{booktitle}{\emph{Proc. Nordic HCI '04}}. \bibinfo{pages}{149--160}.
\newblock


\bibitem[Cover and Thomas(2006)]%
        {Cover:2006:book}
\bibfield{author}{\bibinfo{person}{T.~M. Cover} {and} \bibinfo{person}{J.~A.
  Thomas}.} \bibinfo{year}{2006}\natexlab{}.
\newblock \bibinfo{booktitle}{\emph{Elements of Information Theory}
  (\bibinfo{edition}{2nd} ed.)}.
\newblock \bibinfo{publisher}{John Wiley \& Sons}.
\newblock


\bibitem[Diaper and Stanton(2003)]%
        {Diaper:2003:book}
\bibfield{editor}{\bibinfo{person}{D. Diaper} {and} \bibinfo{person}{N.
  Stanton}} (Eds.). \bibinfo{year}{2003}\natexlab{}.
\newblock \bibinfo{booktitle}{\emph{The Handbook of Task Analysis for
  Human-Computer Interaction}}.
\newblock \bibinfo{publisher}{CRC Press}.
\newblock


\bibitem[Dix and Ellis(1998)]%
        {Dix:1998:AVI}
\bibfield{author}{\bibinfo{person}{A. Dix} {and} \bibinfo{person}{G. Ellis}.}
  \bibinfo{year}{1998}\natexlab{}.
\newblock \showarticletitle{Starting simple: adding value to static
  visualisation through simple interaction}. In \bibinfo{booktitle}{\emph{Proc.
  Working Conference on Advanced Visual Interfaces}}.
  \bibinfo{pages}{124–134}.
\newblock


\bibitem[Dix et~al\mbox{.}(2003)]%
        {Dix:2003:book}
\bibfield{author}{\bibinfo{person}{A. Dix}, \bibinfo{person}{J. Finlay},
  \bibinfo{person}{G.~D. Abowd}, {and} \bibinfo{person}{R. Beale}.}
  \bibinfo{year}{2003}\natexlab{}.
\newblock \bibinfo{booktitle}{\emph{Human-Computer Interaction}
  (\bibinfo{edition}{3rd} ed.)}.
\newblock \bibinfo{publisher}{Prentice Hall}.
\newblock


\bibitem[Donoghue(2002)]%
        {Donoghue:2002:book}
\bibfield{author}{\bibinfo{person}{K. Donoghue}.}
  \bibinfo{year}{2002}\natexlab{}.
\newblock \bibinfo{booktitle}{\emph{Built for Use: Driving Profitability
  Through The User Experience}}.
\newblock \bibinfo{publisher}{McGraw-Hill, New York}.
\newblock


\bibitem[Fitts and Posner(1967)]%
        {Fitts:1967:book}
\bibfield{author}{\bibinfo{person}{P. Fitts} {and} \bibinfo{person}{M.~I.
  Posner}.} \bibinfo{year}{1967}\natexlab{}.
\newblock \bibinfo{booktitle}{\emph{Human Performance}}.
\newblock \bibinfo{publisher}{Brooks/Cole Publishing},
  \bibinfo{address}{Belmont, CA.}
\newblock


\bibitem[Fitts(1954)]%
        {Fitts:1954:JEP}
\bibfield{author}{\bibinfo{person}{P.~M. Fitts}.}
  \bibinfo{year}{1954}\natexlab{}.
\newblock \showarticletitle{The information capacity of the human motor system
  in controlling the amplitude of movement}.
\newblock \bibinfo{journal}{\emph{Journal of Experimental Psychology}}
  \bibinfo{volume}{47}, \bibinfo{number}{6} (\bibinfo{year}{1954}),
  \bibinfo{pages}{381–391}.
\newblock


\bibitem[Friedman(1996)]%
        {Friedman:1996:I}
\bibfield{author}{\bibinfo{person}{B. Friedman}.}
  \bibinfo{year}{1996}\natexlab{}.
\newblock \showarticletitle{Value-Sensitive Design}.
\newblock \bibinfo{journal}{\emph{Interactions}} \bibinfo{volume}{3},
  \bibinfo{number}{6} (\bibinfo{year}{1996}), \bibinfo{pages}{16--23}.
\newblock


\bibitem[Giaccardi(2011)]%
        {Giaccardi:2011:I}
\bibfield{author}{\bibinfo{person}{E. Giaccardi}.}
  \bibinfo{year}{2011}\natexlab{}.
\newblock \showarticletitle{Things We Value}.
\newblock \bibinfo{journal}{\emph{Interactions}} \bibinfo{volume}{18},
  \bibinfo{number}{1} (\bibinfo{year}{2011}), \bibinfo{pages}{17--21}.
\newblock


\bibitem[Gilmore et~al\mbox{.}(2008)]%
        {Gilmore:2008:CHI}
\bibfield{author}{\bibinfo{person}{D. Gilmore}, \bibinfo{person}{G. Cockton},
  \bibinfo{person}{E. Churchill}, \bibinfo{person}{A. Henderson},
  \bibinfo{person}{Sari Kujala}, {and} \bibinfo{person}{M. Hammontree}.}
  \bibinfo{year}{2008}\natexlab{}.
\newblock \showarticletitle{Values, Value and Worth: Their Relationship to
  {HCI}?}. In \bibinfo{booktitle}{\emph{Proc. CHI '08 Extended Abstracts on
  Human Factors in Computing Systems}}. \bibinfo{pages}{3933--3936}.
\newblock


\bibitem[Gilmore(1995)]%
        {Gilmore:1995:book}
\bibfield{author}{\bibinfo{person}{D.~J. Gilmore}.}
  \bibinfo{year}{1995}\natexlab{}.
\newblock \showarticletitle{Interface Design: Have we got it wrong?}
\newblock In \bibinfo{booktitle}{\emph{Human-Computer Interaction (Interact
  '95)}}. \bibinfo{publisher}{Springer}, \bibinfo{pages}{173--178}.
\newblock


\bibitem[Gori et~al\mbox{.}(2018)]%
        {Gori:2018:TOCHI}
\bibfield{author}{\bibinfo{person}{J. Gori}, \bibinfo{person}{O. Rioul},
  \bibinfo{person}{Y.}, {and} \bibinfo{person}{Guiard}.}
  \bibinfo{year}{2018}\natexlab{}.
\newblock \showarticletitle{Speed-accuracy tradeoff: A formal
  information-theoretic transmission scheme ({FITTS})}.
\newblock \bibinfo{journal}{\emph{ACM Transactions on Computer-Human
  Interaction}} \bibinfo{volume}{25}, \bibinfo{number}{5}
  (\bibinfo{year}{2018}), \bibinfo{pages}{1--33}.
\newblock


\bibitem[Helander et~al\mbox{.}(1997)]%
        {Helander:1997:book}
\bibfield{editor}{\bibinfo{person}{M.~G. Helander}, \bibinfo{person}{T.~K.
  Landauer}, {and} \bibinfo{person}{P.~V. Prabhu}} (Eds.).
  \bibinfo{year}{1997}\natexlab{}.
\newblock \bibinfo{booktitle}{\emph{Handbook of Human-Computer Interaction}}.
\newblock \bibinfo{publisher}{Elsevier}.
\newblock


\bibitem[Hornbæk and Oulasvirta(2017)]%
        {Hornbaek:2017:CHI}
\bibfield{author}{\bibinfo{person}{K. Hornbæk} {and} \bibinfo{person}{A.
  Oulasvirta}.} \bibinfo{year}{2017}\natexlab{}.
\newblock \showarticletitle{What is interaction?}. In
  \bibinfo{booktitle}{\emph{Proc. ACM CHI Conference on Human Factors in
  Computing Systems}}. \bibinfo{pages}{5040--5052}.
\newblock


\bibitem[Huffman(1952)]%
        {Huffman:1952:IRE}
\bibfield{author}{\bibinfo{person}{D.~A. Huffman}.}
  \bibinfo{year}{1952}\natexlab{}.
\newblock \showarticletitle{A Method for the Construction of Minimum-Redundancy
  Codes}.
\newblock \bibinfo{journal}{\emph{Proceedings of the IRE}}
  \bibinfo{volume}{40}, \bibinfo{number}{9} (\bibinfo{year}{1952}),
  \bibinfo{pages}{1098--1101}.
\newblock


\bibitem[Jacko(2012)]%
        {Jacko:2012:book}
\bibfield{editor}{\bibinfo{person}{J.~A. Jacko}} (Ed.).
  \bibinfo{year}{2012}\natexlab{}.
\newblock \bibinfo{booktitle}{\emph{Human Computer Interaction Handbook:
  Fundamentals, Evolving Technologies, and Emerging Applications}
  (\bibinfo{edition}{3rd} ed.)}.
\newblock \bibinfo{publisher}{CRC Press}.
\newblock


\bibitem[Jankun-Kelly et~al\mbox{.}(2007)]%
        {Jankun-Kelly:2007:TVCG}
\bibfield{author}{\bibinfo{person}{T. Jankun-Kelly}, \bibinfo{person}{K.-L.
  Ma}, {and} \bibinfo{person}{M. Gertz}.} \bibinfo{year}{2007}\natexlab{}.
\newblock \showarticletitle{A model and framework for visualization
  exploration}.
\newblock \bibinfo{journal}{\emph{IEEE Transactions on Visualization and
  Computer Graphics}} \bibinfo{volume}{13}, \bibinfo{number}{6}
  (\bibinfo{year}{2007}), \bibinfo{pages}{357–369}.
\newblock


\bibitem[Jansen and Dragicevic(2013)]%
        {Jansen:2013:TVCG}
\bibfield{author}{\bibinfo{person}{Yvonne Jansen} {and} \bibinfo{person}{Pierre
  Dragicevic}.} \bibinfo{year}{2013}\natexlab{}.
\newblock \showarticletitle{An Interaction Model for Visualizations Beyond The
  Desktop}.
\newblock \bibinfo{journal}{\emph{IEEE Transactions on Visualization and
  Computer Graphics}} \bibinfo{volume}{19}, \bibinfo{number}{12}
  (\bibinfo{year}{2013}), \bibinfo{pages}{2396--2405}.
\newblock


\bibitem[Kijmongkolchai et~al\mbox{.}(2017)]%
        {Kijmongkolchai:2017:CGF}
\bibfield{author}{\bibinfo{person}{N. Kijmongkolchai}, \bibinfo{person}{A.
  Abdul-Rahman}, {and} \bibinfo{person}{M. Chen}.}
  \bibinfo{year}{2017}\natexlab{}.
\newblock \showarticletitle{Empirically measuring soft knowledge in
  visualization}.
\newblock \bibinfo{journal}{\emph{Computer Graphics Forum}}
  \bibinfo{volume}{36}, \bibinfo{number}{3} (\bibinfo{year}{2017}),
  \bibinfo{pages}{73--85}.
\newblock


\bibitem[Lam(2008)]%
        {Lam:2008:TVCG}
\bibfield{author}{\bibinfo{person}{H. Lam}.} \bibinfo{year}{2008}\natexlab{}.
\newblock \showarticletitle{A Framework of Interaction Costs in Information
  Visualization}.
\newblock \bibinfo{journal}{\emph{IEEE Transactions on Visualization and
  Computer Graphics}} \bibinfo{volume}{14}, \bibinfo{number}{6}
  (\bibinfo{year}{2008}), \bibinfo{pages}{1149--1156}.
\newblock


\bibitem[Lazar et~al\mbox{.}(2010)]%
        {Lazar:2010:book}
\bibfield{author}{\bibinfo{person}{J. Lazar}, \bibinfo{person}{J.~H. Feng},
  {and} \bibinfo{person}{H. Hochheiser}.} \bibinfo{year}{2010}\natexlab{}.
\newblock \bibinfo{booktitle}{\emph{Research Methods in Human-Computer
  Interaction}}.
\newblock \bibinfo{publisher}{Wiley}.
\newblock


\bibitem[Light et~al\mbox{.}(2005)]%
        {Light:2005:CHI}
\bibfield{author}{\bibinfo{person}{A. Light}, \bibinfo{person}{P.~J. Wild},
  \bibinfo{person}{A. Dearden}, {and} \bibinfo{person}{M.~J. Muller}.}
  \bibinfo{year}{2005}\natexlab{}.
\newblock \showarticletitle{Quality, Value(s) and Choice: Exploring deeper
  outcomes for HCI products}. In \bibinfo{booktitle}{\emph{Proc. CHI Extended
  Abstracts on Human Factors in Computing Systems}}.
  \bibinfo{pages}{2124--2125}.
\newblock


\bibitem[Liu et~al\mbox{.}(2017)]%
        {Liu:2017:CHI}
\bibfield{author}{\bibinfo{person}{W. Liu}, \bibinfo{person}{R.~L. d'Oliveira},
  \bibinfo{person}{M. Beaudouin-Lafon}, {and} \bibinfo{person}{O. Rioul}.}
  \bibinfo{year}{2017}\natexlab{}.
\newblock \showarticletitle{Bignav: Bayesian information gain for guiding
  multiscale navigation}. In \bibinfo{booktitle}{\emph{Proc. ACM CHI Conference
  on Human Factors in Computing Systems}}. \bibinfo{pages}{5869--5880}.
\newblock


\bibitem[Liu et~al\mbox{.}(2018)]%
        {Liu:2018:CHI}
\bibfield{author}{\bibinfo{person}{W. Liu}, \bibinfo{person}{O. Rioul},
  \bibinfo{person}{J. Mcgrenere}, \bibinfo{person}{W.~E. Mackay}, {and}
  \bibinfo{person}{M. Beaudouin-Lafon}.} \bibinfo{year}{2018}\natexlab{}.
\newblock \showarticletitle{BIGFile: Bayesian information gain for fast file
  retrieval}. In \bibinfo{booktitle}{\emph{Proc. ACM CHI Conference on Human
  Factors in Computing Systems}}. Article \bibinfo{articleno}{385},
  \bibinfo{numpages}{13}~pages.
\newblock


\bibitem[MacKenzie(1992)]%
        {MacKenzie:1992:CHI}
\bibfield{author}{\bibinfo{person}{I.~S. MacKenzie}.}
  \bibinfo{year}{1992}\natexlab{}.
\newblock \showarticletitle{Fitts' law as a research and design tool in
  human–computer interaction}.
\newblock \bibinfo{journal}{\emph{Human–Computer Interaction}}
  \bibinfo{volume}{7} (\bibinfo{year}{1992}), \bibinfo{pages}{91--139}.
\newblock


\bibitem[MacKenzie(2013)]%
        {MacKenzie:2013:book}
\bibfield{author}{\bibinfo{person}{I.~S. MacKenzie}.}
  \bibinfo{year}{2013}\natexlab{}.
\newblock \bibinfo{booktitle}{\emph{Human-Computer Interaction: An Empirical
  Research Perspective}}.
\newblock \bibinfo{publisher}{Morgan Kaufmann}.
\newblock


\bibitem[Micallef et~al\mbox{.}(2012)]%
        {Micallef:2012:TVCG}
\bibfield{author}{\bibinfo{person}{L. Micallef}, \bibinfo{person}{P.
  Dragicevic}, {and} \bibinfo{person}{J.-D. Fekete}.}
  \bibinfo{year}{2012}\natexlab{}.
\newblock \showarticletitle{Assessing the Effect of Visualizations on Bayesian
  Reasoning Through Crowdsourcing}.
\newblock \bibinfo{journal}{\emph{IEEE Transactions on Visualization and
  Computer Graphics}} \bibinfo{volume}{18}, \bibinfo{number}{12}
  (\bibinfo{year}{2012}), \bibinfo{pages}{2536--2545}.
\newblock


\bibitem[Mosca et~al\mbox{.}(2021)]%
        {Mosca:2021:CHI}
\bibfield{author}{\bibinfo{person}{A. Mosca}, \bibinfo{person}{A. Ottley},
  {and} \bibinfo{person}{R. Chang}.} \bibinfo{year}{2021}\natexlab{}.
\newblock \showarticletitle{Does Interaction Improve Bayesian Reasoning with
  Visualization?}. In \bibinfo{booktitle}{\emph{Proc. ACM CHI Conference on
  Human Factors in Computing Systems}}. Article \bibinfo{articleno}{609},
  \bibinfo{numpages}{14}~pages.
\newblock


\bibitem[Norman and Kirakowski(2018)]%
        {Norman:2018:book}
\bibfield{editor}{\bibinfo{person}{K. Norman} {and} \bibinfo{person}{J.
  Kirakowski}} (Eds.). \bibinfo{year}{2018}\natexlab{}.
\newblock \bibinfo{booktitle}{\emph{The Wiley Handbook of Human Computer
  Interaction}}.
\newblock \bibinfo{publisher}{Wiley-Blackwell}.
\newblock


\bibitem[Ottley et~al\mbox{.}(2016)]%
        {Ottley:2016:TVCG}
\bibfield{author}{\bibinfo{person}{A. Ottley}, \bibinfo{person}{E.~M. Peck},
  \bibinfo{person}{L.~T. Harrison}, \bibinfo{person}{D. Afergan},
  \bibinfo{person}{C. Ziemkiewicz}, \bibinfo{person}{H.~A. Taylor},
  \bibinfo{person}{P.~K.~J. Han}, {and} \bibinfo{person}{R. Chang}.}
  \bibinfo{year}{2016}\natexlab{}.
\newblock \showarticletitle{Improving Bayesian Reasoning: The Effects of
  Phrasing, Visualization, and Spatial Ability}.
\newblock \bibinfo{journal}{\emph{IEEE Transactions on Visualization and
  Computer Graphics}} \bibinfo{volume}{22}, \bibinfo{number}{1}
  (\bibinfo{year}{2016}), \bibinfo{pages}{529--538}.
\newblock


\bibitem[Oulasvirta et~al\mbox{.}(2018)]%
        {Oulasvirta:2018:book}
\bibfield{editor}{\bibinfo{person}{A. Oulasvirta}, \bibinfo{person}{P.~O.
  Kristensson}, \bibinfo{person}{X. Bi}, {and} \bibinfo{person}{A. Howes}}
  (Eds.). \bibinfo{year}{2018}\natexlab{}.
\newblock \bibinfo{booktitle}{\emph{Computational Interaction}}.
\newblock \bibinfo{publisher}{Oxford University Press}.
\newblock


\bibitem[Preece et~al\mbox{.}(1994)]%
        {Preece:1994:book}
\bibfield{author}{\bibinfo{person}{J. Preece}, \bibinfo{person}{Y. Rogers},
  \bibinfo{person}{H. Sharp}, \bibinfo{person}{D. Benyon}, \bibinfo{person}{S.
  Holland}, {and} \bibinfo{person}{T. Carey}.} \bibinfo{year}{1994}\natexlab{}.
\newblock \bibinfo{booktitle}{\emph{Human-Computer Interaction: Concepts And
  Design}}.
\newblock \bibinfo{publisher}{Addison Wesley}.
\newblock


\bibitem[Preece et~al\mbox{.}(2015)]%
        {Preece:2015:book}
\bibfield{author}{\bibinfo{person}{J. Preece}, \bibinfo{person}{H. Sharp},
  {and} \bibinfo{person}{Y. Rogers}.} \bibinfo{year}{2015}\natexlab{}.
\newblock \bibinfo{booktitle}{\emph{Interaction Design: Beyond Human-Computer
  Interaction} (\bibinfo{edition}{4th} ed.)}.
\newblock \bibinfo{publisher}{John Wiley}.
\newblock


\bibitem[Purchase(2012)]%
        {Purchase:2012:book}
\bibfield{author}{\bibinfo{person}{H.~C. Purchase}.}
  \bibinfo{year}{2012}\natexlab{}.
\newblock \bibinfo{booktitle}{\emph{Experimental Human-Computer Interaction: A
  Practical Guide with Visual Examples}}.
\newblock \bibinfo{publisher}{Cambridge University Press}.
\newblock


\bibitem[Rotondo and Freier(2010)]%
        {Rotondo:2010:CHI}
\bibfield{author}{\bibinfo{person}{A. Rotondo} {and} \bibinfo{person}{N.
  Freier}.} \bibinfo{year}{2010}\natexlab{}.
\newblock \showarticletitle{The Problem of Defining Values: A Lack of Common
  Ground Between Industry \& Academia?}. In \bibinfo{booktitle}{\emph{Proc. CHI
  '2010 (Work-in-Progress)}}. \bibinfo{pages}{4183--4188}.
\newblock


\bibitem[Rydow et~al\mbox{.}(2023)]%
        {Rydow:2023:TVCG}
\bibfield{author}{\bibinfo{person}{E. Rydow}, \bibinfo{person}{R. Borgo},
  \bibinfo{person}{H. Fang}, \bibinfo{person}{T. Torsney-Weir},
  \bibinfo{person}{B. Swallow}, \bibinfo{person}{T. Porphyre},
  \bibinfo{person}{C. Turkay}, {and} \bibinfo{person}{M. Chen}.}
  \bibinfo{year}{2023}\natexlab{}.
\newblock \showarticletitle{Development and evaluation of two approaches of
  visual sensitivity analysis to support epidemiological modelling.}
\newblock \bibinfo{journal}{\emph{IEEE Transactions on Visualization and
  Computer Graphics}} \bibinfo{volume}{29}, \bibinfo{number}{1}
  (\bibinfo{year}{2023}).
\newblock


\bibitem[Sacha et~al\mbox{.}(2019)]%
        {Sacha:2019:TVCG}
\bibfield{author}{\bibinfo{person}{D. Sacha}, \bibinfo{person}{M. Kraus},
  \bibinfo{person}{D.~A. Keim}, {and} \bibinfo{person}{M. Chen}.}
  \bibinfo{year}{2019}\natexlab{}.
\newblock \showarticletitle{{VIS4ML}: An ontology for visual analytics assisted
  machine learning}.
\newblock \bibinfo{journal}{\emph{IEEE Transactions on Visualization and
  Computer Graphics}} \bibinfo{volume}{25}, \bibinfo{number}{1}
  (\bibinfo{year}{2019}).
\newblock


\bibitem[Saffer(2009)]%
        {Saffer:2009:book}
\bibfield{author}{\bibinfo{person}{Dan Saffer}.}
  \bibinfo{year}{2009}\natexlab{}.
\newblock \bibinfo{booktitle}{\emph{Designing for Interaction: Creating
  Innovative Applications and Devices} (\bibinfo{edition}{2nd} ed.)}.
\newblock \bibinfo{publisher}{New Riders}.
\newblock


\bibitem[Sedig and Parsons(2013)]%
        {Sedig:2013:THCI}
\bibfield{author}{\bibinfo{person}{K. Sedig} {and} \bibinfo{person}{P.
  Parsons}.} \bibinfo{year}{2013}\natexlab{}.
\newblock \showarticletitle{Interaction design for complex cognitive activities
  with visual representations: A pattern-based approach}.
\newblock \bibinfo{journal}{\emph{AIS Transactions on Human-Computer
  Interaction}} \bibinfo{volume}{5}, \bibinfo{number}{1}
  (\bibinfo{year}{2013}), \bibinfo{pages}{84–133}.
\newblock


\bibitem[Shannon(1948)]%
        {Shannon:1948:BSTJ}
\bibfield{author}{\bibinfo{person}{C.~E. Shannon}.}
  \bibinfo{year}{1948}\natexlab{}.
\newblock \showarticletitle{A mathematical theory of communication}.
\newblock \bibinfo{journal}{\emph{Bell System Technical Journal}}
  \bibinfo{volume}{27} (\bibinfo{year}{1948}), \bibinfo{pages}{379--423}.
\newblock


\bibitem[Shannon(1951)]%
        {Shannon:1951:BSTJ}
\bibfield{author}{\bibinfo{person}{C.~E. Shannon}.}
  \bibinfo{year}{1951}\natexlab{}.
\newblock \showarticletitle{Prediction and Entropy of Printed {English}}.
\newblock \bibinfo{journal}{\emph{Bell System Technical Journal}}
  \bibinfo{volume}{30} (\bibinfo{year}{1951}), \bibinfo{pages}{50--64}.
\newblock


\bibitem[Shilton(2018)]%
        {Shilton:2018:FT}
\bibfield{author}{\bibinfo{person}{K. Shilton}.}
  \bibinfo{year}{2018}\natexlab{}.
\newblock \showarticletitle{Values and Ethics in Human-Computer Interaction}.
\newblock \bibinfo{journal}{\emph{Foundations and Trends in Human-Computer
  Interaction}} \bibinfo{volume}{12}, \bibinfo{number}{2}
  (\bibinfo{year}{2018}), \bibinfo{pages}{107--171}.
\newblock


\bibitem[Shneiderman(1996)]%
        {Shneiderman:1996:VL}
\bibfield{author}{\bibinfo{person}{B. Shneiderman}.}
  \bibinfo{year}{1996}\natexlab{}.
\newblock \showarticletitle{The Eyes Have It: A Task by Data Type Taxonomy for
  Information Visualizations}. In \bibinfo{booktitle}{\emph{Proc. IEEE
  Symposium on Visual Languages}}. \bibinfo{pages}{336--343}.
\newblock


\bibitem[Shneiderman(2022)]%
        {HCAI:2022:web}
\bibfield{author}{\bibinfo{person}{Ben Shneiderman}.} \bibinfo{year}{Accessed
  in September 2022}\natexlab{}.
\newblock \bibinfo{title}{Human-Centered-AI}.
\newblock
  \bibinfo{howpublished}{https://groups.google.com/g/human-centered-ai}.
\newblock


\bibitem[Shneiderman et~al\mbox{.}(2010)]%
        {Shneiderman:2010:book}
\bibfield{author}{\bibinfo{person}{B. Shneiderman}, \bibinfo{person}{C.
  Plaisant}, \bibinfo{person}{M. Cohen}, {and} \bibinfo{person}{S. Jacobs}.}
  \bibinfo{year}{2010}\natexlab{}.
\newblock \bibinfo{booktitle}{\emph{Designing the user interface : strategies
  for effective human-computer interaction}}.
\newblock \bibinfo{publisher}{Pearson}.
\newblock


\bibitem[Smith et~al\mbox{.}(2014)]%
        {Smith:2014:OzCHI}
\bibfield{author}{\bibinfo{person}{W. Smith}, \bibinfo{person}{G. Wadley},
  \bibinfo{person}{S. Webber}, \bibinfo{person}{B. Ploderer}, {and}
  \bibinfo{person}{R. Lederman}.} \bibinfo{year}{2014}\natexlab{}.
\newblock \showarticletitle{Unbounding the Interaction Design Problem: the
  Contribution of HCI in Three Interventions for Well-being}. In
  \bibinfo{booktitle}{\emph{Proc. OzCHI '14}}. \bibinfo{pages}{392--395}.
\newblock


\bibitem[Soukoreff et~al\mbox{.}(2011)]%
        {Soukoreff:2011:INTERACT}
\bibfield{author}{\bibinfo{person}{R.~W. Soukoreff}, \bibinfo{person}{J. Zhao},
  {and} \bibinfo{person}{X. Ren}.} \bibinfo{year}{2011}\natexlab{}.
\newblock \showarticletitle{The Entropy of a Rapid Aimed Movement: Fitts' Index
  of Difficulty versus Shannon's Entropy}.
\newblock \bibinfo{journal}{\emph{Proc. 13th IFIP TC International Conference
  on Human-Computer Interaction (INTERACT'11)}}  \bibinfo{volume}{Part IV},
  \bibinfo{pages}{222–239}.
\newblock


\bibitem[Tam et~al\mbox{.}(2017)]%
        {Tam:2017:TVCG}
\bibfield{author}{\bibinfo{person}{G.~K.~L. Tam}, \bibinfo{person}{V. Kothari},
  {and} \bibinfo{person}{M. Chen}.} \bibinfo{year}{2017}\natexlab{}.
\newblock \showarticletitle{An analysis of machine- and human-analytics in
  classification}.
\newblock \bibinfo{journal}{\emph{IEEE Transactions on Visualization and
  Computer Graphics}} \bibinfo{volume}{23}, \bibinfo{number}{1}
  (\bibinfo{year}{2017}), \bibinfo{pages}{71--80}.
\newblock


\bibitem[Tan and Nijholt(2010)]%
        {Tan:2010:book}
\bibfield{editor}{\bibinfo{person}{D.~S. Tan} {and} \bibinfo{person}{A.
  Nijholt}} (Eds.). \bibinfo{year}{2010}\natexlab{}.
\newblock \bibinfo{booktitle}{\emph{Brain-Computer Interfaces: Applying our
  Minds to Human-Computer Interaction}}.
\newblock \bibinfo{publisher}{Springer}.
\newblock


\bibitem[Tsai et~al\mbox{.}(2011)]%
        {Tsai:2011:PHFESAM}
\bibfield{author}{\bibinfo{person}{J. Tsai}, \bibinfo{person}{S. Miller}, {and}
  \bibinfo{person}{A. Kirlik}.} \bibinfo{year}{2011}\natexlab{}.
\newblock \showarticletitle{Interactive Visualizations to Improve Bayesian
  Reasoning}.
\newblock \bibinfo{journal}{\emph{Proceedings of the Human Factors and
  Ergonomics Society Annual Meeting}} \bibinfo{volume}{55}, \bibinfo{number}{1}
  (\bibinfo{year}{2011}), \bibinfo{pages}{385--389}.
\newblock


\bibitem[Tweedie(1997)]%
        {Tweedie:1997:CHI}
\bibfield{author}{\bibinfo{person}{L. Tweedie}.}
  \bibinfo{year}{1997}\natexlab{}.
\newblock \showarticletitle{Characterizing interactive externalizations}. In
  \bibinfo{booktitle}{\emph{Proc. ACM SIGCHI Conference on Human Factors in
  Computing Systems}}. \bibinfo{pages}{375–382}.
\newblock


\bibitem[{von Landesberger} et~al\mbox{.}(2014)]%
        {Landesberger:2014:book}
\bibfield{author}{\bibinfo{person}{T. {von Landesberger}}, \bibinfo{person}{S.
  Fiebig}, \bibinfo{person}{S. Bremm}, \bibinfo{person}{A. Kuijper}, {and}
  \bibinfo{person}{D.~W. Fellner}.} \bibinfo{year}{2014}\natexlab{}.
\newblock \showarticletitle{Interaction taxonomy for tracking of user actions
  in visual analytics applications}.
\newblock In \bibinfo{booktitle}{\emph{Handbook of Human Centric
  Visualization}}, \bibfield{editor}{\bibinfo{person}{W.~Huang},
  \bibinfo{person}{P.~Parsons}, {and} \bibinfo{person}{K.~Sedig}} (Eds.).
  \bibinfo{publisher}{Springer}, \bibinfo{pages}{653–670}.
\newblock


\bibitem[Wybrow et~al\mbox{.}(2014)]%
        {Wybrow:2014:book}
\bibfield{author}{\bibinfo{person}{M. Wybrow}, \bibinfo{person}{N. Elmqvist},
  \bibinfo{person}{J.-D. Fekete}, \bibinfo{person}{T. {von Landesberger}},
  \bibinfo{person}{J.~J. {van Wijk}}, {and} \bibinfo{person}{B. Zimmer}.}
  \bibinfo{year}{2014}\natexlab{}.
\newblock \showarticletitle{Interaction in the visualization of multivariate
  networks}.
\newblock In \bibinfo{booktitle}{\emph{Multivariate Network Visualization}}.
  Vol.~\bibinfo{volume}{LNCS 8380}. \bibinfo{publisher}{Springer},
  \bibinfo{pages}{97–125}.
\newblock


\bibitem[Ye and Chen(2023)]%
        {Ye:2023:TVCG}
\bibfield{author}{\bibinfo{person}{Z. Ye} {and} \bibinfo{person}{M. Chen}.}
  \bibinfo{year}{2023}\natexlab{}.
\newblock \showarticletitle{Visualizing ensemble predictions of music mood}.
\newblock \bibinfo{journal}{\emph{IEEE Transactions on Visualization and
  Computer Graphics}} \bibinfo{volume}{29}, \bibinfo{number}{1}
  (\bibinfo{year}{2023}).
\newblock


\bibitem[Yi et~al\mbox{.}(2007)]%
        {Yi:2007:TVCG}
\bibfield{author}{\bibinfo{person}{J.~S. Yi}, \bibinfo{person}{Y. a. Kang},
  \bibinfo{person}{J.~T. Stasko}, {and} \bibinfo{person}{J.~A. Jacko}.}
  \bibinfo{year}{2007}\natexlab{}.
\newblock \showarticletitle{Toward a Deeper Understanding of the Role of
  Interaction in Information Visualization}.
\newblock \bibinfo{journal}{\emph{IEEE Transactions on Visualization and
  Computer Graphics}} \bibinfo{volume}{13}, \bibinfo{number}{6}
  (\bibinfo{year}{2007}), \bibinfo{pages}{1224--1231}.
\newblock


\bibitem[Zhao et~al\mbox{.}(2022)]%
        {Zhao:2022:UIST}
\bibfield{author}{\bibinfo{person}{H. Zhao}, \bibinfo{person}{S. Gu},
  \bibinfo{person}{C. Yu}, {and} \bibinfo{person}{X. Bi}.}
  \bibinfo{year}{2022}\natexlab{}.
\newblock \showarticletitle{Bayesian Hierarchical Pointing Models}. In
  \bibinfo{booktitle}{\emph{Proc. 35th Annual ACM Symposium on User Interface
  Software and Technology}}. Article \bibinfo{articleno}{87},
  \bibinfo{numpages}{13}~pages.
\newblock


\end{thebibliography}

\newpage

\noindent
\textsf{\Large{Supplementary Materials} \normalsize{(They are not part of the paper and are included here for the review process)}\\[5mm]
\LARGE{\textbf{The Value of Interaction in Data Intelligence}}\\[2mm]
\Large{MIN CHEN, University of Oxford}\\[10mm]
}

\section*{Supplement A: Author Feedback and Revision Actions}
Dear Kasper Hornbæk (18 October 2023),

\vspace{2mm}
\noindent \textbf{Re:} \emph{The Value of Interaction in Data Intelligence},
\noindent Min Chen, University of Oxford, UK
\vspace{2mm}

Thank you for your review comments and suggestions on 3 September 2023. I have revised the paper according to your review comments and suggestions:

\color{violet}%
\begin{itemize}
    \item[] ``Unfortunately, my reading of the paper is that it does not succeed. One problem is that existing work on interaction is not well covered. For example, I missed the work of Yi and Stasko. Heidi Lam's work was also missing, as well as Yvonne Jansen's work on pipelines in information visualization. The work on interaction by Oulasvirta and myself also seems relevant to this discussion. Another issue is that the proposed frameworks (e.g., in Figure 8) do not cover data intelligence tasks in general in my reading. So my reading is that there are too many loose ends and too much related work that is not well covered currently to warrant consideration for publication in TOCHI.''
\end{itemize}
\color{black}%

In particular, I have included brief discussions on the work by Yi, Kang, Stasko \& Jacko; Lam; Jansen \& Dragicevic, and Hornbæk \& Oulasvirta as well as six newly-added references.

Figure 8 shows examples of ML processes as indicated in the caption and body-text. I am not aware that there is a general diagram for covering all data intelligence workflows. For example, Chen and Golan \cite{Chen:2016:TVCG} provided a few workflows for visual analytics. In terms of whether the proposed cost-benefit measure can be applied to all HCI processes in data intelligence workflows, my answer is yes (see also the first statement in Section 7). I am confident about this because the measure was proposed in Chen and Golan \cite{Chen:2016:TVCG} for covering processes that featuring statistics, algorithms, visualization, and interactions. It was extended further in Chen \cite{Chen:2020:book}. I am not aware of any counter-example so far.  In science, it is often not easy to prove "covering all". If one can point out a counter-example, i.e., a type of HCI processes that are not covered. It will be easy for me to discuss such a counter-example.

From your comments, I cannot see any fundamental issues about the main contributions of this work. I will therefore be updating the arXiv report for this paper (arXiv:1812.06051). I hope that I have addressed your concerns, and hopefully the paper is ready for entering a peer-review process.

\vspace{2mm}
\noindent Sincerely yours,\\
Min Chen

\begin{center}
    $\ast \qquad \ast \qquad \ast$
\end{center}

\vspace{4mm}
\noindent Dear ACM TOCHI Editors and Reviewers (30 August 2023),

\vspace{2mm}
\noindent \textbf{Re:} \emph{The Value of Interaction in Data Intelligence},
\noindent Min Chen, University of Oxford, UK
\vspace{2mm}

Thank you in advance for your consideration of this work. My main research association is with the IEEE VIS community, though I have been working on HCI since my PhD years in the 1980s. My PhD work was on graphical user interface, but somehow I became part of the graphics and then VIS communities. In 2010, we first showed that we could mathematically confirm the usefulness of interaction and visualization in a TVCG paper (10.1109/TVCG.2010.132). After we developed a mathematical formula that could not only explain the usefulness but also sometimes the drawbacks of visualization in 2016 (10.1109/TVCG.2015.2513410), I always wanted to use the formula to explain the usefulness and sometimes the drawbacks of interaction.

A UK colleague invited me to give a keynote at HCI 2018 in Belfast, and I took the opportunity to formulate my thought on this (https://sites.google.com/site/drminchen/slides/HCI2018-MC-P.pdf), which led to the first version of this work that was submitted to ACM CHI2019 in 2018. As I will discuss later, though the paper was not accepted, the reviews did not indicate any serious issue. Hence I made the paper available at arXiv (https://arxiv.org/abs/1812.06051v1). I did think about resubmitting it to TOCHI. After I created a submission at ScholarOne, I realized that TOCHI required anonymized submission at that time, and my arXiv report made such a submission impossible.

During the COVID-19 period, I did not manage to find time to work on this paper, largely due to my commitment to a large VIS project for supporting epidemiological modeling in the UK. At the beginning of 2022, we managed to improve the mathematical formula for measuring the cost-benefit of visualization processes (10.3390/e24020228, 10.3390/e24020282), which would also improve the mathematical underpinning for the empirical studies in HCI as presented in the first version of the work. This motivated me to provide the second version of this work, which was submitted to ACM CHI 2023. The paper was not accepted though the reviewers did not identify any serious issue, as I will discuss later. After I updated my arXiv report (https://arxiv.org/abs/1812.06051), I noticed that TOCHI no longer requires anonymized submission. I hope that TOCHI reviewers will have a bit more time and knowledge to appreciate this work. Importantly, I do wish that HCI colleagues will gradually recognize the merits of this work, as the work offers a mathematical formula that can measure the cost-benefit of the human-to-computer aspects of the HCI. The formula is quantitative by nature and can be estimated in qualitative analysis of an interface design. The paper shows that the formula can be applied to a broad spectrum of HCI activities, from the device level to empirical studies and to HCI in machine learning.

At the end of this document, I am including the reviews from ACM CHI2019 and ACM CHI2023. Just below this letter, I am providing my feedback to the main questions or concerns raised in these reviews, while indicating my revision actions.

\vspace{2mm}
\noindent Sincerely yours,\\
Min Chen

\subsection*{General Merits or Demerits}

\begin{enumerate}
    \color{violet}%
    \item[Q1.] \textbf{Novelty and Significance.}
    There are opposite views on this criterion. CHI2019-2AC2 commented the work positively as ``\emph{addressed an interesting as well as important topic ...}'' and suggested that ``\emph{a better structured or more concise writing could make the paper more understandable in terms of its significant contributions.}'' 
    CHI2023-2AC and CHI2023-R2 indicated ``High originality''.
    CHI2023-R2 commented ``\emph{it is fascinating, thought-provoking, rigorous work and of high relevance to CHI}.''
    CHI2023-R3 commented ``\emph{it is sufficiently different from prior art to warrant a full paper}.''
    Meanwhile, CHI2023-2AC and CHI2023-R2 indicated ``Low significance''.
    \color{blue}%
    \item[A1.] \textbf{Clarification.} The main contributions of this work are: (1) a framework for measuring human knowledge that is made available to computers during HCI, and (2) the methods for using this framework to measure such knowledge in low-level HCI operations, in empirical studies, and in high-level HCI operations in data intelligence workflows (e.g., machine learning). The author is not aware of any previous attempt to place all such measurements under the same framework. Hence the work is both novel and significant. In addition, this work can provide HCI researchers with more confidence about the necessity and importance of HCI. Especially in the current trend of great investment in AI and machine learning, many HCI colleagues, such as Ben Shneiderman, are making a huge effort in reassuring the necessity and importance of HCI. Mathematical reasoning and measurable evidence can be valuable tools for supporting such effort.
    
    \hspace{2mm} Some reviewers discussed at some length about several related or potentially-related topics. The author would like to emphasize that the paper does not claim novelty on any of the followings:
    \begin{itemize}
        \item HCI breaks the conditions of Data Process Inequality, which was discovered and mathematically reasoned by Chen and J\"anicke \cite{Chen:2010:TVCG};
        \item Basic information-theoretic measures \cite{Shannon:1948:BSTJ,Cover:2006:book}, and the relatively new measures \cite{Chen:2016:TVCG,Chen:2022:E1,Chen:2022:E2};
        \item Using information theory for modeling \textbf{how} humans interact with computers.
    \end{itemize}

    \textbf{Action.} The author has revised the last paragraph in Section 1 to improved the description of the main contributions of the paper.
\end{enumerate}

\begin{enumerate}
    \color{violet}%
    \item[Q2.] \textbf{Structure and Organization.}
    Similarly, there are opposite views on this criterion. CHI2019-2AC commented positively. CHI2019-2AC2 indicated that ``\emph{a better structure}'' may help.
     \color{blue}%
     \item[A2.] \textbf{Clarification.} The author has considered carefully how the materials can be organized differently but failed to find a better alternative to the current structure. The author should appreciate more detailed suggestions from the reviewers, such as the section structure or content ordering.  
\end{enumerate}

\begin{enumerate}
    \color{violet}%
    \item[Q3.] \textbf{Pace and Detail Level of Explanations.}
    Similarly, there are opposite views on this criterion. CHI2019-2AC noticed the dense content and appreciated the pace of explanation and commented that ``\emph{this paper is a nice read and very informative}''. CHI2019-2AC2 also commented positively on the pace and examples, which suggesting ``\emph{more concise writing may help}''. CHI2023-R3 commented that ``\emph{The approach is rigorous and well-defined. It is also well-explained.}'' Meanwhile, CHI2019-R2 indicated too much ``\emph{straightforward information theoretic definitions.}''
    \color{blue}%
    \item[A3.] \textbf{Clarification.} The author and his co-authors made a parallel effort to introduce an information-theoretic framework to visualization, that is, information transfer from computers to humans (cf. the opposite direction considered in this paper). That parallel effort took more than a decade, including
    \begin{itemize}
    \item In the 2010 paper \cite{Chen:2010:TVCG}, Chen and J\"anicke proposed the basic framework and used examples to demonstrate that mathematical measurements could explain a number of phenomena, such as Shneiderman's information-seeking mantra, the merits of logarithmic plot, the impact of interaction and visualization on DPI, and so on.
    \item In the 2016 paper \cite{Chen:2016:TVCG}, Chen and Golan introduced a new information-theoretic measure to explain the phenomena that statistics, algorithm, visualization, and interaction all seem to lose information, and use the measure to reason the benefit and cost of visualization, and the need to optimize the trade-off among data intelligence workflows. The paper also solved a theoretical puzzle outlined in \cite{Chen:2016:TVCG} by bringing in the measurement of cost.
    \item In the two-part paper published in 2022 \cite{Chen:2022:E1,Chen:2022:E2}, Chen and his co-authors proposed a revision to the original formula in \cite{Chen:2016:TVCG}, which facilitates more consistent and intuitive measurements of knowledge used in visualization, which can also be used to improve the original version of this paper [https://arxiv.org/abs/1812.06051v1].   
    \end{itemize}
    This paper attempts to introduce an information-theoretic framework to the opposite direction of HCI, i.e., information transfer from humans to computers while capturing the theoretical developments in the three visualization-focused papers. Naturally, the content is dense. Although I do not know the expertise of the reviewers, based on my observation of the reviews, the colleagues in HCI with less knowledge of information theory might appreciate a relatively slow pace of explanation, while the colleagues who are familiar with information theory would consider such a pace unnecessary. In general, this paper is intended for an HCI audience and I consider that CHI2019-2AC's view may be more representative. In the future, it may be helpful to provide more tutorial materials on the topic. 
    \color{black}%
\end{enumerate}

\subsection*{Individual Comments in the Reviews for the CHI2019 Submission}

\begin{enumerate}
    \color{violet}%
    \item[Q4.] CHI2019-1AC: need ``\emph{a more realistic application};'' CHI2019-R2: why ``\emph{this implies to data intelligence};'' CHI2019-R3: ``\emph{need a more convincing application which can be controlled empirically}.''  
    \color{blue}%
    \item[A4.] \textbf{Action.} Section 7 DATA INTELLIGENCE in the CHI2023 submission has already addressed this requirement. It is also included in this submission.
    
    \hspace{2mm} For CHI2019-R3's comment, the author assumes that this refers to empirical study. The study by Kijmongkolchai et al. \cite{Kijmongkolchai:2017:CGF} is such an example. The method described in Section 6 is a theoretical generalization of the experimental and analytical procedures described in that study, and is improved methodologically by the new mathematical formula presented in \cite{Chen:2022:E1,Chen:2022:E2}. This paper focuses on theoretical discourse, which hopefully will stimulate new empirical studies in the future. Section 6 MEASURING COST-BENEFIT EMPIRICALLY has already been improved in the CHI2023 submission with an additional figure and its accompanying text to illustrate how experimental results can be used to make new hypotheses numerically.
\end{enumerate}

\begin{enumerate}
    \color{violet}%
    \item[Q5.] CHI2019-1AC: ``\emph{Clarify the `costs' in the denominator of the key formula};'' CHI2019-2AC2: ``\emph{The measurement of “costs” ... is vague for the audience to understand};'' CHI2019-R2: ``\emph{the introduced `costs' in the denominator of the key formula are vague}.''
     \color{blue}%
    \item[A5.] \textbf{Clarification.} The definition of costs was given by Chen and Golan \cite{Chen:2016:TVCG} as follows:

    \vspace{2mm}\color{black}%
    ``The cost of executing $F_s$ as part of a visualization process can be measured in many ways. Perhaps the most generic cost measure is energy since energy would be consumed by a computer to run an algorithm or to create a visualization, as well as by a human analyst to read data, view visualization, reason about a possible relationship, or make a decision. We denote this generic measurement as a function $\mathcal{C}(F_s)$. While measuring energy usage by computers is becoming more practical [58], measuring that of human activities, especially cognitive activities may not be feasible in most situations. A more convenient measurement is time, $\mathcal{C}_\text{time}(F_s)$, which can be considered as an approximation of $\mathcal{C}(F_s)$. Another is a monetary measurement of computational costs or employment costs, which represent a subjective approximation from a business perspective.''

    \vspace{2mm}\color{blue}%
    The definition given in the submissions to CHI2019 and CHI2023 is a concise version of this original definition, while avoiding repeating the same text or the use the mathematical notations such as $\mathcal{C}(F_s)$. From a theoretical perspective, such a definition should not be considered as ``vague'' as one can find many definitions of a similar or higher level of conciseness in various textbooks and theoretical papers.

    \hspace{2mm} \textbf{Action.} While the author has been struggling to work out what kind of definition that the reviewers were looking for, three new sentences have been added to accompany the definition in Section 4.
\end{enumerate}

\begin{enumerate}
    \color{violet}%
    \item[Q6.] CHI2019-R3: ``\emph{Is that possible? Labelling this as `HCI breaks the conditions of DPI' is just nonsense - the human is just another source of information and processing!}''
     \color{blue}%
    \item[A6.] \textbf{Clarification.} As discussed in A1, the theoretical reasoning that interaction and visualization breaks the condition of DPI was offered by Chen and J\"anicke in \cite{Chen:2010:TVCG}. This is not a new contribution. If the reviewer disagrees with this already-published reasoning, perhaps an appropriate approach is to publish a commentary article through TVCG where \cite{Chen:2010:TVCG} appeared, or arXiv. In such a way, Chen and J\"anicke can response formally and openly.

    \hspace{2mm} In general, I am encouraging editors and reviewers not to judging a paper X based on a private doubt about a previous paper Y cited by paper X unless there is sufficient scholarly evidence confirming Y is incorrect. 
    For the benefit of the reviewers of this paper, the author offers his view here briefly as to why CHI2019-R3 was incorrect.
    \begin{itemize}
        \item The main conditions for DPI to be correct as a theorem are: (a) the process pipeline has 2 or more processes, (b) the whole process pipeline is closed (i.e., only external input to the pipeline is to the first process $P_1$, (c) the processes in the pipeline are memoryless, and (d) a process $P_i$ can only receive input from $P_{i-1}$ cannot transfer its knowledge.
        \item CHI2019-R3 did not seem to have studied these conditions carefully, and made several arguments against the notation that HCI can break the above conditions. One argument is to make humans as processes in the pipeline. This itself breaks the above conditions (b, c, d) as (b) humans can normally receive external information dynamically, e.g., as illustrated in Fig. 5 (previously Fig. 4)., (c) humans are not memoryless, and (d) humans can gain provenance information from processes preceding $P_{i-1}$, e.g., using visualization. Another argument made by CHI2019-R3 is that human inputs can be considered as data that is available at the beginning of the pipeline. Most HCI colleagues can see the problem of this argument. If this would be correct, all HCI interactions could be done at the very beginning of an HCI session. Using the example in Fig. 4, a user would have to anticipate that at a precise moment, the system would ask the user a question about saving a file, and the user would have already knew the answer to such a question.
        \item Golan and Sbert who wrote books on information theory confirmed the correctness of the reasoning when they worked with the author on other theoretical papers \cite{Chen:2016:TVCG,Chen:2016:book,Chen:2022:E1,Chen:2022:E2}.
        \item Since 2010, the author gave a good number of seminars featuring the DPI discussions. In a few seminars (e.g., in MIT), colleagues with some knowledge of DPI raised a what-if question by considering human users as part of the processes. As soon as the author mentioned the memoryless condition, these informed colleagues were easily convinced.
        \item Occasionally, some colleagues mistook that Chen and J\"anicke stated DPI being incorrect. This was never the case. The DPI theorem is correctly defined and proved. Cover and Thomas clearly defined the conditions for the theorem to be correct \cite{Cover:2006:book}. The conditions are there to be broken! As DPI points to a bottleneck problem, the fact that HCI can break the conditions of DPI indicates the merits of HCI. Surprisingly and ironically, this rather straightforward reasoning was labeled as ``nonsense'' in a CHI review process.     
    \end{itemize}
\end{enumerate}

\begin{enumerate}
    \color{violet}%
    \item[Q7.] CHI2019-1AC: ``\emph{Discuss how the work differs from related attempts in the specific domains mentioned by R2 ...}'' CHI2019-R2: ``\emph{... Posner and Fitts in investigating the capacity limitations ... The papers of MacKenzie, Zhai, Seow come to mind.}''
     \color{blue}%
    \item[A7.] \textbf{Clarification.} To the author, the answer seems to be obvious. The previous work mentioned in the question is about modeling, while this work is about measuring while making no proposal or assumption about any specific model for describing the actions of humans and machines. It took a while for the author to guess why the reviewers seemed to see a stronger connection. Possible reasons are:

    \begin{itemize}
        \item Terminology. Information theory is fundamentally a mathematical framework that is more similar to algebra and calculus than Newtonian physics, neural networks, and cognitive theories. The mathematical framework is a more complex version of set theory, such that each set element (each letter) has a probability value. The framework provides operations over such sets (referred to as alphabets). In principle, all such operations can be realized using conventional algebra, except the formulae would be too complex to write. Using the framework, mathematicians derive theorems (such as the aforementioned DPI), which can confusingly be referred to as theories.
        \item Communication Model vs. Information Theory Model. Shannon developed information theory as a framework, but not a model. Shannon used phenomena in a simple communication model (i.e., message-encoder-signal/noise/channel-decoder-message) to demonstrate what information theory can measure and calculate. The simple communication model was widely known before information theory was proposed. Information theory nowadays is applied to much complex communication models as well as non-communication models such as economics models. Semantically, the models by Posner and Fitts, and others are related to the basic communication model, rather than \textbf{the} information theory model since there is no such a model.
        \item Similar mathematical formula. Fitts's law has a form featuring a logarithmic term that is similar to (not exactly the same as) the most basic informative measure, i.e., self-information. This simulated a hypothesis that the control of low-level human motoring skills might largely be done with simple communication. Because there was not sufficient evidence then, it was considered as an "information analogy" (as described in the Wikipedia page on Fitts's law). More precisely, it is an analogy of a noisy basic communication model. Today, it might still considered by some as an analogy. Note that using the term "information analogy", "information theory analogy", "communication analogy", and "communication model analogy" might easily lead to a different interpretation.
        \item low-level vs high-level HCI operations. It is widely accepted that the basic communication model may be too simple for modeling any high-level cognitive process such as thinking and memorizing. This somehow explains why Fitts's law has been applied to a small set of HCI operations. Any slightly complicated HCI such as Figs. 2, 5 (previously 4), and 8 (new) cannot be modeled easily using Fitts's law because they involve cognition with more complicated processing and communication in the mind.
    \end{itemize}

    Finding appropriate models for modeling cognition may take some time for hypothesizing and evaluating such models. This is out of the scope of this work. In fact, when Chen and Golan proposed the cost-benefit measure in 2016 \cite{Chen:2016:TVCG}, one intention was to be able to confirm the impact of human knowledge in visualization processes without the need to confirm how such knowledge was formulated, represented, or deployed in the mind. With the definition of the potential distortion in information reconstruction, one can detect and confirm the impact (as done in \cite{Kijmongkolchai:2017:CGF} without the need to model the cognitive process in the mind. This work extends this approach. \textbf{By looking at the inputs received from the perspective of the computer, we can detect and confirm the impact of HCI operations without the need to model the cognitive process in the mind.}
    
    \textbf{Action.} We have now mentioned such previous research works in Section 2, and state clearly this paper is not about modeling.
\end{enumerate}

\begin{enumerate}
    \color{violet}%
    \item[Q8.] CHI2019-1AC: ``\emph{Discuss how the work differs from ... Bayesian models of interactive systems mentioned by R3.}'' CHI2019-R3: ``\emph{Many of the same principles are there in standard Bayesian models where the prior distributions for input movements are coupled with the sensor model distribution. More of this literature should have been included.}''
     \color{blue}%
    \item[A8.] \textbf{Classification.} Following the same line of the reasoning in Q7-A7, this work is about measuring, but not about modeling the human thought process during HCI. This work does not assumes either the correctness or incorrectness of the hypothesis that the human thought process in HCI may be a Bayesian process. This work does not propose any new model to replace the hypothesis of a Bayesian process.

    \hspace{2mm} There are a number of hypotheses about human thought processes, including variants of neural network models and Bayesian models. At least, they cannot be all correct. Some experiments in HCI indicated that humans' Bayesian reasoning is much weaker than computational Bayesian models in terms of accuracy and network size, casting a doubt on the hypothesis that humans thought like Bayesian models in HCI. Future research in HCI and psychology will offer more insight about such hypotheses. Evaluating such hypotheses is beyond the scope of this work. As mentioned in A7, this work provides a way to detect and confirm the impact of HCI while avoiding the need for ascertaining a model of human motion or cognitive processes.
    \textbf{Action.} We have now mentioned such previous research works in Section 2, and state clearly this paper is not about modeling.
\end{enumerate}

\begin{enumerate}
    \color{violet}%
    \item[Q9.] CHI2019-R3: ``\emph{The claims around the benefit of HCI to the computer are also a red herring. The computer being able to reduce its uncertainty of the user's intentions is an essential feature of the computer ...}''
     \color{blue}%
    \item[A9.] \textbf{Classification.} I do not agree with this one-sided machine-centric view. I am sure that most HCI colleagues will not agree with this. Otherwise, it would be a rather dark future for HCI. The information-transfer are in both directions, hence the benefits are mutual. This paper mathematically evidences the information transfer from humans to computers via HCI, hence re-ascertaining the value of HCI.
\end{enumerate}

\begin{enumerate}
    \color{violet}%
    \item[Q10.] CHI2019-2AC: ``\emph{When the "cost-benefit metric" is first mentioned ... it would be good to give readers an intuitive idea of its meaning. Does it measure `the capacity and efficiency for a computer to receive knowledge from users?'}'' ``\emph{I also find the paragraph of `Some HCI tasks require additional display space or other resources for providing users with additional information...' a bit confusing.}'' ``\emph{What is the theoretic upper bound of the cost-benefit metric (like entropy is 0~1), so that one could assess is a design is satisfactory in terms of this measure?}''
     \color{blue}%
    \item[A10.] \textbf{Action.} These queries are helpful to the author. Thank you. The author added new sentences in the relevant paragraphs to improve the clarity of the explanation.
\end{enumerate}

\subsection*{Individual Comments in the Reviews for the CHI2023 Submission}

\begin{enumerate}
    \color{violet}%
    \item[Q11.] CHI2023-1AC: ``\emph{Reviewers found the link to information theory weak, and also the implementation of “information theory” in the modeling proposed in the submission could benefit from further details and careful consideration}'' 
     \color{blue}%
    \item[A11.] \textbf{Clarification.} I guess that this might be due to the confusion between information theory and communication as mentioned previously in A7. In comparison with Fitts's law and subsequent works in HCI (collectively referred to as ``Fitts and others'' in the list below), this work is more rooted in information theory because:
    \begin{itemize}
        \item In terms of information-theoretic measures, Fitts' law and others used only a formula, $a + b \log_2(2D/W)$ that features a logarithmic term similar to the measure of self-information $\log_2(p)$. Note that $a$ and $b$ does not have an information-theoretic definition, while $D/W$ is analogical definition of probability $p$. Meanwhile, in this work uses the measure of Shannon entropy $\mathcal{H}$, divergence $\mathcal{D}_\text{cs}$ (replacing $\mathcal{D}_\text{kl}$ in version 1), and benefit (introduced in 2016 \cite{Chen:2016:TVCG}) throughout the paper.
        \item In terms of information-theoretic concepts, Fitts' law and others did not consider the fundamental notion of alphabet (i.e., a set of all valid values for a variable) explicitly. This work considers this notion throughout the paper, and uses its to provide a unified abstraction of the low-level human inputs to interaction devices, various interactions to be captured in empirical studies, and high-level inputs to data intelligence workflows.
    \end{itemize}
    So the work utilized information theory substantially, and much more than Fitts' law.

    \hspace{2mm} In order to avoid tangling into the question about how to model human thought and actions in HCI, this work intentionally avoids hinging itself on the basic communication model as done by Fitts' law and others. It is possible that some reviewers might have considered the absence of the basic communication model as the absence of information theory. This should not be the case.
\end{enumerate}

\begin{enumerate}
    \color{violet}%
    \item[Q12.] CHI2023-1AC: ``\emph{A convincing argument for another “metric for value” needs to be presented.}''.
     \color{blue}%
    \item[A12.] \textbf{Clarification and Action.} Of course, all HCI colleagues know very well about the value of HCI (perhaps except the ``nonsense'' and ``red herring'' arguments in Q6 and Q9 respectively). The author has appreciated the value of HCI since he was a PhD student working on graphical user interfaces in the 1980s. This work is not intended to re-aurge the value of HCI, and is absolutely not for teaching HCI colleagues to recognize the value of HCI.
    
    \hspace{2mm} Nevertheless, the author is not aware of an existing quantitative metric or measure that is similar to what is described in this paper or has a similar broad capability for detecting, estimating, and evaluating the benefits of HCI operations. This point has been mentioned in several places in the paper, and the author has added a further statement at the end of Section 1. 
\end{enumerate}

\begin{enumerate}
    \color{violet}%
    \item[Q13.]  CHI2023-1AC: ``\emph{perhaps in relation to the `real world HCI dataset'.}''.
     \color{blue}%
    \item[A13.] \textbf{Clarification.} As mentioned in Q3-A3, this paper is already very dense because it corresponds to four papers over a decade in the field of visualization. Applications of the cost-benefit measure to real world data has already been reported in \cite{Kijmongkolchai:2017:CGF} and \cite{Chen:2022:E2}, each of which is a full-length paper. In general, the author hopes that the reviewers will appreciate the needs for many papers in the lifecycle of a theoretical development.    
\end{enumerate}

\begin{enumerate}
    \color{violet}%
    \item[Q14.]  CHI2023-2AC: ``\emph{the writing of the paper is poor. The text is very poorly structured; sections like 5, 6 and 7 are extremely indigestible.}''.
     \color{blue}%
    \item[A14.] \textbf{Clarification.} See Q2-A2 and Q3-A3.    
\end{enumerate}

\begin{enumerate}
    \color{violet}%
    \item[Q14.]  CHI2023-2AC: ``\emph{The link with information theory is very loose: I provide many details about misuses of information-theoretic vocabulary and notions in my detailed review, but the biggest issue is that no channel is ever introduced.}''.
     \color{blue}%
    \item[A14.] \textbf{Clarification.} The word of ``channel'' (as an information-theoretic term) was mentioned twice in Section 3.1 in the 2022 submission. The notion of channel is part of the basic communication model, and it was not heavily used because of the reasons mentioned in Q7-A7 and Q11-A11. Similarly, we did not mention the terms ``messages'', ``encoder'', ``decoder'', ``signal'', and ``noise'', which are all terms of communication rather than information theory. The related information-theoretic term used in the paper is ``transformation'', which is more generic.    

    \textbf{Action.} However, the reviewer's comment indicates a need to go back to explain the generic term ``transformation'' better in the paper, and explain how the notion of ``channel'' is included in the generic term. We have added additional text and a figure (new Fig. 3) in Section 3.4. Meanwhile, the author hopes that the reviewers can appreciate that in applications other than communication, the concepts and terminologies have evolved since information theory emerged in the 1940s. In some cases, the basic communication model can still be used as an analogy. In other cases, it could be a distraction.     
\end{enumerate}

\begin{enumerate}
    \color{violet}%
    \item[Q15.]  CHI2023-2AC: ``\emph{Not considering errors, as well as not considering the dependence between consecutive symbols in user input (e.g. for a sequence of mouse positions, two consecutive samples are not independent) leads to gross overestimations of entropies.}''
    \color{blue}%
    \item[A15.] \textbf{Clarification.} I guess that CHI2023-2AC might have drawn the conclusions in a bit rush. The correct conclusions are the opposite. (i) The user errors are measured using the component of potential distortion. The potential distortion is a more powerful measure as it takes into account not only user errors but also the ability of the software to correct errors automatically. Such automated error correction mechanisms are nowadays common in user interfaces.
    
    \hspace{2mm} (ii) The probabilistic dependence of multiple HCI operations is also covered by the cost-benefit measure. The TV listing example demonstrated this exactly. I guess that the CHI2023-2AC might have scanned for some terms such as conditional probability. When one did not see desired terms, one could draw a ``cannot'' conclusion. In information theory, as a weighted average, Shannon entropy (Eq. 1) considers the individual probability values of different letters in an alphabet. When the probability of every letter in an alphabet is the same, the Shannon entropy $\mathcal{H}$ reaches the maximum, and is denoted as $\mathcal{H}_\text{max}$. In most cases, computer software has to encode every alphabet based on $\mathcal{H}_\text{max}$. That is why the action capacity usually corresponds to $\mathcal{H}_\text{max}$.
    
    \hspace{2mm} If one types ``a8\#$\pi$'', this is defined as four transformations from "<null>" to ``a'', then ``a8'', ``a8\#'' and ``a8\#$\pi$''. All possible variations of four letters in four elementary alphabets (can be the same or different alphabets) are encompassed in a composite alphabet, to which ``a8\#$\pi$'' is just a letter and has its own probability. The probability of ``a8\#$\pi$'' may take into account of the probability of individual letters, their ordered dependence, as well as conditions that is encoded in the transformation or the HCI inputs received by the transformation dynamically. Such multi-letter alphabets were the major contributions of Shannon when information theory was formulated. This work asks readers to consider alphabets in a generalized manner. For example, ``a8\#$\pi$'' could be four ordered HCI operations, e.g., <select a button> <draw a path> <press a mouse button> <select an option in a pop-up menu>, or <select an ML framework><select training data><define all learning parameters><press a button to start training>. Each operation is a letter in an alphabet that includes all HCI operations available to the user at that moment. This shows that the concepts of alphabet transformation and composite alphabet can represent a sequence of low-level HCI operations as well as high-level ones.
    \textbf{Action.} For this paper, the author has added a few sentences in a few places to remind readers of these basic properties of information theory. Beyond this paper, if it is welcomed by HCI colleagues, the author can offer additional tutorial materials.
\end{enumerate}

\begin{enumerate}
    \color{violet}%
    \item[Q16.] CHI2023-2AC: ``\emph{there is little that is information-theoretic about the core of this paper. ... mutual information, conditional mutual
information, conditional entropy, cross entropy.}''
    \color{blue}%
    \item[A16.] \textbf{Clarification.} If the HCI colleagues are looking for a paper that showing how different information-theoretic measures are or can be used in HCI. This work will not meet the requirement for these colleagues. If the HCI colleagues are welcoming a theoretical approach for measuring, estimating, and evaluating the cost-benefit of HCI and for confirming the merits of HCI numerically, hopefully this work offers something new. See also Q11-A11.
\end{enumerate}

\begin{enumerate}
    \color{violet}%
    \item[Q17.]  CHI2023-2AC: ``\emph{An alternative to the main formula Eq 3. could have been proposed without recourse to information theory (it would just not have any log transforms, and not be expressed in bit).}''
    \color{blue}%
    \item[A17.] \textbf{Clarification.} I guess that this is a private speculation of the CHI2023-2AC. Of course, all colleagues are encouraged to transform such a speculation to a publication. I guess that the fact that Entropy published  two papers (a two-part paper) about  Eq 3 \cite{Chen:2022:E1,Chen:2022:E2} in its anniversary volume, there must be some worthy contribution in the formula.
\end{enumerate}

\begin{enumerate}
    \color{violet}%
    \item[Q18.] CHI2023-2AC: ``\emph{The use of Cdev is misleading in my opinion, and this value can never be reached.}''
    \color{blue}%
    \item[A18.] \textbf{Clarification.} This is related Q15-A15. $\mathcal{C}_\text{dev}$ reflects the perspective of a computer as the computer has to allocate the memory space without knowing the probability. So in most cases, this value is always reached. If an HCI designer can utilize this measure, the designer can potentially optimize a design with less $\mathcal{C}_\text{dev}$. The author has added a simple example in Section 3.2 to illustrate the usefulness of this measure. 
\end{enumerate}

\begin{enumerate}
    \color{violet}%
    \item[Q19.]  CHI2023-2AC: ``\emph{The use of the term "bandwidth" is in line with its misuse in HCI but contrary to information-theoretic vocabulary (where bandwidth is expressed in Hz.}''
    \color{blue}%
    \item[A19.] \textbf{Classification.} CHI2023-2AC might have taken a narrow view that information theory = tele-communication. Today, Hz is rarely used in many applications of information theory, e.g., data compression, data encryption, economics, bio-informatics, and so on.   
\end{enumerate}

\begin{enumerate}
    \color{violet}%
    \item[Q20.]  CHI2023-2AC: ``\emph{Interestingly, 0 bit information alphabets can be used to convey information in practice.}''
    \color{blue}%
    \item[A20.] \textbf{Classification.} I do not think that CHI2023-2AC is correct. If a signal (or a letter) is useful to the receiver, the ``no signal'' state is a letter in the alphabet. In the sender and receiver's alphabet, both options of ``a signal'' or ``no signal'' must be available. CHI2023-2AC might have looked the signal only in the communication channel, assuming that if it is not encoded explicitly, it is not a letter. This would be an incorrect interpretation of information theory. 
\end{enumerate}

\begin{enumerate}
    \color{violet}%
    \item[Q21.]  CHI2023-2AC: ``\emph{How would this even support an argument for less HCI.}''
    \color{blue}%
    \item[A21.] \textbf{Classification.} CHI2023-2AC must have misread the statement. The author was not arguing for less HCI. The author was using this statement to prepare readers for a better argument for the merits of HCI in the section immediately after that statement.
    \textbf{Action.} The author has added a statement to point reader to the next section.
\end{enumerate}

\begin{enumerate}
    \color{violet}%
    \item[Q22.]  CHI2023-2AC: ``\emph{Alphabet compression:
    labeling H(Z\_i) - H(P\_i(Z\_i)) as alphabet compression is not in line with vocabulary from IT.
    The fact that H(P\_i(Z\_i)) <= H(Z\_i) is a well known property.}''
    \color{blue}%
    \item[A22.] \textbf{Classification.} CHI2023-2AC used a different notation in Cover and Thomas \cite{Cover:2006:book} without checking the definitions. Chen and his co-authors \cite{Chen:2016:TVCG,Chen:2016:book,Chen:2022:E1,Chen:2022:E2} used $\mathbb{Z}_i$ as an alphabet that is equivalent to $X$ used by Cover and Thomas as a variable.  Chen and his co-authors used $P_i(\mathbb{Z}_i)$ to describe the probability mass function of $\mathbb{Z}_i$, while Cover and Thomas used $X$ as a proxy for $P(X)$. In information theory, authors use $X$, $P(X)$, and $P$ exchangeably when using them as part of a measure, e.g., $\mathcal{H}(X) = \mathcal{H}(P(X)) = \mathcal{H}(P)$. Similarly, Chen and Golan stated clearly in their paper for the VIS audience that $\mathbb{Z}_i$, $P_i(\mathbb{Z}_i)$, $P_i$ can be used exchangeably, where $P_i$ is a probability mass function.

    \hspace{2mm} In Cover and Thomas's formula $H(g(X)) \leq H(X)$, $g$ is not a probability mass function, but a function that transforms $X$ to a new variable $Y = g(X)$. CHI2023-2AC rewrote $g$ as $P_i$ incorrectly, leading to an incorrect conclusion.
    Meanwhile, the term ``alphabet compression'' was introduced by Chen and Golan \cite{Chen:2016:TVCG}. It is clearly defined mathematically. Borrowing Cover and Thomas' notation, it would be $H(X) - H((Y)$, where $Y = g(X))$.
    It has never been criticized by information theory experts, including Entropy's editors and reviewers in addition to Chen's coauthors (Gloan and Sbert). CHI2023-2AC did not provide an alternative term to replace the term ``alphabet compression''.

    \hspace{2mm} It is also important to note that $Y = g(X)$ in Cover and Thomas's book has an assumption that was not stated in Problem 2.4. That is, $g(X)$ is a transformation that does not include any other information in addition to $X$. In other words, no extra information can be added to the transformation of $Y = g(X)$, or no HCI is allowed in the transformation. This is related to the aforementioned DPI (see Q8-A8). If HCI can influence the transformation with another variable $Z$ (or alphabet $\mathbb{Z}$) as $Y = g(X, Z)$, the inequality $H(g(X)) \leq H(X)$ does not stand. This shows the powerful impact of HCI, which the HCI colleagues are expected to welcome and celebrate.
\end{enumerate}

\begin{enumerate}
    \color{violet}%
    \item[Q23.]  CHI2023-2AC: ``\emph{There are almost no tasks which end in 0 entropy.}''
    \color{blue}%
    \item[A23.] \textbf{Classification.} I could not understand how CHI2023-2AC could make such an incorrect observation. For example, if a user has the task to choose one option from an alphabet (a, b, c, d), there was uncertainty defined as entropy. The maximum value is 2 bits when (a, b, c, d) have equal probability 0.25. When the selection is done, (e.g., selecting (c)), the uncertainty become the certainty of (c). The probability mass function changes to (0, 0, 1, 0). The certainty thus has zero entropy at least to the computer. The author has been wondering if CHI2023-2AC was making a philosophical comment rather than a mathematical comment.
\end{enumerate}

\begin{enumerate}
    \color{violet}%
    \item[Q24.]  CHI2023-2AC: ``\emph{Dcs is not defined.}''
    \color{blue}%
    \item[A24.] \textbf{Classification.} It is defined in Eq. 3. 
\end{enumerate}

\begin{enumerate}
    \color{violet}%
    \item[Q25.]  CHI2023-2AC: ``\emph{sign problem in Eq. 3.}''
    \color{blue}%
    \item[A25.] \textbf{Action.} Thank you for spotting this typo. Much appreciated.   
\end{enumerate}

\begin{enumerate}
    \color{violet}%
    \item[Q26.]  CHI2023-2AC: ``\emph{The definition of Z'\_i is unclear.}''
    \color{blue}%
    \item[A26.] \textbf{Clarification.} This was defined just below Eq. 3.  CHI2023-2AC seems to have swapped input and output in the comment.
\end{enumerate}

\begin{enumerate}
    \color{violet}%
    \item[Q27.]  CHI2023-2AC: ``\emph{Dcs ... where does it come from?}''
    \color{blue}%
    \item[A27.] \textbf{Clarification.} The references were given just before Eq. 3.
\end{enumerate}

\begin{enumerate}
    \color{violet}%
    \item[Q28.]  CHI2023-R2: ``References on ML in HCI publications''
    \color{blue}%
    \item[A28.] \textbf{Action.} A few example references have been added in Section 7.
\end{enumerate}

\begin{enumerate}
    \color{violet}%
    \item[Q29.]  CHI2023-R3: ``\emph{It would be necessary to see more extensive guidance, trials and demonstrations for `value' as well.}''
    \color{blue}%
    \item[A29.] \textbf{Clarification.} Numerous HCI researchers have already provided an extensive number of papers featuring guidance, trials and demonstrations of the value of HCI. Perhaps what lacks is a mathematical formulation that can provide a sound theoretical abstraction of these guidance, trials and demonstrations. This is exactly what this paper can provide. See also Q13-A13.
    
    In general, a theoretical development takes long time, and it takes many years and many papers to evaluate or falsify a theoretical postulation. In the history, many historical postulations could not be validated easily but could be falsified by finding only one counter-example. In the field of visualization, a number of published theoretical postulations were later confirmed to be incomplete (i.e., only partially correct) by finding one counter-example that the postulations could not explain. To the best knowledge of the author, no counter-example has yet been found to falsify the cost-benefit postulation by Chen and Golan \cite{Chen:2016:TVCG}. The author welcomes any further theoretical, experimental, and practical work that can help improve our understanding of the cost-benefit of HCI, including finding better theoretical frameworks than this work. 
\end{enumerate}

\begin{enumerate}
    \color{violet}%
    \item[Q30.]  CHI2023-R3: ``\emph{I am not convinced that the metric works in a reasonable way in different representative conditions in HCI. One example is given, but it remains cursory (to say the least).}''
    \color{blue}%
    \item[A30.] \textbf{Clarification.} The examples cited or discussed in the paper include:

    \textbf{Cited:}
    \begin{itemize}
        \item The original proof of Information Seeking Mantra \cite{Chen:2010:TVCG} (based on mutual information) and the re-proof (based on the cost-benefit ratio) and the explanation of ``occasional'' details-first phenomena \cite{Chen:2016:book}.
        \item An empirical study by Kijmongkolchai et al. \cite{Kijmongkolchai:2017:CGF}.
        \item Application of the measure to two real-world datasets (volume visualization and viewing London underground maps) \cite{Chen:2022:E2}.
        \item Application of Eq. 3 qualitatively to the analysis of cost-benefit of visualization applications in VR environments \cite{Chen:2019:TVCG}.
        \item An information-theoretical analysis of two machine-learning problems \cite{Tam:2017:TVCG}.
        \item A systematic method for improving data intelligence workflows that consist of machine-centric processes (e.g., statics, algorithms) and human-centric processes (e.g., visualization, interaction) \cite{Chen:2019:CGF}.
    \end{itemize}
    \textbf{Discussed:}
    \begin{itemize}
        \item Application of the measure in low-level HCI operations (Section 3, Section 5).
        \item Application of the measure in medium-level HCI operations in a word-processing example (Section 4, Section 5). It is a real example as the author used to work this way before spell-checking became available in LaTex editing tools.
        \item Application of the measure in high-level HCI operations in machine learning workflows (Section 7).
        \item Application of the measure in empirical studies (Section 6).
    \end{itemize}
\end{enumerate}

For the work presented in this paper, at the end, it is a scholarly decision by the editors and reviewers as to whether the work offers new contributions that will likely benefit the field, i.e., HCI in this case, even if the work may be improved in the future. So far, no serious technical issue (except some typos) has been discovered in the review processes. Meanwhile, the parallel work was published in the field of VIS, and no counter-example that may falsify the parallel work has been reported. As an author, all I can say is that the potential benefit to HCI is rather high while the risk is rather low.

\color{black}%
\section*{Supplement B: Reviews for the ACM CHI2019 Submission}

\small
\begin{Verbatim}[breaklines]

Dear Min Chen,

You should have received an email earlier this week indicating that your submission, The Value of Interaction in Data Intelligence (number 2115), was not accepted for inclusion in the CHI 2019 Papers Program. The final reviews for your submission are included below and are also available to you in the conference submission system: https://new.precisionconference.com/sigchi

We encourage you to consider other CHI 2019 venues, many of which have deadlines in January or February: see http://chi2019.acm.org/authors/ for details.

Best regards,

Anind K. Dey and Shengdong Zhao
CHI 2019 Papers Chairs
----------------------------------------------------------------
AC review
score 2/5

  Expertise
    Knowledgeable
  Recommendation
    Possibly Reject: The submission is weak and probably shouldn't be accepted, but
    there is some chance it should get in; 2.0

  1AC: The Meta-Review

    Three reviewers assessed the submission. While they all see some value in the
    submission, they come to somewhat differing final conclusions. The submission is
    very well structured [2AC]. Despite the varying scores, all reviewers seem to
    appreciate the submissions general direction. R3, for example, states that the
    paper addresses an important but challenging topic and that there are
    inspirational elements to the paper.

    2AC found the content very dense and appreciated how the authors walk the reader
    through the different steps. R2 & R3, however, criticize that "straightforward
    information theoretic definitions" [R2] are described in length. This seems like a
    general problem for this kind of submission to me. The authors did a good job in
    making the submission accessible to readers that are not experts in this domain
    like myself. Just as 2AC, I found the paper very easy to read but just like R2 &
    R3 I also wonder about the novelty provided by the work. Finding the right balance
    when addressing different audiences is a major challenge for this kind of work.

    In their rebuttal, I would suggest that the authors address the following aspects
    discussed by the reviewers:
    1. Highlight what the authors believe is the grand contribution of the submission
    (see R2) and provide an example of a more realistic application (see R3 & R2).
    2. Discuss how the work differs from related attempts in the specific domains
    mentioned by R2 and Bayesian models of interactive systems mentioned by R3.
    3. Respond to the additional aspects criticized in the reviews and describe how
    they could be addressed in a revised version of the paper.
       + Briefly clarify aspects unclear to 2AC
       + Clarify the "costs" in the denominator of the key formula (see R2).
       + Address additional aspects raised by the reviewers.
       
    When preparing the rebuttal, I would recommend to make it as tangible as possible.
    I.e. CHI has a very tight review process. Thus, it should be as clear as possible
    how a refined version would look like. The authors' can, for example, provide
    short versions of paragraphs that could be almost directly copied into the paper.

    Post PC Meeting
    --
    The submission got repeatedly discussed at the PC meeting. The committee generally
    wants to highlight that we appreciate this type of work. We need more theoretical
    work at CHI and work such as this submission could ultimately make a significant
    contribution. Unfortunately, the submission requires at least another iteration.
    We recommend that the authors take the reviews carefully into account to improve
    the submission.
----------------------------------------------------------------
2AC review 
score 3.5/5

  Expertise
    Knowledgeable
  Recommendation
    . . . Between neutral and possibly accept; 3.5

  Review

    This paper proposes an information-theoretic approach to measure the cost-benefit
    of HCI in data intelligence workflows. It has in-depth explanation of the
    theories, the metric, and applications.

    I really enjoy reading this paper. It is very-well structured. The content is very
    dense, but the examples walk readers through the concepts, math, and ideas.

    There are a couple of places that could be improved for an audience without
    information theory background.

    When the "cost-benefit metric" is first mentioned in Line 178, it would be good to
    give readers an intuitive idea of its meaning. Does it measure "the capacity and
    efficiency for a computer to receive knowledge from users?" It would also be nice
    to present the intuition behind "the lower the action capacity Cact and thereby
    the lower the DU," although we can see it mathematically (for example, low entropy
    means high certainty).

    I am confused by the definition of value presented in Related Work. It would be
    nice to provide references to each of value referents. What does value mean in
    "value-centered designs?" How does the value definition differ when it is from the
    perspective of computer versus human? This paper examines value in terms of (a)
    and (c), but they are different concepts. Under what circumstance does the paper
    refer to (a) and when it means (c) (in the series of equation for the cost-benefit
    metric)?

    I also find the paragraph of "Some HCI tasks require additional display space or
    other resources for providing users with additional information..." a bit
    confusing. At first I thought it means a bigger screen space so that the mouse has
    more potential positions, but the yes/no example seems to indicate more radio
    buttons (interactions). What does the "the varying nature of the information"
    mean?

    Some statements may need further clarification. For example, "On the contrary, it
    is less common to discuss the usefulness and benefits of HCI to computers" -- but
    human computation and human-in-the-loop machine learning are all studying this
    topic. Why is it necessary to assume in the freehand path example that "though the
    computer does not store the time of each sample?"

    What is the theoretic upper bound of the cost-benefit metric (like entropy is
    0~1), so that one could assess is a design is satisfactory in terms of this
    measure? In the current example, we could only compare if one action is better
    than the other or one design is better than before.

    In the conclusion, it is said that the proposed approach "an addition to the
    existing toolbox for supporting the design and evaluation of HCI devices." I would
    like to see more discussion on what is the existing toolbox, and in what way the
    cost-benefit metric complements it. Do they measure different things? How could
    they be used together?

    Some minor issues:
    (Line 352) 'z3: "View HD Alternatives".' should be 'a3'
    Figure 2 is too small to see
    It would be nice to explain the meaning of i in Eq. (2) or refer it to Figure 3.

    Overall, this paper is a nice read and very informative.
----------------------------------------------------------------
2AC2 review 
score 3/5
[author note: this review was added after the rebuttal stage. Likely the reviewer was invited to provide additional opinions to the review panel. The reviewer was originally labelled as 2AC. This label was replaced with 2AC2 to differentiate it from the 2AC above.]

  Expertise
    Knowledgeable
  Recommendation
    Neutral: I am unable to argue for accepting or rejecting this paper; 3.0

  Review

    In this paper, the authors addressed an interesting as well as important topic of
    quantifying the value of human computer interaction in data intelligence via an
    information-theoretic framework.

    A few concerns that I have with this submission even after reading the rebuttal
    are listed below:

    1.      The measurement of “costs” in the denominator of equation 2 (L550-L552) is
    vague for the audience to understand, and the authors failed to clarify this point
    in their rebuttal. Although I understand that the “cost” here must be evaluated on
    a case-by-case basis, a more general, but clear measurement of  cost needs to be
    proposed, like what the authors did in measuring the “benefits” in the numerator.

    2.      I get the point of “HCI breaks the conditions of DPI” as the authors
    mentioned in their rebuttal. However, I found there is still a gap between the
    cost-benefit analysis and the measurement of DPI. The authors mentioned that
    “Human computer interaction (HCI) can provide cost-beneficial means to alleviate
    the problems due to DPI” and they need to quantify this point in their writings.

    3.      The authors mentioned a few practical implications as the grand
    contribution of this work as well as its connection with data intelligence, such
    as “selecting a set of statistical measures, changing the parameters of a
    clustering algorithm, choosing keys for sorting, selecting an area to zoom in in
    visualization, reconfiguring layers in machine learning, accepting or discarding
    an automatically generated recommendation, and so on.” However, I am a little bit
    confused about the context of “changing the parameters of a clustering algorithm”,
    and “reconfiguring layers in machine learning”, given that those two settings are
    quite different from the ones the authors mentioned in their submission (which
    have a very limited set of conditions, whereas parameter settings might be via a
    grid search, which contains an infinite or very large of number conditions; also
    the parameter distribution varies along with the changes of training data
    distribution, so the probability of occurrence would be relatively hard to
    measure). The authors need to reconsider if “data intelligence” is an appropriate
    context of this study or they want to change it to some other more specific
    context.

    Regarding the presentation of this work, although the authors spent a lot of space
    and effort explaining about the straightforward information theoretic definitions,
    I actually found it easier for audience without relevant knowledge to understand
    and follow. I also like the examples that the authors adopted to walk readers
    through the concepts, math, and ideas.  But a better structured or more concise
    writing could make the paper more understandable in terms of its significant
    contributions.
----------------------------------------------------------------
reviewer 2 review
score 2/5

  Expertise
    Knowledgeable
  Recommendation
    Possibly Reject: The submission is weak and probably shouldn't be accepted, but
    there is some chance it should get in; 2.0

  Review

    This paper aims to contribute to theoretical understanding of interaction, and
    especially information theoretical analyses of communication between humans and
    computers. Inspired by the work of Posner and Fitts in investigating the capacity
    limitations of the human motor control system, HCI researchers have looked at the
    relationship between information theoretic variables like throughput and
    interactive human performance in domains like pointing and forced choice. The
    papers of MacKenzie, Zhai, Seow come to mind.

    The submitted paper aims to contribute by proposing to quantify the "value of
    interaction" especially in the area of data intelligence. Data intelligence was
    here defined as all processes that transform data to decisions.

    The starting point to the paper is the DPI theorem of Cover and Thomas, according
    to which post-processing of data can only lose but not increase information. This
    is intuitive: the decoder cannot add information. The authors refer to a source
    pointing out (or proving?) that HCI relaxes some assumptions of the Markov Chain
    in the DPI theorem that change the game. This point is already published, and I
    see the aimed contribution of this paper in the quantification of the added value.

    The paper proceeds to relatively straightforward information theoretic definitions
    that base on alphabet, entropy, and distortion, in order to characterize the
    capacity of input devices. This is neat, but the insight obtainable with these is
    never made very clear. Then, tutorial-like exposition follows with simple
    examples. I'm afraid that similar treatments have been proposed in studies of
    pointing, choice, and more recently in intelligent text entry (e.g., probabilistic
    decoding). I’d like to hear how the paper differs from those.

    After this part, I’m afraid, the paper falls short from the promised goals in
    three ways.

    First, the introduced ”costs” in the denominator of the key formula are vague. It
    is not clear what it means to divide bits with, say, task completion time or
    workload, as the authors suggest. Perhaps this could be clarified in the rebuttal.

    Secondly, in the end, the grand contribution remains ambiguous. I am not sure what
    the obtained scores imply of ”value of humans to computers” . What would have been
    obtained by other means? I wish the authors would return to develop a broader,
    general point about ”data intelligence”, or what this work means for applications
    of information theory in HCI.

    Third , the application of this framework remains unclear. The given examples in
    the "empirical" section are hand-crafted and I failed to see neither interesting
    findings nor a general procedure that would be replicable and rigorous. The
    examples that are given are often related to simple input interactions. I’m at
    loss what this all implies to ”data intelligence”.

    In sum, while I commend the general ambitious of this paper, I believe that more
    work is needed in theory, application procedure, and empirical work. I encourage
    the authors to keep working on this topic.

    Post rebuttal: I thank the authors for a thorough rebuttal. Reading through it, I
    believe that more work is needed to articulate the theoretical insight and the
    practical value of this work which goes beyond existing expository accounts of
    InfoT. The authors offer many directions to improve the paper in the rebuttal,
    many of which remain cursory. I believe that a major revision is needed.
----------------------------------------------------------------
reviewer 3 review
score 2/5

  Expertise
    Expert
  Recommendation
    Possibly Reject: The submission is weak and probably shouldn't be accepted, but
    there is some chance it should get in; 2.0

  Review

    Significance of the paper's contribution to HCI and the benefit that others can
    gain from the contribution: ?
    The paper addresses an important but challenging topic, how to measure the value
    of the information provided by human users. The formal information-theoretic
    representation of the problem is a sensible, and valuable approach.  I would like
    to see more of this style of analysis being accepted at CHI, and there are
    inspirational elements to the paper. However, once the authors try to get more
    detailed, the formal clarity dissipates. The discussion around the Data processing
    inequality appears convoluted. Of course the human is a source of external data
    for the system, and that needs to be taken into account.  This is in no way
    'breaking the DPI'.

    The claims around the benefit of HCI to the computer are also a red herring. The
    computer being able to reduce its uncertainty of the user's intentions is an
    essential feature of the computer being able to mean the user's desires, so
    reducing the uncertainty in the computer enables us to maximise the utility for
    the user.

    The authors step us through basic inference of statistics at great length, for a
    relatively simple trial system, but the authors do not demonstrate any convincing
    examples of practical measurement of human input in a realistically complex
    setting.

    Presentation clarity;
    The figures are often unclear. E.g. in Figure 1 the arrow into P1 claims to be
    interaction but there is only an arrow from the user, not one to them - so is P1
    only affected in an open-loop manner? Is that possible? Labelling this as 'HCI
    breaks the conditions of DPI'  is just nonsense - the human is just another source
    of information and processing! If you replace/simulate the human with a separate
    computational model, what would that look like?

    The term 'data intelligence' although used as a buzz word in industrial sales
    pitch, appears imprecise in an academic context. This looks like basic statistical
    inference - why not keep the terminology clean and standard?

    The paper has many spelling and grammatical errors, and would have been better
    proofread thoroughly. At times this becomes distracting for the reader

    Originality of the work, and relevant previous work:
    The paper comes over as an textbook introduction to the application of information
    theory, and builds on recent work by Chen and colleagues, but the review and
    terminology tend to be limited to an information theoretic vocabulary. Many of the
    same principles are there in standard Bayesian models where the prior
    distributions for input movements are coupled with the sensor model distribution.
    More of this literature should have been included.  (along with recognition of the
    challenges involved in specifying such distributions)

    In conclusion, I think that the authors have got quite a few interesting elements
    in the paper, but that it is not yet mature enough for publication at CHI. There
    needs to be a tightening up of the argumentation and justification of the work,
    clearer figures which represent how the information processing aspects of the user
    are coupled into the computational elements. I think you then need a more
    convincing application which can be controlled empirically, and test the validity
    of the proposed measures. (You could do this with a computational model replacing
    the user, where you knew theoretically how many bits were being provided by that
    agent, then seeing if your estimates from empirical observations are in line with
    the capacity of the simulated agent).


    REBUTTAL RESPONSE
    After reading the rebuttal I am if anything less keen to see the paper published
    in this form. My comment on the DPI was really about it being either 1. obvious,
    given the nature of human input, and the DPI being about 'fully automated' input,
    as mentioned in the paper, so why discuss it? or 2. a misspecification of the
    sources of information in the first place. In either case I just didn't find it a
    particularly helpful setup for the paper.

    The response on the Bayesian aspects was not particularly informative

    There was no response to the issues around the benefit to the computer being a
    red-herring. It is important when formalising these things, to make clear that the
    benefit to the computer in reducing the uncertainty about user intentions will
    feed back to benefit to the user in better performance, so it is completely
    aligned with benefit for the user.
----------------------------------------------------------------
\end{Verbatim}
\normalsize

\section*{Supplement C: Author Rebuttal for the ACM CHI2019 Submission}

\small
\begin{Verbatim}[breaklines]
Thank you for comments. We will improve the paper accordingly regardless if it would appear in CHI or arXiv. R2 & R3 are knowable about information theory (InfoT). We are pleased that they did not point out any serious theoretical flaw except that R3 disagrees with the DPI statement (see E below). R2 is uneasy about the explanatory style of writing. This is understandable from the InfoT perspective. The paper was meant for an HCI audience. If possible, 1AC and 2AC may advise the style and content from the HCI perspective.

Below we focus on the main points of 1AC.

1.

A. The novelty of the work can be seen from what is missing in the HCI literature, e.g.,
(a) the need for a simple and effective theoretical framework for analysing humans’ contributions to machine processes; and
(b) the need for methodologies for estimating human knowledge quantitatively in practice and for measuring such qualities in empirical studies.
Addressing such lacks will of course take years or decades. This work is a non-trivial step towards this goal. The mathematical simplicity is a merit rather than demerit. See also L30-38, L120-132.

B. The work is built on the theoretical advances in VIS, but focus on HCI. Sec.3 relates to [6] and Sec.4 relates to [5]. [34] (VAST 2016 best paper) is a realistic application that supports Sec.5 for estimating human knowledge in interactive ML. A recent lab. study [16] supports Sec.6. This work brings these together to provide HCI with a coherent InfoT framework. The examples in the paper are designed to be easily understandable without application-specific explanations.

R2 considers Sec.6 hand-crafted, possibly due to a mix-up between applications and lab. studies. A lab. study is meant to be "hand-draft" or controlled in order to study a phenomenon with statistical significance. 1AC and 2AC may revisit this comment.

C. While it is not easy to conduct another substantial lab. study, it is relatively easy to describe another real-world application by (a) replacing the LaTex example in Sec.5 or (b) including one or two in the supplementary material.

To improve the connection with data intelligence, we will add in Sec.7:

"Although this work uses relatively simple HCI examples to illustrate the concepts of measuring information, it is not difficult to extrapolate such interactions to those in a variety of data intelligence processes. For instances, one may estimate the "values" of selecting a set of statistical measures, changing the parameters of a clustering algorithm, choosing keys for sorting, selecting an area to zoom in in visualization, reconfiguring layers in machine learning, accepting or discarding an automatically generated recommendation, and so on."

2.

D. The works inspired by Fitts & Posner focus on psychological or perceptual responses to some stimuli (i.e., without thinking). These models cannot be used to estimate human knowledge used in HCI processes.

E. R3 is uneasy with the statement "HCI breaks the conditions of DPI", possibly due to the overlook of the word “conditions”. The DPI theorem is correct but has conditions. Any defined condition must be breakable. If HCI could not in general break such conditions, we should seriously ask when these conditions would be broken. If such conditions could not be broken at all, DPI would be incorrectly defined and proved.

Further, (a) DPI is formulated based on a definable input space, while the human knowledge and ad hoc sensing of new variables is rarely included as part of the input space for any DP application. (b) If a human is replaced by an algorithmic model, and if this model can access the information discarded by earlier processes in the pipeline, the proof of DPI cannot be obtained. (c) If humans are allowed to add any arbitrarily new information into any process in a workflow, the corresponding DPI has to assume that the input space consists of infinitely all possible variables. This renders DPI meaningless in practice. These arguments have been validated by a number of InfoT experts.

F. R3 correctly noticed some structural resemblance with Bayesian models used in ML. A Bayesian network assumes that knowing the probability distribution of an input is good enough. Whenever a computer requests a user input in HCI, it assumes that it does not know the answer. There are also hypotheses that human mind might be similar to a Bayesian network, CNN, RNN, etc. The proposed framework is not biased towards any of such hypotheses.

3

G. 2AC on cost-benefit. Very good question as we have not found an appropriate intuitive description so far. The suggested wording with "capacity” and "efficiency" is good as long as the term "capacity" is defined as the capacity of alphabet transformation rather than just communication. Action capacity is a simplified case assuming no reconstructive distortion. It is good for F2 in Fig. 3(c), but not for F1. That is why Action capacity and DU do not capture the amount of knowledge accurately. See also L617 - L657.

H. Additional display space. We will add "e.g., textual instructions for aiding multiple choices."

I. The ideal measure of cost is energy (unit: Joule). See also L605-616.

J. Intelligent text entry is a mean for improving the cost-benefit of HCI using the underlying probabilistic distribution. It also relates to L450-453.

K. The term of data intelligence is defined at the beginning of the paper. If there is a better encompassing term, please suggest.

L. 2AC is right that human computation and interactive ML follow the same thinking. The method in Sec. 5 can be applied to these applications.

M. In CompSc, we like to emphasize the benefit of computing. R3 refers to user inputs as users' intentions/desires. Intentions are just one category of variables that the computer do not have. HCI can reduce the entropy of any variables that the computer do not have or are out of date. See also C. If we in HCI cannot give humans a bit more credit, who else can? (cf. Darwin's confession.)

\end{Verbatim}
\normalsize
\section*{Supplement D: Reviews for the ACM CHI2023 Submission}

\small
\begin{Verbatim}[breaklines]

Dear Min Chen,

Unfortunately, your paper: The Value of Interaction: An Information-Theoretic Perspective (number 9602) has been rejected from CHI2023. 

All papers that met a minimum threshold (at least one reviewer recommending: RR, Accept or RR, or Accept) have been put through to round 2. Any paper that did not receive at least one of these recommendations did not meet the threshold for entering round 2. This typically means that, although there could be merit for this paper in the future, reviewers thought that the paper could not be sufficiently improved within the five-week revise and resubmit period. Your reviews are below, and also will show in PCS in a few days. If you are interested in future publication, you could consider submitting to CHI2024 (or another conference) after making substantial changes to your paper.

CHI2023 received 3182 submissions. 234 papers (7.4%) were withdrawn or quick rejected before going out to review. 1530 (48.1%) received at least one R&R (or above) recommendation and were put forward to round 2. 1384 papers (43.5%) did not receive at least one R&R (or above) recommendation, and did not meet the threshold for entering round 2.

We wish you the best of luck with future submissions relating to this work.

Stefanie Mueller, Julie Williamson, and Max L. Wilson
CHI 2023 Papers Chairs
----------------------------------------------------------------
1AC review (reviewer 4)

  Expertise: Knowledgeable
  Originality: Medium originality
  Significance: Medium significance
  Rigor: Medium rigor
  Recommendation: I recommend Reject

  1AC: The Meta-Review

    This submission presents an information theoretic approach to explain the value of
    human input in an interaction.

    Reviewers found the general direction of the research interesting and relevant to
    the community. However, reviewers have raised a number of issues to be addressed
    and further clarified.

    #information theory
    Reviewers found the link to information theory weak, and also the implementation
    of “information theory” in the modeling proposed in the submission could benefit
    from further details and careful consideration.

    #Contextualizing the work
    Reviewers have pointed out that the motivation, specifically a convincing argument
    for another “metric for value” needs to be presented. Also reviewers think
    empirical evidence of how the proposed methods work for diverse real HCI datasets
    would strengthen the submission. Also, perhaps in relation to the “real world HCI
    dataset”, reviewers point out that human errors are an important factor in models
    for HCI, which the model does not take into account yet.

    #Missing related work
    Please refer to the individual reviews

    Reviewers have provided detailed feedback in their individual reviews, and I
    encourage the authors to read them carefully. I hope the authors find them useful
    for the revision.
----------------------------------------------------------------
2AC review (reviewer 1)

  Expertise: Expert
  Originality: High originality
  Significance: Low significance
  Rigor: Low rigor
  Recommendation: I recommend Reject

Review

    The authors propose an information-theoretic perspective on the value of
    information in HCI pipelines. After introducing some important measures for their
    work (based on Information-theory (IT)) which would allow them to quantify the
    costs and benefits of interaction, the authors explain how these measures could be
    applied analytically, empirically, and on any HCI pipeline including ML workflows.
    The work is original, and in line with recent applications of information theory
    to HCI.

    Overall, I am negative about this paper:
            1) the writing of the paper is poor. The text is very poorly structured;
    sections like 5, 6 and 7 are extremely indigestible.
            2) The link with information theory is very loose: I provide many details
    about misuses of information-theoretic vocabulary and notions in my detailed
    review, but the biggest issue is that no channel is ever introduced. Capacity is
    defined as the maximum value of mutual information between input and output of the
    channel --- meaning it is affected by errors (the more errors in the channel ,the
    lower the capacity). Considering that in HCI a human operates devices and makes
    many errors (see e.g. [R2]), it is very curious that the authors introduce
    capacities without ever mentioning errors (or even output). The errors are
    introduced in an ad-hoc manner in Eq 3, via a KL divergence.
            3) Not considering errors, as well as not considering the dependence
    between consecutive symbols in user input (e.g. for a sequence of mouse positions,
    two consecutive samples are not independent) leads to gross overestimations of
    entropies.
            4) On the whole, there is little that is information-theoretic about the
    core of this paper. Yes, the content is wrapped in IT vocabulary, but the same
    arguments could be made by simply "counting" things (I.e. counting the number of
    pixels on a screen, counting the number of radio buttons or boxes). This shows
    because only the definition-level metrics, such as entropy and divergence are
    used. Fundamental results, like DPI, channel capacity theorem, or even more
    complex metrics than entropy such as mutual information, conditional mutual
    information, conditional entropy, cross entropy etc. are not used. Conversely,
    most of the IT terms are actually misused.
            5) An alternative to the main formula Eq 3. could have been proposed
    without recourse to information theory (it would just not have any log transforms,
    and not be expressed in bit).

    To summarize, while applying IT promises a principled way of handling errors and
    various input prior probability distributions, Eq.3 and the work in general does
    not actually follow an information-theoretic derivation -- hence the handling of
    errors and input distributions ends up being rather arbitrary.
    For these reasons, I recommend a reject of this paper.
    	
    # Introduction

    * a lot of focus on the DPI, which is not used afterwards.

    # Related work

    * Authors could cite other successful applications of information theory in HCI
    context [R1--R6], but there are more.

    # Fundamental measures

    * "From every human action during when a user interface or an HCI device, a
    computer receives some data" word(s) missing?

    3.2 & 3.3:

    * It would be more clear if the authors could refer to input and outputs from the
    computer-to-user channel or to the user-to-computer channel. Also, it would be
    clearer if the authors drew a schematic of the various components, as done
    typically in IT (what is the source of information, what are the various encoding
    processes, what is the channel etc.).

    * The authors introduce capacity in 3.2 and 3.3, yet no channel has been
    introduced.

    3.2 Input device alphabet

    * The use of Cdev is misleading in my opinion, and this value can never be
    reached, because consecutive readings of the mouse are not independent (Chain rule
    for entropy, 2.48 in Cover and Thomas 2006 edition: H(X_1, X_2, ... X_n) =
    sum_{i=1}^n(H(X_i|X_{i-1}, ..., X_1) where H(X2|X1) != H(X2). The fact that the
    computer does not actually exploit that dependence, and encodes data as if the
    received samples were in fact independent does not increase the capacity of the
    channel.

    * "While the notion of instantaneous device capacity may be useful for
    characterizing input devices with which a sampling process has to be triggered by
    a user’s action, it is not suited for input devices with a continuing sampling
    process". This is actually incorrect; in fact the first definition of capacity in
    Shannon's seminal work occurs for an analysis of a teletype channel which is a
    discrete space continuous time input, yet the capacity is expressed as lim
    log(N(T))/T i.e. in bit/second.

    * The use of the term "bandwidth" is in line with its misuse in HCI but contrary
    to information-theoretic vocabulary (where bandwidth is expressed in Hz, as the
    width of bands of frequencies that can be used to transmit information, in line
    with vocabulary of signal processing from which the capacity in bit/s draws). In
    IT, capacity is both used in bit or bit/second. This issue was already addressed
    in [R1]. Given the fact that the paper claims an IT perspective, the vocabulary of
    IT should be used.

    * inconsistency between bit and/or bits. Should be bit, but bits is also accepted
    in practice. Either way consistency is key.

    3.3 Input action alphabet

    * "As shown in Fig. 2(b), if a TV offers only one optional answer, the device
    capacity of the corresponding alphabet, C act , is of 0 bits". Yet 1-letter
    alphabets are commonly used to signal information to the user and wait for its
    acknowledgment. Interestingly, 0 bit information alphabets can be used to convey
    information in practice. This is a case where the analogy between IT and
    interaction doesn't necessarily work well.

    * While I appreciate the comparison with radio buttons, the comparison is not as
    meaningful as the authors pertain it to be. Because errors are not accounted for,
    the capacity is not meaningful. Yes there are potentially about 2^18m possible
    paths, yet no human could reliably draw any of those if it were mapped to an
    actual command. To make this feasible, the space of possible paths would have to
    be cut into sub-spaces of some sort which the user could navigate reliably, but
    this would necessary mean a fewer paths, hence lower capacity. For example, the
    different paths could be N different tunnels through which the user could steer
    through, but that would mean only a capacity of log2(N) bit/second. There are
    probably better ways, but the best one can do is related to the skill of the user,
    in particular by how much it is able to reproduce reliably some signals, and
    likely much much (much!) lower than 18m.

    3.4 Input device utilization

    * DU: at first sight, the quantity could be mistaken for spectral efficiency (see
    [R1], Subs 2.5), which is expressed as the ratio between a rate ( in bit/s divided
    by time and bandwidth). But this is due to the fact the authors use non standard
    vocabulary. What the authors define as DU is simply the ratio between two entropy
    rates (which follow each other). Again, a schematic would be much appreciated
    here, to show the serial nature of Input Action --> Input Device.

    "However, there may also be an uncomfortable sense that the device utilization is
    typically poor for many input devices and HCI tasks. One cannot help wonder if
    this would support an argument about having less HCI." Again, the comparisons here
    are misleading. That being said, how would this even support an argument for less
    HCI. The point here is that the rate of communication between a user and a
    computer is limited. So either the solution is to remove the human (but while this
    is possible for some cases, this does not make sense in many situations: e.g. one
    wants to write a diary on a computer, how could the user be removed in that
    case?), or the solution is to find better ways to communicate between users and
    computers. So by all means, more (or perhaps better) HCI is needed.

    * part of the schematic I am asking for is Figure 3, but it is not alluded to in
    this section unfortunately.
    	
    Note that almost any point in this section can be made without recourse to
    information theory (except for the part where different probability distributions
    are assumed, but the authors don't exploit that anyways), and is simply a matter
    of estimating cardinalities of various discrete spaces.

    # Cost-benefit of HCI
    Overall this section is not very clear, and imprecise.

    * "the measure first considers the transformation from an alphabet before the
    processing to the alphabet after the processing" --> hard to parse

    * Alphabet compression: labeling H(Z_i) - H(P_i(Z_i)) as alphabet compression is
    not in line with vocabulary from IT. The fact that H(P_i(Z_i)) <= H(Z_i) is a well
    known property (See Problem 2.4 in Cover and Thomas, 2006 edition) that has
    nothing to do with compression. Indeed, data compression is about reducing the
    average description length of the input characters, by assigning small
    descriptions to the most frequent symbols. For a sequence of random variables
    (X_0, X_1, ..., X_n) the average minimum description length is lim (\lim{n \to
    \infty}H((X_i)^{i=1}^n)/n). So compression will not reduce the entropy of a
    sequence, it will just make sure its description length converges to its entropy.
    I think the authors should use a different word than compression to signal that
    the entropy of the vocabulary decreases, considering compression has a precise
    meaning in IT.

     * "At the end of the HCI task, the computer receives an answer from the user, the
    subsequent alphabet Z_i+1 usually consists of only one letter (e.g., selecting a
    radio button). Therefore the entropy H(Z_i+1) is 0, and the alphabet compression
    H(Z_i) - H (Z_i+1) = H(Z_i) = C_act": There are almost no tasks which end in
    0 entropy (i.e. a situation where the user can only, say, select one item, such as
    an ackowledgment box like in Fig.2b.

    * "D_cs(Z'_i || Z_i)" has not been defined. I assume it is the KL divergence
    between sequences Z'_i and Z_i. Rather than calling it distortion, which again has
    a precise meaning in IT, consider using the term divergence instead.

    * sign problem in Eq3: H(p(x)) = -\sum p(x)\log p(x) --> implies the alphabet
    compression part should be \sum (k) theta_k log theta_k - \sum_j phi_j log phi_j

    * The definition of Z'_i is unclear: is it the predicted (inferred) user input Z_i
    based on observation Z_{i+1}, something else?

    * writing Dcs as 1/2 \sum (phi + psi) log (|phi-psi|^2 + 1) is not obvious; where
    does it come from?

    # Estimating cost-benefit analytically

    This section applies Eq 3 to a hypothetical scenario.

    "On the other hand, when given any one of 10 file types (e.g., .doc, .tex, .txt,
    .htm, etc.), the user has the knowledge about whether formatting styles matter.
    There are 10 binary variables or 10 bits of knowledge available." --> I fail to
    see the link between the first and second sentence.

    # Measuring cost-benefit empirically

    This section unpacks how the measures introduced in the manuscript could be
    computed in practice.

    # Data Intelligence

    This section explains that the previously developed metrics are applicable in
    various contexts, including complex pipelines of information.

    [R1] Gori, J., Rioul, O., & Guiard, Y. (2018). Speed-accuracy tradeoff: A formal
    information-theoretic transmission scheme (fitts). ACM Transactions on Computer-
    Human Interaction (TOCHI), 25(5), 1-33.
    [R2] Gori, J., Rioul, O., & Guiard, Y. (2017, May). To Miss is Human: Information-
    Theoretic Rationale for Target Misses in Fitts' Law. In Proceedings of the 2017
    CHI Conference on Human Factors in Computing Systems (pp. 260-264).
    [R3] Liu, W., d'Oliveira, R. L., Beaudouin-Lafon, M., & Rioul, O. (2017, May).
    Bignav: Bayesian information gain for guiding multiscale navigation. In
    Proceedings of the 2017 CHI Conference on Human Factors in Computing Systems (pp.
    5869-5880).
    [R4] Liu, W., Rioul, O., Mcgrenere, J., Mackay, W. E., & Beaudouin-Lafon, M.
    (2018, April). BIGFile: Bayesian information gain for fast file retrieval. In
    Proceedings of the 2018 CHI Conference on Human Factors in Computing Systems (pp.
    1-13).
    [R5] Williamson, J., & Murray-Smith, R. (2004, April). Pointing without a pointer.
    In CHI'04 Extended Abstracts on Human Factors in Computing Systems (pp.
    1407-1410).
    [R6] Williamson, J. H., Quek, M., Popescu, I., Ramsay, A., & Murray-Smith, R.
    (2020). Efficient human-machine control with asymmetric marginal reliability input
    devices. Plos one, 15(6), e0233603.
----------------------------------------------------------------
reviewer review (reviewer 2)

  Expertise: Knowledgeable
  Originality: High originality
  Significance: Low significance
  Rigor: High rigor
  Recommendation: I can go with either Reject or Revise and Resubmit

  Review

    Originality: Though I'm not an expert in this area, I believe the development of a
    generalized approach for measuring the knowledge received through interactions is
    highly original.

    Significance: The paper is highly relevant to the CHI community as a theory-based
    contribution. However, the significance of this contribution is my primary concern
    with this submission. I'd like to see the authors expand on the practical value of
    this approach. To what end are we measuring the information gained from
    interactions? What should our goal be? To maximize the value of the information
    gained at the lowest cost? If so, why is that more important than other
    interaction goals? I recognize that theoretical contributions are valuable and we
    cannot always predict their long-term significance. However, without a clear point
    of view from the authors on what sort of things their measurements will be used
    for, it becomes very easy

    Rigor: The authors offer a thorough explanation of their ideas, that, to a non-
    expert like myself, seemed grounded in information theory.

    Validity: At times I found the author's definitions of HCI or its goals to be
    somewhat simplistic. As I state below, I believe this paper likely passes the
    validity bar for a CHI paper.

    Previous work: Work in information theory and principles of HCI are well cited.
    However, while the authors state their goal is to "understand and measure how much
    a computer needs to be assisted by its users before becoming useful, usable, and
    used", the related work section does not cite or discuss approaches in
    reinforcement learning, interactive machine learning / machine teaching, or active
    learning. Techniques in these fields all have their own mathematical ways of
    defining and measuring information gain (e.g., loss functions), and I think it
    necessary that the authors discuss how their work is related to or builds off of
    those approaches. Later in the paper, the authors discuss how the process of
    training AI models can result in information gain under the framework presented by
    authors, but do not explicitly discuss how their approach builds on or is
    different from, for instance, the way active learning measures this.

    Recommendation: Whether or not one agrees with the paper's claims, it is
    fascinating, thought-provoking, rigorous work and of high relevance to CHI.
    Unfortunately, I believe the lack of a clear point of view on what their work
    might be used for means it is not yet ready for publication at CHI. I also think a
    discussion of how their approach differs from the way in which active learning
    models value from interactions and its cost-benefit analysis of gathering
    additional labels must be addressed.
----------------------------------------------------------------
reviewer review (reviewer 3)

  Expertise: Knowledgeable
  Originality: Medium originality
  Significance: Medium significance
  Rigor: High rigor
  Recommendation: I recommend Reject

  Review

    This paper presents an information-theoretical approach to quantify and talk about
    the value of knowledge given by a (human) user in an interactive system. The
    approach is rigorous and well-defined. It is also well-explained.

    The approach builds on previously studied concepts in HCI like capacity and
    bandwidth. New formulae are derived and demonstrated. While the formulation of
    “value” is not entirely novel in the context of HCI (ref [9] in the paper), it is
    sufficiently different from prior art to warrant a full paper.

    While I appreciate the aims of this paper, I have three concerns with this paper.
    The first one is a minor issue, the two latter ones are major.

    1) (minor)

    First, related work: Recently an alternative (Bayesian) view to a very similar
    problem was proposed based on Bayesian statistics. Steyvers et al. (2022)
    presented a method for assessing human-AI complementarity. Their argument is that
    humans can be valuable in labeling tasks even in conditions where ML algorithms
    are objectively better (i.e., have higher accuracy). In conditions where
    correlation between the latent representations of humans and AI is low, there is a
    possibility for complementarity, i.e. the benefiting from working together over
    the condition of humans or AI working separately. I would have liked to learn how
    the information-theoretical view is better/worse/different from this one.

    2) (major) 

    Second, I'm not entirely sure what the practical value of the new value metric
    might be in HCI. Although the authors demonstrate how it might be computed in
    practical settings, they do not show that something insightful can be learned by
    using it. They give little empirical guidance and demonstrations for collecting
    high quality datasets. By contrast, throughput became popular in HCI, thanks to
    two facts: 1) it was associated with an experimental paradigm and 2) it allowed
    comparison of input devices even across conditions that had different target sizes
    and distances. It would be necessary to see more extensive guidance, trials and
    demonstrations for 'value' as well.

    3) (major)

    Third, evidence: I am not convinced that the metric works in a reasonable way in
    different representative conditions in HCI. One example is given, but it remains
    cursory (to say the least). I am also not convinced that it can be computed for
    realistic HCI datasets. More rigorous and extensive empirical evidence is needed.

    Taken together, I believe that more work is needed for publication. I hope that
    the authors keep working on this promising idea.

    References

    Steyvers, Mark, et al. "Bayesian modeling of human–AI complementarity."
    Proceedings of the National Academy of Sciences 119.11 (2022): e2111547119.
----------------------------------------------------------------
\end{Verbatim}
\normalsize

\end{document}